\documentclass[a4paper,fleqn]{cas-sc}

\usepackage{stfloats}
\usepackage{graphicx} 
\usepackage{float} 
\usepackage{amssymb}
\usepackage{amsmath}
\usepackage{mathtools}
\usepackage[thmmarks,amsmath]{ntheorem} 
\usepackage{bm} 
\usepackage{extarrows}
\usepackage{booktabs} 
\usepackage{tocloft}
\usepackage{epstopdf}
{
    \theoremstyle{nonumberplain}
    \theoremheaderfont{\bfseries}
    \theorembodyfont{\normalfont}
    \theoremsymbol{\mbox{$\Box$}}
    
}

\usepackage{theorem}

{
    \theoremheaderfont{\bfseries}
    \theorembodyfont{\normalfont}
    
}
\graphicspath{{figure/}}                                 
\usepackage{algorithmic}
\usepackage{graphicx}
\usepackage{textcomp}
\usepackage{xcolor}
\usepackage{diagbox}
\usepackage{booktabs}
\usepackage{multirow} 
\usepackage{makecell}
\usepackage{color}
\usepackage[ruled,vlined]{algorithm2e}
\usepackage{float}
\usepackage{url}
\usepackage[numbers,sort&compress]{natbib}

\usepackage[section]{placeins}

\begin{document}

\let\WriteBookmarks\relax
\def\floatpagepagefraction{1}
\def\textpagefraction{.001}
\shorttitle{Electrical peak demand forecasting-- A review}
\shortauthors{Shuang Dai et~al.}

\title[mode = title]{Electrical peak demand forecasting-- A review}                      



\author[1]{Shuang Dai}

\ead{sd19628@essex.ac.uk}


\author[1]{Fanlin Meng}[orcid=0000-0002-4866-0011
]
\cormark[1]
\ead{fanlin.meng@essex.ac.uk}


\author[1]{Hongsheng Dai}
\ead{hdaia@essex.ac.uk}


\address[1]{Department of Mathematical Sciences, University of Essex, Colchester, UK}

\author[2]{Qian Wang}
\ead{qian.wang173@hotmail.com}

\address[2]{Department of Computer Science, Durham University, Durham, UK}

\author[3]{Xizhong Chen}
\ead{xizhong.chen@ucc.ie}


\address[3]{School of Engineering, University College Cork, Cork, Ireland}






  \cortext[cor1]{Corresponding author}


\begin{abstract}
The power system is undergoing rapid evolution with the roll-out of advanced metering infrastructure and local energy  applications (e.g. electric vehicles) as well as the increasing penetration of intermittent renewable energy at both transmission and distribution level, which characterizes the peak load demand with stronger randomness and less predictability and therefore poses a threat to the power grid security. Since storing large quantities of electricity to satisfy load demand is neither economically nor environmentally friendly, effective peak demand management strategies and reliable peak load forecast methods become essential for optimizing the power system operations. To this end, this paper provides a timely and comprehensive overview of peak load demand forecast methods in the literature. To our best knowledge, this is the first comprehensive review on such topic. In this paper we first give a precise and unified problem definition of peak load demand forecast. Second, 139 papers on peak load forecast methods were systematically reviewed where methods were classified into different stages based on the timeline. Thirdly,  a comparative analysis of peak load forecast methods are summarized and different optimizing methods to improve the forecast performance are discussed. The paper ends with a comprehensive summary of the reviewed papers and a discussion of potential future research directions.

\end{abstract}



\begin{keywords}
peak demand management \sep peak load forecast \sep time series  \sep machine learning \sep deep learning \sep smart grid
\end{keywords}

\maketitle


\section{Introduction}\label{S1}

With the roll-out of advanced metering infrastructure (AMI) \cite{AR-5}, the power system is undergoing rapid evolution. The Office for Gas and Electricity Markets (Ofgem) has announced that the UK plans to install more than 50 million smart meters by 2020 \cite{AR-6}. On the one hand, the installation of smart meters enables real-time information exchange between power suppliers and end-users and therefore increases the efficiency of the electric power supply and encourages different smart energy applications such as demand response and demand side management \cite{Areview-1}. On the other hand, the high temporal resolution energy consumption data coupled with intermittent energy resources such as wind energy make electricity demand present unprecedented diversity and complexity.

Different electricity generation units have been adopted in the power plants to meet the specific electrical demand/ load types. Among all the units, peak load units have the lowest efficiency and the highest cost. It is estimated that a 5\%-15\% reduction in peak load would bring substantial benefits in saving resources and decreasing real-time electricity tariffs \cite{reason-6}, which calls for effective peak load management strategies. 

To realize that, being able to accurately predict the magnitude and occurring time of peak load/ demand, which can not only give the power plants sufficient start-up time to avoid grid congestion but also is fundamental in ensuring
the economic benefits and the security and stability of the power grid. With the increasing penetration of large-scale intermittent energy such as wind and solar as well as energy storage power station, it has given rise to new characteristics of peak loads and a more challenging task for peak load/ demand
forecast.

In such a context, it is evident that accurate peak load demand forecast becomes essential element to the power grid operations \cite{reason-10}. Although optimizing smart grid operation based on standard load forecast has been a long-held principle \cite{reason-4}, the new digital and smart grid era calls for more attention to build flexible peak load forecast frameworks to adapt to the rapid development of the power system.

\subsection{Motivation and contributions of this review} \label{II}

Instead of the continuous and stable power generation, peak load power plants only run for a short time over a year, which is neither economical nor environmentally friendly. Peak load management strategies were therefore proposed to reduce peak load generation costs based on the incentive and punishment mechanism and programs, such as interruptible load control, demand-side bidding, and emergency demand response \cite{saebi2010demand} \cite{cappers2010demand}. Moreover, it is important to know reasonably well the future peak load demand in order to plan and trigger relevant peak demand management strategies and mechanisms. Therefore, accurate and reliable peak load forecast is crucial for materializing any peak demand management strategy.  

In general, a mature peak load forecast can help the system operator to manage the peak load demand effectively in advance, and thus to achieve demand response to help reduce greenhouse gas emissions and decrease non-renewable fuel reliance. Moreover, the ultimate goal of peak demand management and forecast is to balance electricity supply and demand \textcolor{black}{to} maximize the benefits of system. Therefore, to further highlight the motivation of the peak demand forecast, TABLE \ref{t1} lists key stakeholders (grid operators, electricity retailers, electricity end-users, government) and the impact of peak load demand forecast on them \cite{koh2015evaluating} \cite{sardi2017multiple} \cite{mao2009short}. 

\begin{table*}[!h]
    
    \caption{The importance of peak load demand forecast for electricity stakeholders}
    \begin{center}
    \resizebox{\textwidth}{0.13\textwidth}{
    \begin{tabular}{|c|c|l|}
        \hline
        \multicolumn{2}{|c|}{Electricity market stakeholders} &
        \multicolumn{1}{c|}{Importance of peak load demand forecast} \\ \hline
        \multicolumn{2}{|c|}{Grid operators} &
        \begin{tabular}[c]{@{}l@{}} $\bullet$ Improve the utilization rate of power generation equipment\\ $\bullet$ Reduce the cost of power generation and investment in power facilities \\ $\bullet$ Alleviate  the supply pressure of the grid during peak hours\end{tabular} \\ \hline
        \multicolumn{2}{|c|}{Electricity retailers} &
        \begin{tabular}[c]{@{}l@{}}$\bullet$ Make reasonable tariff schemes so as to maximize profits \\ $\bullet$ Offer energy-efficiency rebates to encourage customers to reduce peak load demand \end{tabular} \\ \hline
        \multirow{2}{*}{End-users} &
        \begin{tabular}[c]{@{}c@{}}Commercial and industrial \end{tabular} &
        \begin{tabular}[c]{@{}l@{}}$\bullet$ Improve the economic benefits and save production resources \\ $\bullet$ Alleviate environmental pollution by distributing emissions concentrated in the peak hours \end{tabular} \\ \cline{2-3} 
        &
        \begin{tabular}[c]{@{}c@{}}Residential \end{tabular} &
        Save electricity bills and improve their living standard \\ \hline
        \multicolumn{2}{|c|}{Government} &
        \begin{tabular}[c]{@{}l@{}} $\bullet$ Enable a reliable power supply system \\  $\bullet$ 
            Ensure economic growth and social welfare \end{tabular} \\ \hline
    \end{tabular}}
    \end{center}
    \label{t1}
 \end{table*}

Based on the above analysis, the objective of this review is to provide a clear and comprehensive overview of the peak load demand forecast methods in the literature. To the best of our knowledge, this should be the significant review on the topic of peak load forecast. In particular, the key contributions of this review are described as follows. 
\begin{itemize}
    
    \item{We give \textcolor{black}{precise} definitions of key factors relevant to the peak load forecast framework, which could serve as a useful standardization and guidance for future research in this area.}
    
    \item{We conduct a thorough review of peak load demand forecast methods and explores hybrid forecast models from a historical and systematic point of view.}
    
    \item{We provide a comparative analysis based on existing studies and discuss potential improving methods for peak load demand forecast. A comprehensive summary \textcolor{black}{regarding} the application scopes of the reviewed methods is also presented, which could provide useful insights for future research directions.}
    
\end{itemize}

\subsection{Literature retrieval strategy} \label{LRS}

\textcolor{black}{Before} a detailed overview of peak load demand forecast methods, a necessary initial step is to follow the standard criteria and protocols to select highly related and high-quality sources and publications. 

The literature retrieval databases \textcolor{black}{selected} are ScienceDirect (SD) and the Institution of Electrical and Electronics Engineers (IEEE). SD is a famous academic database provided by Elsevier, in which more than a billion articles are downloaded every year, making it the most downloaded academic search platform among academic databases \cite{SD}. IEEE publishes a wide range of peer-reviewed journals, and the criteria defined are of recognized authoritative influence in the field of electrical power analysis \cite{7-IEEE}.

The following key phrases are used during the literature retrieval process (searching range of the year: 1872-2020):

\begin{itemize}
    \item Peak load forecasting/estimation/prediction
    \item Peak load demand forecast/estimation/prediction
    \item Maximum load forecasting/estimation/prediction
\end{itemize}

The keywords in each key phrase utilize the Boolean operator 'AND'\textcolor{black}{;} each key phrase \textcolor{black}{is} connected with the Boolean operator 'OR'.

After excluding low-relevance articles without key phrases in the title and abstract, a total of 139 highly related and high-quality papers form the basis of this review through a preliminary analysis. The obtained studies consist of 67 journal papers and 72 conference papers. The subsequent discussions of peak load demand forecast are all based on the literature obtained in this section.

\subsection{Systematic overview of literature based on time line}

\begin{figure*}[htbp]
    \centering
    \includegraphics[width=1\textwidth]{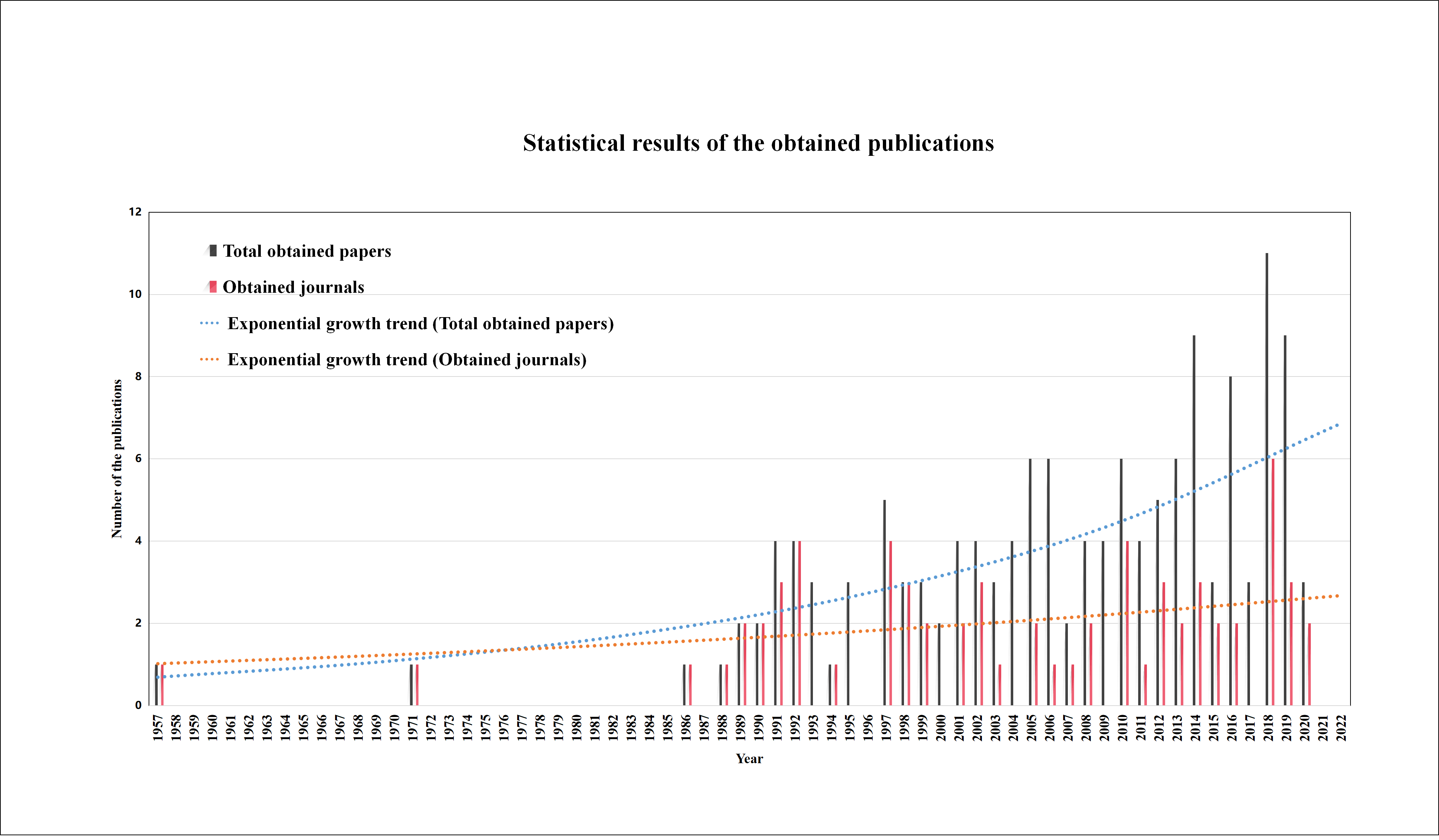}
    \caption{The number of papers published every year from 1956.}
    \label{f1}
\end{figure*}

\textcolor{black}{To} understand the historical development trend of peak load demand forecast, important to follow the timeline to conduct a systematic review. Figure. \ref{f1} shows the the number of total publications and journal publications published every year from 1956 to 2020.

Built on the exponential trend of obtained publications, the exploration of peak load forecast can be roughly categorized into three stages following the \textcolor{black}{timeline}: the initial stage, the developing stage, and the developed stage. The initial stage was from 1956 to 1990, during which the research on peak load demand forecast was in its infancy with a small number of publications. The strengthening phase was from 1991 to 2003, during which the number of publications began to increase gradually, with three or four articles published every year. The developed period is from 2004 to 2020 with a large number of publications on peak load forecast. 

The large number of publications over the past decade reveal that there are increasing interests on peak demand forecast. This could be explained by the fact that with the economic development, there are increasing electricity consumption. As a result, peak load forecast becomes increasingly important for safe and reliable energy systems operation. \textcolor{black}{Moreover}, considering increasing integration of modern and clean energy technologies such as electric vehicles (EVs) and wind energy, it would be become more challenging for the peak demand forecast and the research interests on the topic will continue to grow in the future.

It should be noted that the number of journal publications on peak load forecast each year over the last decade is usually within the range of 1 to 4, which may indicate there still lacks sufficient efforts from the researchers but on the other hand indicate more research opportunities. This paper will provide a timely review on the important topic of peak load forecast with a clear definition of the research problem, comprehensive review of existing methods and a comparative and forward-looking analysis of future research directions.

\subsection{Structure of the review}

The remainder of this paper is organized as follows: Section \ref{S2} provides comprehensive summaries and precise definitions for the peak load forecast problem including the forecast period, influential variables, general outputs, and evaluation metrics. Section \ref{S3}  describes peak load demand forecast methods following the timeline by dividing them into manual/human expert stage, classic peak load demand forecast stage, and advanced peak load demand forecast stage. Section \ref{S4} firstly gives a comparative analysis and \textcolor{black}{explores} possible improving methods for the peak load demand forecast framework. Then, a comprehensive summary of existing studies on the peak load forecast will be presented. In Section \ref{S5}, a conclusion is given with possible future research directions discussed.

\section{Peak load demand forecast problem definition}\label{S2}

\begin{figure*}[htbp]
    \centering
    \includegraphics[width=1\textwidth]{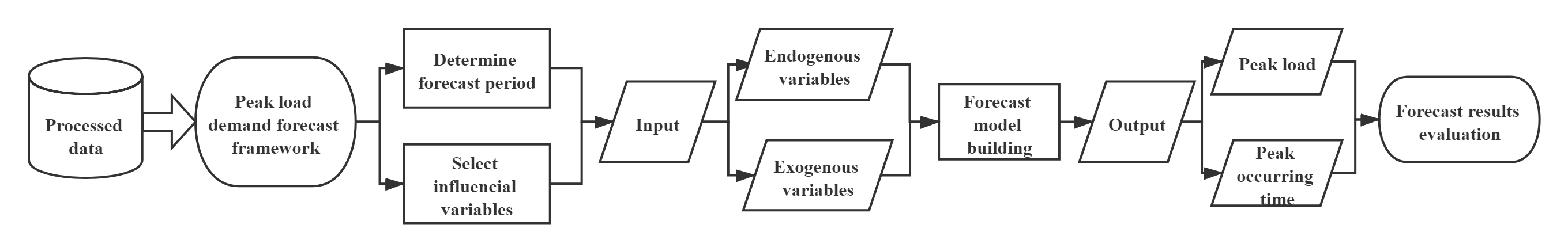}
    \caption{A general peak load demand forecast framework.}
    \label{f2}
\end{figure*}

A general peak load demand forecast framework is shown in Figure \ref{f2}. Intuitively, the general framework for peak load demand forecast is similar to standard load forecast. However, peak load demand forecast has its particularity when it comes to specific sub-processes, such as input variables and output results. To our best knowledge, many terms that have been defined in the standard load forecast have not been well defined in the peak load demand forecast, which leads to different understanding of the same terms in different studies. To this end, for the first time, this paper will provide an unification of relevant terms to accurately define peak load demand forecast methods and to provide generalized guidance for future research on this topic.

The following subsections will first summarize the commonly used time horizon for short-term, medium-term, and long-term peak load demand forecast according to the reviewed literature. Secondly, influential variables used in peak load demand forecast models will be discussed. Thirdly, the outputs of the forecasting model will be summarised. Finally, special evaluation indicators for peak load demand forecast results are presented.

\subsection{Peak load demand forecast time period}
Although the forecast horizon of standard load forecast has been well known, there is no such summary for \textcolor{black}{peak load demand forecast.}

Therefore, through analysing the reviewed literature, we classify \textcolor{black}{the} time horizon of peak load demand forecast into following categories:

\begin{itemize}
    \item Short-term peak load demand forecast (STPLF), to forecast peak load from several hours to days (days$<$7) \cite{46-2005Probabilistic,56newd-2006Developed,70-2008Density,81-6121762}.
    \item Medium-term peak load demand forecast (MTPLF), to forecast peak load from per week to months  (months $<$ 12) \cite{46news-2005Short,46new1,81-6121762}.
    \item Long-term peak load demand forecast (LTPLF), to forecast peak load from more than a year ahead \cite{46news-2005Short,46new1,70-2008Density,81-6121762}.
\end{itemize}

It is worth noting that based on the reviewed papers: 1) daily peak load demand forecast is mainly studied among STPLF; 2) weekly and monthly peak load demand forecast is mainly studied among MTPLF; and 3) annual peak load demand forecast is mainly studied for LTPLF.   

\subsection{Influential variables of peak load demand forecast} \label{Input}

\subsubsection{\textbf{Endogenous variables}}

The endogenous variables used in peak load demand forecast differ from those used in standard load forecast. For example, assume that the training data are hourly load consumption for one year. A standard load demand forecast model will use the hourly load data, i.e. $24\times365$ data points, as the endogenous variables. However, a peak load demand forecast model will use the daily peak load value (sometimes also with the daily peak time), i.e. $1\times365$ data points ($1\times365$ data pairs if with the daily peak time), as the endogenous variables.

The endogenous variables used by peak load demand forecast models are generally peak load data in similar days, which can often reflect the internal structure similar to the peak load in the forecast period, making it easier for the algorithm to capture the characteristics of the predicted target. \cite{126-2019Deep} proposed novel algorithms to identify the recent days that are similar to the days before the forecast, and the peak load close to the predicted date is then deduced as historical training data by analogy with the rule of thumb to improve the prediction accuracy. 

Furthermore, since the input data only need the peak load in a specific period, the input variable dimension of peak load demand forecast is much lower than that of standard load forecast. \textcolor{black}{The} advantage of this is reflected in the high computational efficiency of peak load demand forecast model. \cite{30-Amjady2001Short} compared both the number of input features and the computation time of hourly load forecast and daily peak load forecast based on the same historical data. The statistical analysis showed that only \textcolor{black}{six} input features were needed for the daily peak load demand forecast, while 171 input features were necessary for the hourly load forecast.

\subsubsection{\textbf{Exogenous variables}}

TABLE \ref{t2} summarizes the exogenous variables that are frequently used in peak load demand forecast models.

\begin{table}[htbp]
    \caption{Popular exogenous variables for peak load demand forecast models}
    \begin{center}
            \begin{tabular}{cc}
                \hline
                \multicolumn{1}{l}{}                                                     & Detail                                                                                                                                                                                                                                                                                                                     \\ \hline
                \begin{tabular}[c]{@{}c@{}}Weather variables\\ \cite{44-1412874} \end{tabular} & \begin{tabular}[c]{@{}c@{}}Maximum dry-bulb temperature, average dry-bulb temperature, \\ minimum dry-bulb temperature,  average relative humidity, average pressure, \\ average amount of cloud,  rainfall volume, duration of bright sunshine, \\ daily global solar radiation,  average wind speed\end{tabular} \\ \hline
                \begin{tabular}[c]{@{}c@{}}Calendar variables\\ \cite{36-2002Artificial}\end{tabular}    & \begin{tabular}[c]{@{}c@{}}Time of the day, day of the week,  week of the month, month of the year, \\ season, year number,  holidays,  special events
                \end{tabular}                                                                                                                                                     \\ \hline
                \begin{tabular}[c]{@{}c@{}}Economic variables\\ \cite{40-1372805},\cite{44-1412874}\end{tabular} & \begin{tabular}[c]{@{}c@{}}Gross National Product (GNP),  Gross Domestic Product (GDP), \\ population growth rate,  consumer growth rate, tariff structure, electricity price\end{tabular}                                                                                                                          \\ \hline
                \begin{tabular}[c]{@{}c@{}}Other variables\\ \cite{26-1997Estimating},\cite{60-2008Special}\end{tabular} & \begin{tabular}[c]{@{}c@{}}Customer type  (commercial, residential, industrial, etc.)\end{tabular}                                                                                                                                                                                                                    \\ \hline
        \end{tabular}
        \label{t2}
    \end{center}
 \end{table}

The selection of exogenous variables of peak load demand forecast models is similar to yet different from standard load forecast. According to the table, it can be seen that the input variables of peak load are similar to load prediction on the macro level, namely temperature, humidity, etc. Moreover, the selection of input variables of peak load is closely related to the forecasting period, which is also similar to that of the standard load forecast. The commonly used variables of STPLF are weather and calendar factors. For MTPLF and LTPLF, in addition to the weather and calendar variables, it is necessary to capture socio-economic development and population growth trends. Besides, the acquisition method and accuracy of long-term weather data are also thorny problems that must be considered wisely for MTPLF and LTPLE.

On the other hand, the difference between peak load demand forecast and standard load forecast is that, since the prediction target is a series of extreme values under most conditions, the variables that most closely related to a peak load demand forecast model are the extreme variables with \textcolor{black}{the} ability to indicate the change degrees of the weather, such as the maximum and minimum temperature. In addition, some weather variables are internally related and can affect each other. For example, \cite{56-4075966} pointed out that high relative humidity in months with \textcolor{black}{apparent} seasonality (summer/winter) would lead to an increase in demand for refrigeration or heating, thus affecting the forecast accuracy of peak load demand. Therefore, in their study, relative humidity was quantified as \textcolor{black}{temperature change} to correct the inaccurate input variables, which significantly improved the forecasting accuracy of peak load demand.

Calendar variables have \textcolor{black}{a} significant influence on areas with \textcolor{black}{rare} special events and regular holidays. In \cite{36-2002Artificial}, the influence of lunar calendar festivals in Egypt on peak load was considered, and the influence of Ramadan is quantified as a weight factor and input into the expert system. The prediction results showed that models considering special festivals had better performance than others. Moreover, electricity consumption in the weekend and holidays of commercial and industrial sectors is \textcolor{black}{considerably} changed from that of working days, and the peak load may not even occur in these sectors during non-working days for the most time. Therefore, some of the reviewed works modeled historical data separately based on these calendar factors to improve the forecast performance. \cite{26news-1997Short} trained models separately for each hour of a day, and the weekend and weekdays were also considered as \textcolor{black}{criteria} for model training, which resulted in 48 independent \textcolor{black}{models} to predict morning peak and afternoon peak in a day. This time-division modeling method distinguishes between working days and non-working days, which significantly overcome the defect of the traditional model in predicting peak load on weekends. 

\subsection{Outputs of peak load demand forecast}

The main difference between peak load demand forecast and standard load forecast is that the output is usually one value or a pair of values (e.g. a peak load value with its occurring time/date) whereas the output of standard load forecast is generally a set of load values (time series). Existing studies did not make an unified definition for the output of the peak load demand forecast. By reviewing relevant literature, the output of peak load forecast are summarized as follows:

\begin{itemize}
    \item Forecast the total peak consumption on a given peak day \cite{108-7893595}
    \item Forecast load usage pattern during a given peak period \cite{62-2008Electricity}
    \item Forecast peak time \cite{128-8791587}.
    \item Forecast peak value \cite{73aa-GOIA2010700}.
    \item Forecast peak value and its occurring time simultaneously \cite{85aab-2012Building}. 
    \item Forecast peak value and forecast its occurring time separately \cite{20}.
    \item Forecast peak (or together with valley) value as an additional input to produce load profile \cite{26newc-1997Cascaded}.
\end{itemize}

\subsection{Evaluation indicators for peak load demand forecast}
The evaluation indicators of peak load demand forecast models are partly the same as those of standard load forecast models, such as mean absolute error (MAE), mean square error (MSE), root mean square error (RMSE), mean absolute percentage error (MAPE), etc. Since these indicators are well known in standard load forecast, this section will not describe them in detail. Instead, to highlight the particularity of peak load demand forecast accuracy metrics compared with standard load forecast, this section will list some special evaluation indicators used in the existing studies. In addition, the following evaluation indicators that are specific to peak load forecast are given below. 

\begin{itemize}
    \item{{\bf Evaluation indicators for peak value}}
    
    Assuming that $\widehat{y}$ is the predicted peak value, $y$ is the actual peak value, $n$ is the number of the training samples:
    \begin{equation}
    \widehat{y} = \left\{\widehat{y}_1, \widehat{y}_2,..., \widehat{y}_n\right\}
    \end{equation}
    \begin{equation}
    y = \left\{y_1, y_2,..., y_n\right\}
    \end{equation}
    
    {\textit{Peak absolute percentage error ($PAPE$) \cite{85aab-2012Building}} is defined as:}
    \begin{equation}
    PAPE = \frac{100\%}{n}\sum_{i=1}^n\mid\frac{\widehat{y}_i-y_i}{y_i}\mid
    \end{equation}
    The range of $PAPE$ is $[0, +\infty)$. $PAPE$ equals 0\% represents a perfect trained model, while $PAPE$ greater than 100\% indicates an unacceptable model. 
    
    \item{\bf {Evaluation indicators for peak occurring time}}
    
    Assuming that $\widehat{t}$ is the predicted peak occurring time, $t$ is the actual peak occurring time, $n$ is the number of the training samples:
    \begin{equation}
    \widehat{t} = \left\{\widehat{t}_1, \widehat{t}_2,..., \widehat{t}_n\right\}
    \end{equation}
    \begin{equation}
    t = \left\{t_1, t_2,..., t_n\right\}
    \end{equation}
    
    By using $\epsilon$ to represent the tolerance residual for the peak occurring time, and $h$ as a flag to represent whether the predicted occurring time hits the tolerance interval $[t_i-\epsilon, t_i+\epsilon]$, then we have, the {\textit{Hit rate (HR)}} \cite{85aab-2012Building} is defined as::
    \begin{equation}
    HR=\frac{100\%}{n}\sum_{i=1}^nh_i
    \end{equation}
    where
    \begin{equation}
    h_i=\begin{cases}1 & \widehat{t}_i\in[t_i-\epsilon, t_i+\epsilon]\\0 & otherwise\end{cases}
    \end{equation}
    
    The peak time forecasting is usually measured by $HR$ \cite{85aab-2012Building}\cite{20}, which specifies a \textcolor{black}{period} before and after the peak load \textcolor{black}{occurring} as the forecast error tolerance range. The prediction is considered to be correct as long as the predicted time falls within the tolerance interval. 
    
\end{itemize}

\section{Peak load demand forecast methods}\label{S3}

Traditional forecasting methods can be roughly divided into qualitative and quantitative analysis \cite{7-30}. Qualitative analysis refers to the use of expert opinions to develop theoretical insights for prediction, such as curve fitting and extrapolation techniques, on the premise that historical data are not available or technical experiments are not feasible \cite{7-30}. Quantitative analysis, on the other hand, presumes historical data are available and the future development of data still follows the changing trend of historical data within a reasonable range. In quantitative analysis, mathematical or statistical methods are usually used. 

\begin{figure*}[htbp]
	\centering
    \includegraphics[width=1\textwidth]{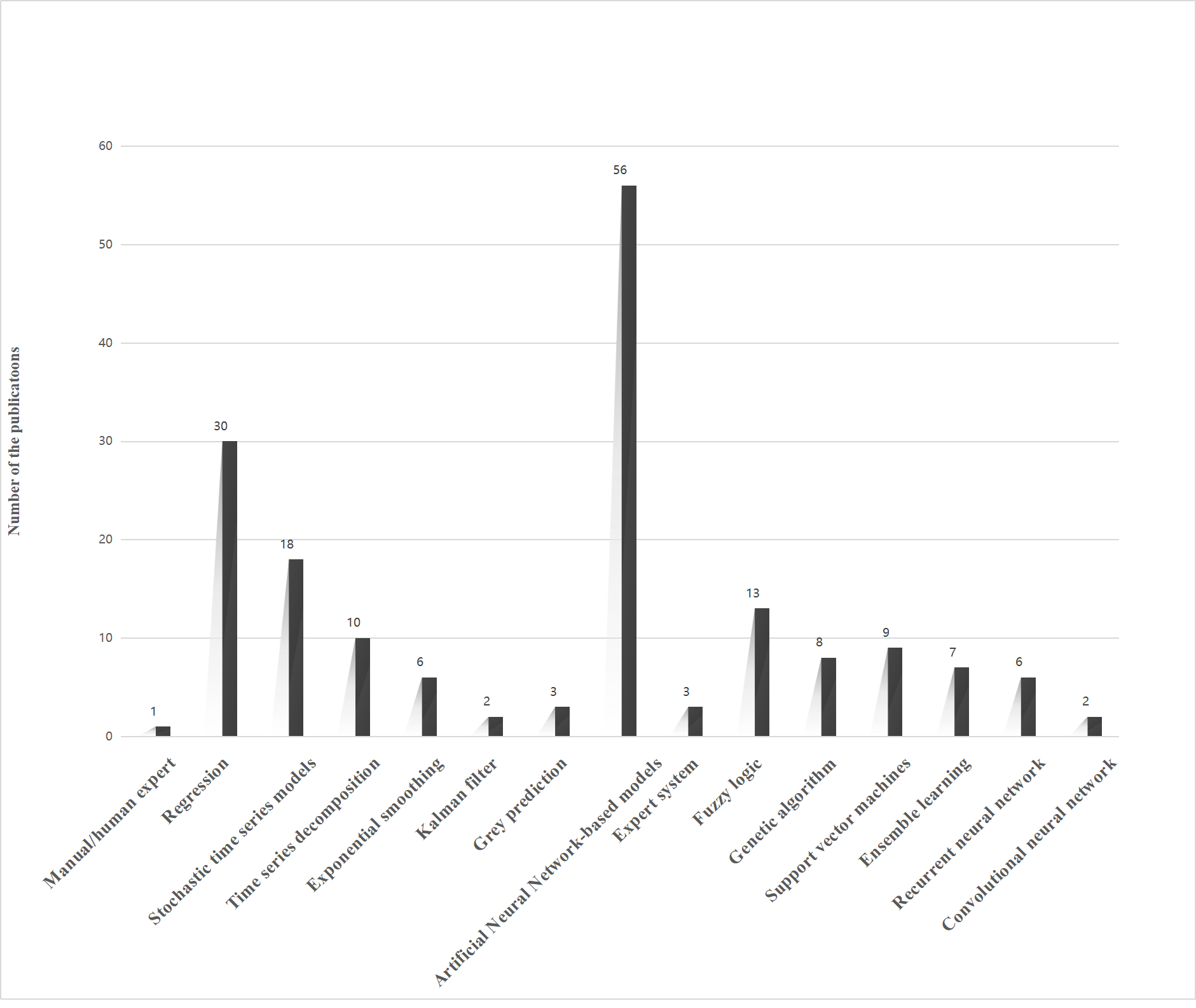}
    \caption{The number of publications for each type of methods.}
    \label{f3}
\end{figure*}

\begin{figure*}[htbp]
    \centering
    \includegraphics[width=0.95\textwidth]{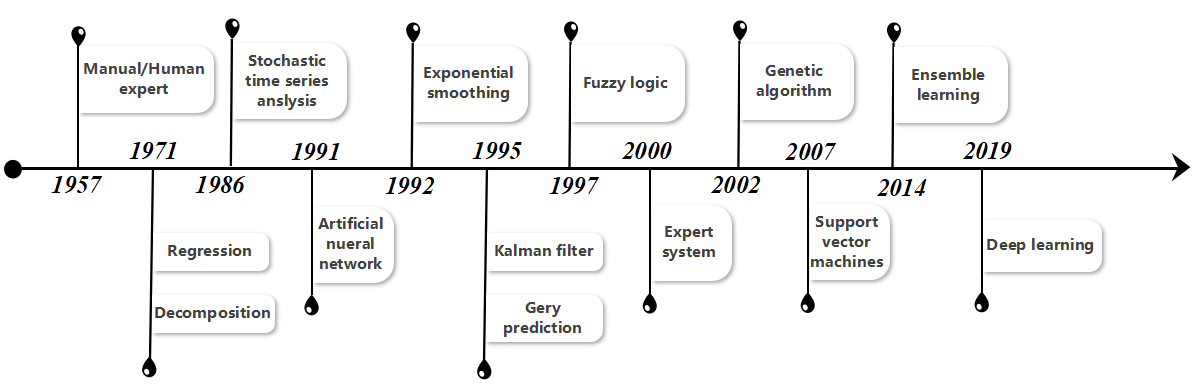}
    \caption{Time line of the first time each method was used for peak load demand forecast.}
    \label{f4}
\end{figure*}

\begin{figure*}[htbp]
    \centering
    \includegraphics[width=1\textwidth]{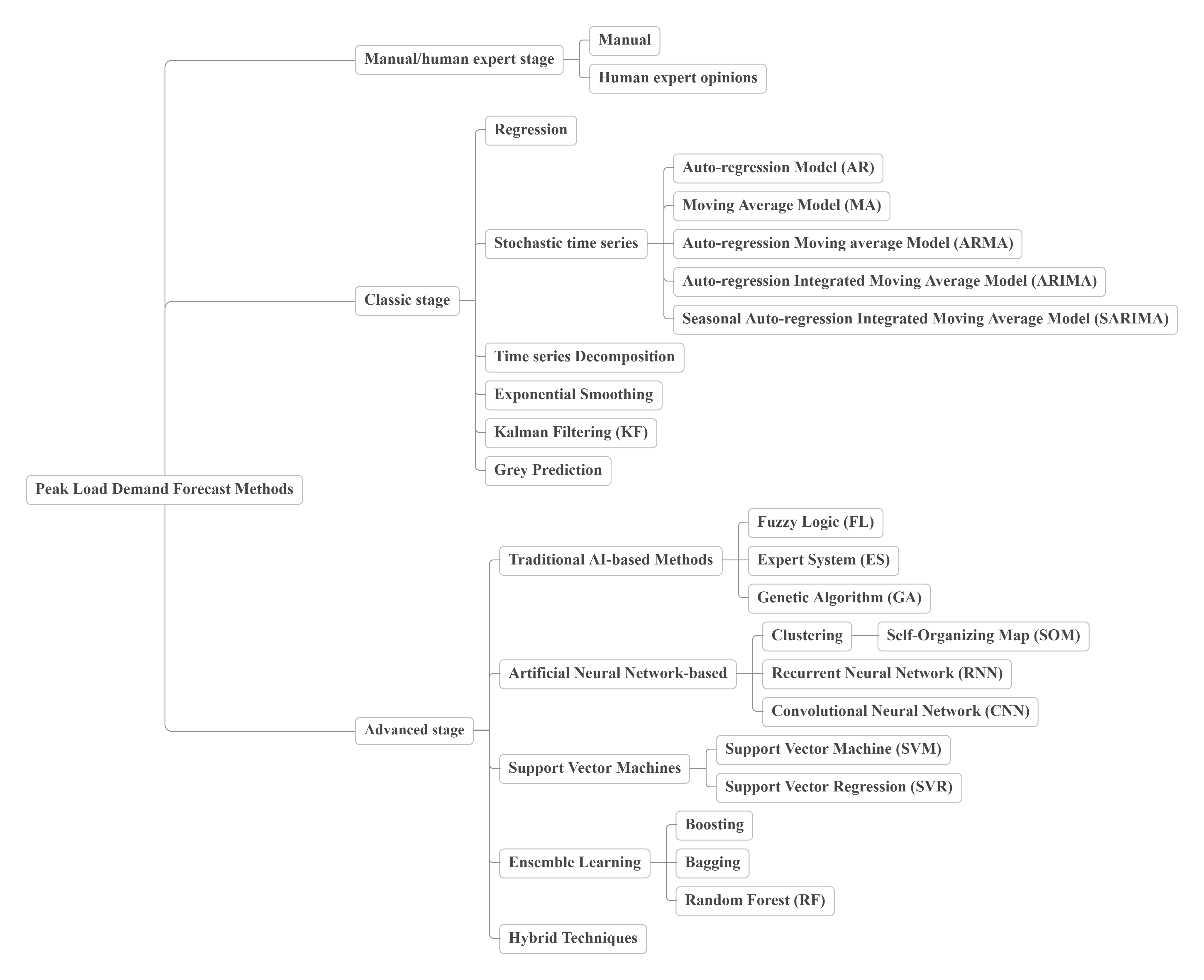}
    \caption{Three stages of peak load demand forecast methods.}
    \label{f5}
\end{figure*}

Following the literature obtained in Section \ref{LRS}, the number of publications for each method is summarized in Figure \ref{f3} while Figure \ref{f4} shows the timeline of each method being first used for peak load forecast. Finally, the peak load demand forecast methods are categorized according to the development timeline and types in Figure \ref{f5}.

According to Figures \ref{f3}, \ref{f4} and \ref{f5}, the development of peak load demand forecast methods can be broadly classified into three stages: the manual/human expert stage starting in the late 1950s, the classic stage starting in the early 1970s, and the advanced stage starting in the early 1990s. There are few studies in the manual/human expert stage. In the classic stage, regression is the most popular method for peak load demand forecast, followed by time series decomposition, stochastic time series models, and exponential smoothing. In the advanced stage, artificial neural network (ANN) based methods are the most favorable choice.

In the following we will provide a detailed review of methods in each stage.

\subsection{Manual/Human expert peak load demand forecast stage}

A manual peak load demand forecast method to transform the forecasted weather into daylight illumination parameters was proposed in \cite{1-1956A}. The obtained results were then combined with the peak load demand table to calculate peak load increment, which was then applied to load curves to forecast daily peak load demand one day ahead. However, most of the studies were based on the simple calculation of variable relations or human experts’ opinions to estimate peak load demand before 1971. Most of the results predicted by over-relying on calculation and human experience were unsatisfying, due to the special characteristics of peak load demand.

\subsection{Classic peak load demand forecast stage}\label{classic}

\subsubsection{\textbf{Regression}}

Regression analysis decides model parameters by function expression based on the historical data, thus obtains the causal relationship between $m$ explanatory variables $\mathbf{x} = \left[ x_1,..., x_m \right]$ and response variables $y$ (peak loads) to produce peak load demand forecast \cite{8-Turner2012Regression}. Mathematically, a general regression model can be represented by Eq (8).
\begin{equation}
y = f(\mathbf{x}, \mathbf{\beta}) + e
\end{equation}
\label{e1} where $\mathbf{\beta}$ represent the model coefficients to be learned from data and $e$ denotes the residual left unexplained by this model.

The above model is said to be univariate regression when $y$ describes a univariate random variable, otherwise, it can be described as multivariate regression. Besides, if there is only one explanatory variable (i.e., $m=1$), the model is simple regression, otherwise, it is multiple regression. A parametric regression model is based on the knowing form of $f(\centerdot)$, in which parameters $\mathbf{\beta}$ are \textcolor{black}{needed} to be estimated. When there is a linear relationship between parameters $\mathbf{\beta}$ and the explanatory variables $\mathbf{x}$, the model is said linear, otherwise, the model is known as non-linear \cite{8-Turner2012Regression}.

Most of the obtained papers used univariate regression and selected multiple explanatory variables to get precise forecast results. \cite{9a-1990A} proposed multiple regression-based approaches that took calendar effects into account to forecast short-term system load. The approach produced forecasts using four models: an initial peak forecast regression model, an initial hourly forecast regression model, an adjusted peak model, and an adjusted hourly model. All four models were based on regression, and the forecasted initial peak load and the maximum hourly load were combined with past errors in the adjusted peak model to produce the new peak forecast. Then the new peak forecast was used as a constraint in the adjusted hourly model to produce the final hourly forecasts. The proposed model was more flexible in handling the effects of special days and avoided causing iterative residuals for multiple-day forecasts since past forecast errors were considered as one of the influential variables. A regression-based model was proposed in \cite{94-6934706}, which considered econometric effects such as GDP, consumer price index, and population as extra explanatory variables to perform LTPLF for Zimbabwe. \textcolor{black}{ \cite{28newr-1999reg} included a new variable, average wind chill, for the winter season, and considered the holidays’ effects by using transformation and reflection techniques to produce better daily peak forecast.} \cite{19-2002Regression} adopted a multiple regression model to linearise the load trend for Tokyo Electricity Power Company. The model is simple and has a promising peak load demand forecast performance with a MAPE of 1.43\%. However, the performance of the model on a dataset with more fluctuated load patterns was not satisfactory. 

There are a few papers utilizing probabilistic forecasts in the regression framework for peak load demand forecast. Instead of providing an estimate of the peak load demand, the probabilistic forecast is capable of predicting the distribution intervals, in which the uncertainty inherent of the demand could be quantified. \cite{70-2008Density} presented a new methodology to forecast annual and weekly peak load demand. This method adopted semi-parametric additive models to estimate the correlations between predictors and the peak load demand. Same as \cite{94-6934706}, weather, calendar, economic, and population variables were also considered in the paper, and the results showed a remarkable improvement \textcolor{black}{in} the forecast accuracy. \cite{90-2014Long} modeled monthly peak load demand as a function by considering the monthly peak time and the monthly peak load as two key variables. In this paper, the Gaussian process was used to forecast peak load demand. It also proposed a method to optimize the hyper-parameters in the kernel function, which was vital to improve forecast accuracy. \cite{95-2014Non} utilized Alternating Conditional Expectation (ACE) to model hourly peak load during a month. \textcolor{black}{Unlike} most multiple linear regression models, the seasonal and trend components in this model did not require a priori decomposition, and the non-parametric transformed functions could be obtained through ACE. In the model, weather variables such as temperature and humidity were analyzed and used as input to perform a probability density peak load demand forecast.

Regression models are also often found being used combined with other techniques to improve forecast accuracy. \cite{27newdaa-Haida1998Peak} extended the model in \cite{19-2002Regression} by using trend cancellation and estimation techniques to minimize the effects of transitional seasons. \cite{9s-Barakat1990Short} performed monthly peak load demand forecast for the central region of Saudi Arabia where three time-series methods (Census-II multiplicative
decomposition, seasonal auto-regressive integrated moving average (SARIMA), Winters’ seasonal smoothing) were combined with the regression model. \cite{73aa-GOIA2010700} considered intra-daily seasonality effect, and proposed a new method to forecast daily peak load based on functional data analysis. They firstly introduced functional clustering to obtain \textcolor{black}{groups} that contains similar load usage patterns. Then each group was assigned a specialized functional regression model. Finally, functional linear discriminant analysis was applied to assign new curves to the classified groups to perform peak load demand forecast. The proposed method demonstrated promising performance.

Some representative regression methods for peak load forecast are summarized in TABLE \ref{t3}. As aforementioned, the advantage of regression analysis lies in the model is usually simple and easy to understand, with fewer parameters and higher forecast efficiency. However, regression analysis usually makes assumptions on the historical data and did not consider the correlation between different time trends, which could limit its applications in some cases. In the following sections, we will review time series models for peak demand forecast to consider the potential correlation between historical data at different time points. \\

\begin{table*}[!ht]
    \caption{Regression analysis for peak load demand forecast}
    \begin{center}
        \resizebox{\textwidth}{0.4\textwidth}{ 
            \begin{tabular}{lllllll}
                \hline
                \multirow{2}{*}{Reference} & \multirow{2}{*}{Model detail}                                                                                                                                                                                                                                & \multirow{2}{*}{Input Variable(s)}                                                                                                                                                                                                                                                          & \multirow{2}{*}{Forecast horizon}                                                & \multirow{2}{*}{Geographic scope}                                                                                                       & \multirow{2}{*}{Forecast output(s)}                                                                                                                                     & \multirow{2}{*}{Performance}                                                                                                       \\
                &                                                                                                                                                                                                                                                              &                                                                                                                                                                                                                                                                                             &                                                                                  &                                                                                                                                         &                                                                                                                                                                         &                                                                                                                                    \\ \hline
                &                                                                                                                                                                                                                                                              &                                                                                                                                                                                                                                                                                             &                                                                                  &                                                                                                                                         &                                                                                                                                                                         &                                                                                                                                    \\
                \cite{8-Barakat1989Forecasting}                    & \begin{tabular}[c]{@{}l@{}}A composite \\ multiregression-\\ decomposition\\ model\end{tabular}                                                                                                                                                              & \begin{tabular}[c]{@{}l@{}}Daily peak load, \\ weather variables \\ (base ambient temperature, \\ mean value of maximum and \\ minimum ambient temperatures), \\ calendar variables \\ (holiday, day of the week) (9 years)\end{tabular}                                                    & \begin{tabular}[c]{@{}l@{}}Monthly peak\\ demand forecast\\ (MTPLF)\end{tabular} & \begin{tabular}[c]{@{}l@{}}Region \\ (Consolidated Electric \\ Company of Central \\ Region of Kingdom \\ of Saudi Arabia)\end{tabular} & \begin{tabular}[c]{@{}l@{}}Load values \\ (monthly peak \\ load value)\end{tabular}                                                                                      & \begin{tabular}[c]{@{}l@{}}MAPE:\\ 7.88\%\end{tabular}                                                                             \\
                &                                                                                                                                                                                                                                                              &                                                                                                                                                                                                                                                                                             &                                                                                  &                                                                                                                                         &                                                                                                                                                                         &                                                                                                                                    \\
                \cite{9a-1990A}                    & \begin{tabular}[c]{@{}l@{}}Modeled temperature, \\ holiday and other special \\ effects using binary variables \\ to build a time-independent \\ peak load demand forecast model\end{tabular}                                                                    & \begin{tabular}[c]{@{}l@{}}daily peak load, \\ weather variables \\ (maximum temperature), \\ calendar variables \\ (holiday, day of week) (6 years)\end{tabular}                                                                                                                           & \begin{tabular}[c]{@{}l@{}}Daily peak \\ demand forecast \\ (STPLF)\end{tabular} & \begin{tabular}[c]{@{}l@{}}Region \\ (Pacific Gas and \\ Electric Company)\end{tabular}                                                 & \begin{tabular}[c]{@{}l@{}}Load values \\ (next day’s peak \\ load value, \\ hourly loads)\end{tabular}                                                                 & \begin{tabular}[c]{@{}l@{}}MPE:\\ 0.51\%-0.04\%\end{tabular}                                                                       \\
                &                                                                                                                                                                                                                                                              &                                                                                                                                                                                                                                                                                             &                                                                                  &                                                                                                                                         &                                                                                                                                                                         &                                                                                                                                    \\
                \cite{19-2002Regression}                         & \begin{tabular}[c]{@{}l@{}}Proposed a transformation \\ function with translation and \\ reflection methods to deal with \\ the seasonal load change, \\ annual load growth and \\ the latest daily load change\end{tabular}                                 & \begin{tabular}[c]{@{}l@{}}peak load of weekdays \\ (except holidays), \\ weather variables \\ (daily maximum temperature, \\ minimum temperature and humidity) \\ (4 years)\end{tabular}                                                                                                   & \begin{tabular}[c]{@{}l@{}}Daily peak \\ demand forecast \\ (STPLF)\end{tabular} & \begin{tabular}[c]{@{}l@{}}Region \\ (Tokyo Electric \\ Power Company \\ (TEPCO))\end{tabular}                                          & \begin{tabular}[c]{@{}l@{}}Load values \\ (next day’s peak \\ load value)\end{tabular}                                                                                  & \begin{tabular}[c]{@{}l@{}}MPE:\\ 1.50\%\end{tabular}                                                                              \\
                &                                                                                                                                                                                                                                                              &                                                                                                                                                                                                                                                                                             &                                                                                  &                                                                                                                                         &                                                                                                                                                                         &                                                                                                                                    \\
                \cite{26news-1997Short}                     & Multiple linear regression                                                                                                                                                                                                                                                          & \begin{tabular}[c]{@{}l@{}}Fall\&winter historical load, \\ weather variables \\ (daily maximum temperature, \\ seven-day moving average \\ of past-midnight temperature), \\ calendar variables \\ (day of week, \\ month of the year, year)\end{tabular}                                  & \begin{tabular}[c]{@{}l@{}}Daily peak\\ demand forecast \\ (STPLF)\end{tabular}  & \begin{tabular}[c]{@{}l@{}}Region \\ (Puget Sound \\ Power and \\ Light Company)\end{tabular}                                           & \begin{tabular}[c]{@{}l@{}}Load values \\ (next day’s hourly \\ load value, \\ next day’s \\ peak load)\end{tabular}                                                    & \begin{tabular}[c]{@{}l@{}}MAPE: \\ 2.45\%-6.40\%\end{tabular}                                                                     \\
                &                                                                                                                                                                                                                                                              &                                                                                                                                                                                                                                                                                             &                                                                                  &                                                                                                                                         &                                                                                                                                                                         &                                                                                                                                    \\
                \cite{28-1999The}                         & \begin{tabular}[c]{@{}l@{}}Combine fuzzy system \\ with MLR, weekdays \\ and weekend days \\ were assigned with \\ different models\end{tabular}                                                                                                             & \begin{tabular}[c]{@{}l@{}}Daily 15-minutes peak load, \\ daily energy consumption \\ (4 months)\end{tabular}                                                                                                                                                                               & \begin{tabular}[c]{@{}l@{}}Daily peak\\ demand forecast \\ (STPLF)\end{tabular}  & Region (substations)                                                                                                                    & \begin{tabular}[c]{@{}l@{}}Load values \\ (daily peak load)\end{tabular}                                                                                                & \begin{tabular}[c]{@{}l@{}}Absolute Error \\ range:\\ work days: \\ 0.33\%-14.05\% \\ weekend days: \\ 0.30\%-18.98\%\end{tabular} \\
                &                                                                                                                                                                                                                                                              &                                                                                                                                                                                                                                                                                             &                                                                                  &                                                                                                                                         &                                                                                                                                                                         &                                                                                                                                    \\
                \cite{73aa-GOIA2010700}                         & \begin{tabular}[c]{@{}l@{}}Built model based on \\ functional data analysis. \\ Functional clustering, \\ functional linear regression \\ and functional linear \\ discriminant analysis were used\end{tabular}                                              & \begin{tabular}[c]{@{}l@{}}Daily load curve (hourly) \\ for heating demand \\ (four discrete periods \\ covering four years)\end{tabular}                                                                                                                                                   & \begin{tabular}[c]{@{}l@{}}Daily peak\\ demand forecast \\ (STPLF)\end{tabular}  & \begin{tabular}[c]{@{}l@{}}Buildings \\ (using \\ district-heating\\ system)\end{tabular}                                               & \begin{tabular}[c]{@{}l@{}}Load values \\ (daily peak load)\end{tabular}                                                                                                & \begin{tabular}[c]{@{}l@{}}MAPE: \\ 3.61\%-32.68\%\end{tabular}                                                                    \\
                &                                                                                                                                                                                                                                                              &                                                                                                                                                                                                                                                                                             &                                                                                  &                                                                                                                                         &                                                                                                                                                                         &                                                                                                                                    \\ \hline
        \end{tabular}}
        \label{t3}
    \end{center}
 \end{table*}

\subsubsection{\textbf{Time series decomposition}}

There are different time series decomposition methods, such as Fourier series analysis, wavelet methods and empirical mode decomposition (EMD). 

A general time-series decomposition model usually adopts the addition or multiplication model to split the original time series into four sub-parts: Secular trend (T), Seasonal Variation (S), Cyclical Variation (C), Irregular Variation, (I). In the context of peak load demand forecast: 

\textcolor{black}{
    \begin{itemize}
        \item{The secular trend refers to the continuous change of peak load demand in a long period.}
        \item{The seasonal variation refers to the regular seasonal change of peak load demand.}
        \item{The cyclical variation is the periodic change in peak load demand over years.}
        \item{The irregular variation refers to the unexpected change of the peak load demand caused by many random factors.}
    \end{itemize}
}

When \textcolor{black}{predicting} the future peak load, each component is calculated separately first, and then the forecasted value of for each sub-part is passed to the addition model or the multiplication model to obtain the final prediction.

The addition model of time series decomposition is defined as:
\begin{equation}
y_t = T_t + S_t + C_t + I_t
\end{equation} where the four components in the addition model are independent of each other, all of which are expressed in absolute quantities and of the same order of magnitude. 

On the other hand, the multiplication model of time series decomposition is:
\begin{equation}
y_t = T_t \times S_t \times C_t \times I_t
\end{equation} Different from the addition model, the four components of the multiplication model are dependent on each other. In general, the secular trend in the multiplication model is expressed in absolute quantity, while other components are expressed in relative quantity.

When $T_t$, $S_t$, $C_t$ do not change over time, addition model is usually selected. Otherwise, the multiplication model could be a better choice. It should, however, be noted that, there is a convertible relationship between the addition and multiplication models where log function is one of the effective converting methods \cite{7-30}.

In \cite{2-Gupta1971A}, the addition model is utilized to produce monthly peak load demand probabilistic forecast. It also utilized Fourier transformation to reduce the non-stationary time series to a stationary series. Moreover, Monte Carlo simulation was adopted to simplify the computation process. \cite{5-Fong2011The} designed a comparative experiment to compare the time-series model using Fourier series with the auto-regressive model. Weekly peak load demand forecasts for one year ahead were produced, which showed that although over-forecasts exist, the model using Fourier expansion could track the dynamic behavior of peak load demand and produce better results than the auto-regressive model. 
 
\cite{68-2009The} utilized a multiplicative model to forecast \textcolor{black}{the} monthly peak load on a regional power grid. \cite{8-Turner2012Regression} utilized the decomposition method to develop a multi regression-decomposition model. The proposed method \textcolor{black}{aims} to forecast monthly peak loads for one year ahead, and the result was promising with a MAPE of 7.88\%. The advantage of this method is that it related the historical load trend with diverse influencing factors, and \textcolor{black}{could} simulate additional cyclic effects. \cite{22-Choi1996A} used the real-world load data from Korea Electric Power Corporation to perform one day ahead daily peak load demand forecast. In this paper, Fourier transformation was adopted to identify the chaotic characteristics of the time series, and \textcolor{black}{the} optimal and embedding dimension and delay time were determined to be used as inputs to train an artificial neural network (ANN) model. MAPE for daily peak load demand forecast of the proposed model was close to 1.4\%. As aforementioned, wavelet transformation is also a traditional time series decomposition method, and it can transform the information from \textcolor{black}{the} time domain to frequency domain, and thus capture the low-frequency and high-frequency components of the peak load signal. In \cite{88aaa-2013A}, wavelet decomposition was introduced to combine with other advanced methods to build a hybrid model. Daily peak load demand forecast for Iran National Grid was conducted based on the proposed model. 

Empirical mode decomposition as a more recent time series decomposition method is also used in peak demand forecast. In \cite{66-2009Long}, empirical mode decomposition was proposed to capture long-run seasonality, short-run effects, and trend effects for the daily peak load. The load decomposition results given by EMD contain physical meaning related to time series characterizes, and thus can improve the forecast accuracy.

Note that time series decomposition uses the deterministic function to extract information. Therefore, it often ignores the stochastic factors of the original time series, resulting in insufficient information extraction, which can be compensated by the stochastic time series models, as we will discuss in the next subsection.

\subsubsection{\textbf{Stochastic time series models}}

Stochastic time series models can be generally divided into: the auto-regression (AR$(p)$) model where $p$ denotes the order of auto-regression;  the moving average (MA$(q)$) model where $q$ is the order of moving average; the auto-regression moving average (ARMA$(p,q)$) model; the auto-regression integrated moving average (ARIMA$(p,d,q)$) model where $d$ denotes the order of integration; and the seasonal auto-regression integrated moving average (SARIMA$(p,d,q)$\\$(P, D, Q)_s$) model where $P, D, Q$ are the seasonal parts of the model corresponding to $p, d, q$. 

A SARIMA model may be written as \cite{80-2011Discrete}:
\begin{equation}
\phi_p(B)\psi_P(B^S)Y_t=\theta_q(B)\tau_Q(B^S)\alpha_t
\end{equation}
\label{e1} where:
$Y_t = \triangledown^d\triangledown^D_Sy_t$, $y_t$ is the peak load demand observed at time $t$. $\triangledown^d$ and $\triangledown^D_S$ denote the non-seasonal and seasonal difference ($S$ is the seasonal length) operators respectively, which transform $y_t$ into stationary time series. $B$ is the backshift operator, which is used to represent the backshift of time. When $B$ is used for $y_t$, it means to reverse by one unit of time ($By_t=y_t-1$). For monthly data, $B_{12}y_t=y_t-12$ represents data of the same month of the last year. $\phi_p(B)$ and $\psi_P(B^S)$ are the non-seasonal and seasonal auto-regression operators, respectively. $\theta_q(B)$ and $\tau_Q(B^S)$ are the non-seasonal and seasonal moving average operators, respectively. $p, P, q, Q$ denote the maximum backshift order for the non-seasonal, seasonal, auto-regression, and moving average operators, respectively. $\alpha_t$ is the white noise at time $t$. The above model can be represented by SARIMA$(p,d,q)(P, D, Q)_s$ where $s=4$ represents the seasonal time series, and $s=12$ represents the monthly time series. When $P=D=Q=0$, the SARIMA model degenerates into an ARIMA model, and when $P=D=Q=p=d=q=0$, the SARIMA model degenerates into the white noise process.

It is worth pointing out that AR, MA, and ARMA models are suitable for weak/wide stationary time series. ARIMA is used for non-stationary time series, and SARIMA can deal with non-stationary and cyclical time series. 

\cite{5-El1986Weekly} developed two models to forecast weekly peak load one year ahead, where MA model for the seasonal-cyclic component is utilized. In \cite{12-1992New}, ARMA was used for monthly peak load demand forecast by considering the seasonal patterns and load fluctuations. Some hybrid methods combining ARMA with other methods such as regression models \cite{63newm-5398613} and ANN \cite{133-2020The} have been seen in the peak demand forecast. 

 \begin{table*}[!ht]
    \caption{Stochastic time series models for peak load demand forecast}
    \begin{center}
        \resizebox{\textwidth}{0.3\textwidth}{ 
            \begin{tabular}{lllllll}
                \hline
                Reference & Model detail                                                                                                                                                                                           & Input variable                                                                                                                                                                                                                                                                                    & Forecast horizon                                                                  & Geographic scope                                                                                                     & Forecast output                                                                                         & Performance                                                                                                                                  \\ \hline
                &                                                                                                                                                                                                        &                                                                                                                                                                                                                                                                                                   &                                                                                   &                                                                                                                      &                                                                                                         &                                                                                                                                              \\
                \cite{30-Amjady2001Short}    & \begin{tabular}[c]{@{}l@{}}Modified ARIMA \\ (incorporate ARIMA \\ with human experience)\end{tabular}                                                                                                 & \begin{tabular}[c]{@{}l@{}}Hourly loads \\ (in the range of \\ peak times), \\ weather variables \\ (current temperature), \\ calendar variables \\ (day of year, which \\ was distinguished based\\ on average temperature), \\ human operator’s estimation \\ of current peak load\end{tabular} & \begin{tabular}[c]{@{}l@{}}Daily peak \\ demand forecast \\ (STPLF)\end{tabular}  & \begin{tabular}[c]{@{}l@{}}Country \\ (Iran’s power \\ network)\end{tabular}                                         & \begin{tabular}[c]{@{}l@{}}Load values \\ (next day’s peak \\ load value, hourly \\ loads)\end{tabular} & \begin{tabular}[c]{@{}l@{}}Hourly loads: \\ MAPE:\\ 1.45\%- 1.99\%;  \\ Daily peak loads: \\ ME:\\ 1.01\%-1.97\%\end{tabular}                \\
                &                                                                                                                                                                                                        &                                                                                                                                                                                                                                                                                                   &                                                                                   &                                                                                                                      &                                                                                                         &                                                                                                                                              \\
                \cite{12-1992New}        & \begin{tabular}[c]{@{}l@{}}ARMA(6,5), use \\ SARIMA(1,1,2)x(1,1,2)12 \\ to extrapolated maximum \\ temperature for \\ forecasted periods\end{tabular}                                                  & \begin{tabular}[c]{@{}l@{}}Monthly peak load demand, \\ weather variables \\ (maximum temperature), \\ calendar variables \\ (time, months)\end{tabular}                                                                                                                                               & \begin{tabular}[c]{@{}l@{}}Monthly peak \\ demand forecast\\ (MTPLF)\end{tabular} & \begin{tabular}[c]{@{}l@{}}Region \\ (Saudi Consolidated \\ Electric Company)\end{tabular}                           & \begin{tabular}[c]{@{}l@{}}Load values \\ (monthly peak load demands \\ for next year)\end{tabular}          & \begin{tabular}[c]{@{}l@{}}Most months \\ were observed to \\ fall between the \\ upper and lower \\ limits of \\ the forecasts\end{tabular} \\
                &                                                                                                                                                                                                        &                                                                                                                                                                                                                                                                                                   &                                                                                   &                                                                                                                      &                                                                                                         &                                                                                                                                              \\
                \cite{82aap-2011Prediction}      & \begin{tabular}[c]{@{}l@{}}SARIMA, SARIMA with \\ generalized autoregressive \\ conditional heteroskedastic \\ errors (SARIMA-GARCH),\\  regression-SARIMA-\\ GARCH \\ (Reg-SARIMA-GARCH)\end{tabular} & \begin{tabular}[c]{@{}l@{}}Daily peak load demand \\ (the maximum hourly \\ demand in a 24-hour period), \\ weather variables \\ (maximum temperature), \\ calendar variables \\ (day of week, holiday, \\ month of a year)\end{tabular}                                                               & \begin{tabular}[c]{@{}l@{}}Daily peak \\ demand forecast\\ (STPLF)\end{tabular}   & \begin{tabular}[c]{@{}l@{}}Region \\ (industrial commercial \\ and domestic sectors \\ of South Africa)\end{tabular} & \begin{tabular}[c]{@{}l@{}}Load values \\ (monthly peak \\ demands for \\ next year)\end{tabular}       & \begin{tabular}[c]{@{}l@{}}MAPE:\\ 1.42\%\end{tabular}                                                                                       \\
                &                                                                                                                                                                                                        &                                                                                                                                                                                                                                                                                                   &                                                                                   &                                                                                                                      &                                                                                                         &                                                                                                                                              \\
                \cite{84aaf-2012Finding}      & \begin{tabular}[c]{@{}l@{}}SARIMA models \\ with different length \\ of historical data\end{tabular}                                                                                                   & Daily peak load demand                                                                                                                                                                                                                                                                                 & \begin{tabular}[c]{@{}l@{}}Daily peak \\ demand forecast\\ (STPLF)\end{tabular}   & \begin{tabular}[c]{@{}l@{}}Region \\ (New South Wales, \\ Australia)\end{tabular}                                    & \begin{tabular}[c]{@{}l@{}}Load values \\ (daily peak load demands)\end{tabular}                             & \begin{tabular}[c]{@{}l@{}}MAPE:\\ 2.921\%-7.946\% \\ (one to seven \\ days ahead)\end{tabular}                                              \\
                &                                                                                                                                                                                                        &                                                                                                                                                                                                                                                                                                   &                                                                                   &                                                                                                                      &                                                                                                         &                                                                                                                                              \\ \hline
        \end{tabular}}
        \label{t4}
    \end{center}
 \end{table*}

When practical conditions are considered such as economic and cultural factors, it often lead to nonstationary time series problems. For such cases, ARIMA and SARIMA, are widely used. \cite{30-Amjady2001Short} utilized ARIMA models to forecast hourly loads and daily peak load demand, \textcolor{black}{incorporating} human experience within the model. The proposed approach adopted ARIMA to produce a \textcolor{black}{raw} output as the initial input of the modified model, and also took temperature into account to distinguish hot and cold days to perform a regression-based analysis. The results of the proposed models were compared with ANN, standard ARIMA, and human operators, and the accuracy of the modified ARIMA models outperformed other models. \cite{57aat-article} used load data from Dubai to build models based on ARIMA and dynamic regression, to forecast monthly peak load where the  $R^2$ of method is 0.997. \cite{57aam-2007Monthly} developed a model based on SARIMA to forecast monthly peak load for Sulaimany Governorate in northern Iraq. The adequate SARIMA model they found was $(1,1,0)(0,2,1)_{12}$, and the forecast results gave better opportunities for the power planners to determine the maximum generating capacity for peak load demand. The paper also pointed out that ARIMA was suitable for short-term forecasting since it \textcolor{black}{prioritized the} closer time series. \cite{83aaf-2012Forecasting} utilized SARIMA to forecast monthly peak load demand for India. The SARIMA model outperformed the official load forecasting provided by the Central Electricity Authority (CEA) for both static and dynamic horizons in all five regional grids in India. \cite{8-78} compared SARIMA with Holt-Winters multiplicative exponential smoothing, and it showed \textcolor{black}{that} the SARIMA model produced better forecast results.

Although ARIMA and SARIMA models have promising forecast results such as in \cite{30-Amjady2001Short}\cite{57aam-2007Monthly}\cite{73-5697708}\cite{99-7117006}\cite{102-7399017}, they usually do not take into account trend fluctuations in the data. Following this, variations based on ARIMA and SARIMA have been proposed in some studies to overcome the limitation and to further improve the forecast accuracy. \cite{56aad-4202249} combine generalized autoregressive conditional heteroskedastic errors (GARCH) with ARIMA to define the maximum peak load demand level by considering the unexpected randomness of the load series. \cite{82aap-2011Prediction} presented a regression-SARIMA model with generalized autoregressive conditional heteroskedastic errors (Reg-SARIMA-GARCH), which could accommodate the volatility of the daily peak load demand and the multiple seasonality of the mid and long term peak demand. The proposed model was used to conduct daily peak forecast for South Africa, and the comparative experiment showed that the proposed model produced better prediction accuracy than the piecewise linear regression model, the SARIMA model, and the SARIMA-GARCH model.

Some representative stochastic time series  methods for peak load forecast are summarized in TABLE \ref{t4}.

As aforementioned, the historical load data for peak load demand forecast are characterized by its high randomness. Therefore, the stochastic time series model is widely used in peak load demand forecast as an effective method to deal with random sequences. However, there are some limitations in the stochastic  time series methods. For instance, for MA model, it gives the same weight to all the time series data, which does not necessarily reflect the actual situation. As such, a more flexible weights assignment of the data needs to be considered.  In the next subsection, we will discuss the exponential smoothing, which can overcome the above limitation.

\subsubsection{\textbf{Exponential smoothing}}

Exponential smoothing is a time series analysis method developed \textcolor{black}{based on} MA. Exponential smoothing predicts the future peak load according to the weighted average of the historical time series. \textcolor{black}{The recent data are given a larger weight} whereas the previous data are given a smaller weight. This is based on the principle that the influence of a certain variable on subsequent behavior is gradually attenuating \cite{7-64-incollection}. 

A general exponential smoothing model for peak load demand forecast can be written as \cite{7-30}:
\begin{equation}
\hat{y}_{t+1} = \alpha y_t + \alpha(1-\alpha)y_{t-1} + \alpha(1-\alpha)^2y_{t-2} + \cdot\cdot\cdot
\end{equation} where $\hat{y}_{t+1}$ is the forecasted peak load demand at time $t+1$, and $\alpha\in\left[0, 1\right]$ is the smoothing parameter that controls the weights decrease (exponentially). 

Exponential smoothing can be divided into several different forms. \cite{26aa-27} provides a comprehensive review of exponential smoothing methods, in which 17 basic methods and some extensions based on these methods are described in detail. In general, single exponential smoothing is applied to sequences without trends or seasonality, and second exponential smoothing is applied to time series that only have trends. The triple exponential smoothing (also known as the Holt-Winters) targets sequences with both trends and seasonality. When modeling the seasonal data, a Holt-Winters model consists of three smoothing equations each having its smoothing parameters: trend, level, and seasonality components \cite{17du12-Yaffee2000Introduction}. When the seasonal variations are constant and uncorrelated with time series, the additive Holt-Winters model can be hired. However, if the seasonal variables change proportionally with time series, the multiplication model can be chosen to predict the seasonal data.

The triple exponential smoothing is the most commonly used method in the reviewed papers. \cite{9a-1990A} used exponential smoothing to correct the forecast values that were consistently too high or too small, and the adjusted model showed good capability to track the \textcolor{black}{fast-changing load demand and produce hourly forecasts with higher accuracy.} \cite{26aa-Masood1997EDSSF} proposed a decision support system based on a variety of time series techniques. The near-optimal monthly peak forecast models were built by exponential smoothing, Box-Jenkins vector, and dynamic regression to perform short-term peak load demand forecast. Moreover, a comprehensive assessment of the models was provided by using several evaluation indicators such as MAPE, Akaike information criterion (AIC), Bayes information criterion (BIC), and MSE. The results of the proposed system showed that different models performed differently towards different regions of the country. 

\textcolor{black}{When trends and seasonal variations dominate the time series}, the Holt-Winters exponential smoothing usually outperforms the ARIMA. This was confirmed by \cite{38aa-J2017Short}, which used the Holt-Winters Smoothing to forecast peak load demand for England and Wales. The model they built described the intra-daily and intra-weekly seasonal cycles, and the comparative results showed that this approach achieved a better accuracy than the ARIMA model. 

Exponential smoothing is often used in combination with other methods to build composite models. \cite{9a-1990A} applied exponential smoothing to a regression model and compared it with the regression model with ARIMA, \textcolor{black}{and the experiment revealed that most of the initial forecasts were corrected after applying the exponential smoothing.} However, the smoothing coefficient of exponential smoothing needs to be artificially selected, and if the time series fluctuates wildly (e.g., the peak load), it will produce unsatisfactory prediction results \cite{billah2006exponential}.

\subsubsection{\textbf{Kalman filter and grey prediction}}\label{others}

\textcolor{black}{Kalman filter (KF) is a linear system state equation that can estimate the system state optimally through the input and output observations of the system.} Since the observed data include the noise, the optimal estimation can also be regarded as the filtering process. KF comprises \textcolor{black}{two} processes: prediction and correction. During the prediction, the filter makes a forecast of the current state using the estimation of the state from the previous timestep. The correction was performed using observations of the current state to correct the predicted value acquired in the prediction phase to obtain an improved estimation. \textcolor{black} {Besides, apart from being known as the recursive state estimator for linear systems, some KF variants are also capable of non-linear systems \cite{Kalman1}.}

\cite{21-1995Short} presented a hybrid learning scheme that consists of unsupervised and supervised learning phases to forecast daily and weekly peak/average load profiles. \textcolor{black}{The KF-based learning algorithm was engaged to find the optimum parameters and functions in the supervised learning phase.} \cite{85aab-2012Building} selected KF as one of the benchmarks to carry out an integrated hybrid model for STPLF.

The grey system is between a white box model and a black box model, where it focuses on learning the internal structure, parameters, and general characteristics, and tries to decipher known information as much as possible. Grey time series prediction model was constructed \textcolor{black}{based on the observed historical time series reflecting the predicted peak load characteristics.} 
 
\cite{61-2008The} used grey correlation theory in sensitivity analysis to select relevant meteorological variables for daily peak load demand forecast. \cite{62-2008Electricity} developed a variable weight combination forecasting model by combining the grey model and ARIMA model, which was used to forecast load consumption in the peak load month for MTPLF. The hybrid model was \textcolor{black}{proved reliable to handle the non-smooth characteristics of monthly load data and achieved satisfactory forecast accuracy.} \cite{20} proposed a hybrid grey model to forecast yearly peak load and its occurring date simultaneously for LTPLF. The model only needed \textcolor{black}{a} small amount of historical annual peak load data to produce the forecast results, and it was claimed the model was highly adaptive to dynamic changes of yearly peak load. 

\subsection{Advanced peak load demand forecast stage}

With the emergence of artificial intelligence (AI) and big data, traditional AI-based techniques such as fuzzy logic (FL), expert system (ES) and genetic algorithm (GA) and modern AI and machine learning based methods such as artificial neural network (ANN) and deep learning, support vector machines (SVMs), and ensemble models have been adopted for peak load demand forecast. 

\subsubsection{\textbf{Traditional AI-based methods}}

As a bridging stage between the classic and advanced methods, there are a few studies in the reviewed literature using traditional AI-based methods for peak load forecast. 

Fuzzy logic imitates the uncertainty concept of judging and reasoning of \textcolor{black}{the human brain. It applies fuzzy rules to the reasoning of the system with the uncertain model to deal with the fuzzy information that is difficult to handle by conventional methods.} \cite{27newdaa-Haida1998Peak} adopted separate fuzzy models to predict the peak and valley load, and the simulation results showed a good prediction accuracy. \cite{28-1999The} proposed a fuzzy regression approach to peak load estimation. The effectiveness of the proposed method was demonstrated by forecasting daily load consumption and daily peak load demand at \textcolor{black}{the} distribution level. 

Expert system has also been used in peak load forecast. An ES is a computer system, which is a knowledge-based programming method that absorbs the domain knowledge and experience of experts and makes intelligent decisions based on the reasoning of such knowledge and experience. A complete ES consists of \textcolor{black}{the knowledge base, the reasoning machine, the knowledge acquisition part, and the interpretation interface.} \cite{36-2002Artificial} implemented a knowledge-based ES to forecast yearly peak load for both typical \textcolor{black}{fast-developing} system and \textcolor{black}{regular} developing system, and the knowledge base of this system was composed of both static and dynamic variables. The results proved that the knowledge-based ES \textcolor{black}{yielded the best performance} among all considered models (time series model, traditional ES model, econometric model). 

Fuzzy logic has been combined with ES to produce better prediction results. \cite{13aa-Hsu1992Fuzzy} built an ES based on fuzzy set theory to forecast hourly load in Taiwan by improving the estimation accuracy of \textcolor{black}{the} peak and trough loads. The proposed ES could handle uncertain weather variables and heuristic rules, and it \textcolor{black}{could update} peak and trough loads iteratively to produce a more accurate forecast. \cite{30newa-30Kiartzis2000A} proposed a fuzzy ES to forecast morning and afternoon valley, noon and evening peak based on weather information and historical load data from the Greek power system. The results showed that fuzzy ES could forecast daily peak and valley loads \textcolor{black}{reasonably} well compared with neural networks.

Genetic Algorithm has also been adopted in building models for peak load forecast. GA is based on natural selection and population genetics, which makes the population evolve to the optimal region in the searching space through selection, crossover, variation, evaluation, and other operations. GA can also be used to optimize the parameters of the forecast models such as initial connection weights of the networks and the threshold values of nodes for neural networks. In \cite{85aab-2012Building}, 13 years \textcolor{black}{of} regional data from France were utilized \textcolor{black}{for training} a real-valued genetic algorithm (RGA)-based neural network with support vector machine (NN-SVM) model. Daily load profile forecast and monthly peak load demand forecast were generated, and the comparative experiments showed that the proposed model was suitable for forecasting long-term peak load. \cite{87aaa-El2013Electric} implemented a comparative analysis for yearly peak load demand forecast based on \textcolor{black}{the unified Egyptian network data.} In the experiment, models based on GA, least-square, and least absolute value filtering were trained separately. The results showed that the model developed based on GA gave the best performance with the \textcolor{black}{lowest} forecast error of 0.70\%.

\subsubsection{\textbf{Artificial neural network and deep learning-based methods}}

ANN was proposed in 1991, and it has attracted much attention in the peak load demand forecast. Many advanced methods based on ANN such as deep learning methods have since been proposed with good performance.

\begin{figure*}[htbp]
    \centering
    \includegraphics[width=0.4\textwidth]{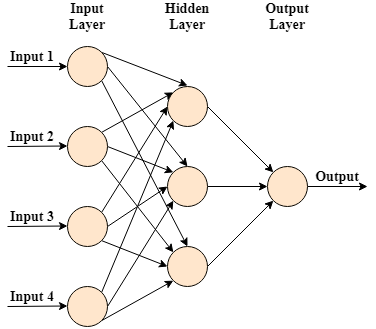}
    \caption{An example structure of ANN.}
    \label{f6}
\end{figure*}

Artificial neural network (ANN) is inspired \textcolor{black}{by} the anatomy of the human brain, and it consists of artificial neurons in multi-layers for information communication. An example structure of ANN is shown in Figure \ref{f6}. 

\begin{figure*}[htbp]
    \centering
    \includegraphics[width=0.4\textwidth]{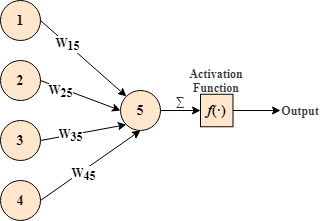}
    \caption{The calculation process of a neuron in ANN.}
    \label{f7}
\end{figure*}

A typical ANN consists of the input layer, the hidden layer, and the output layer. Except for the input layer, each neuron in ANN is connected to neurons of the former layer (i.e. the input neurons), with each connection corresponding to a weight. The sum of the product of all input and the corresponding connection weights are passed to an active function to calculate each neuron's final value, as is shown in Figure \ref{f7}. 

The activation function needs to be selected according to data characteristics, and the Sigmoid function is the most commonly used active function of ANN models \cite{17du20-Agatonovic2000Basic}. One of the well-known ANN is the backpropagation (BP) neural network, a multi-layer neural network with error backward propagation. BP is widely used for its satisfying performance on prediction tasks. It, however, suffers from high computational cost and low computational efficiency, and therefore the radial basis function network (RBFN) was brought up to deal with this. The input variables of RBFN pass directly to the hidden layers without additional weights, and RBFN is proved to be less time-consuming than the traditional multi-layer neural network \cite{69-2010Peak}.

\cite{11e-Saeed1991Electric} collected hourly temperature and load data from Seattle to build a model based on ANN, and the trained ANN was then used to forecast daily peak load and hourly load and \textcolor{black}{total daily} load. The mean error of the peak load forecast model was ranged between 1.55\% and 2.60\%. This ANN model allowed \textcolor{black}{a} more flexible relationship between weather variables and load patterns. However, the model produced higher errors \textcolor{black}{when} people have specific start-up activities, which indicates that the use of additional calendar variables should yield better results. \cite{37aa-2003Regional} utilized ANN to perform annual regional peak load demand forecast of Taiwan. The proposed model had three input neurons corresponding to economic, demographic, and weather variables, two hidden neurons, and one output neuron representing the regional peak load that needs to be estimated. The effectiveness of the proposed model was demonstrated by comparing the forecast accuracy with a regression-based model. \cite{63newp-Saini2008Peak} was presented to forecast daily peak load up to seven days ahead based on a feed-forward neural network with \textcolor{black}{the} steepest descent, Bayesian regularization, resilient and adaptive backpropagation learning methods. \cite{108aaa-2016Artificial} presented a new model based on ANN by employing Bayesian regularized neural network model with Levenberg-Marquart (LM) backpropagation algorithm to forecast daily peak load demand for a commercial building complex. \cite{74aa-article} forecasted annual peak load demand five years ahead for Iran based on RBF. The paper selected variables related to yearly incremental growth rate and pointed out that long-term forecasting should pay more attention to economic factors \textcolor{black}{than weather conditions.}

There are a few studies combining multiple ANNs to improve the forecast accuracy, in which the peak load is usually generated as a by-product to enhance the forecast performance \textcolor{black}{further}. \cite{27} developed a model based on cascaded ANNs (CANNs) to forecast \textcolor{black}{the} load profile one day ahead where the daily peak, valley, and total load consumption were \textcolor{black}{first} estimated by an ANN, and then such forecasted values were used as additional input data for the next day’s load profile forecast. The results revealed that the cascaded structure of ANN could produce more satisfactory forecasting results. \cite{89aai-Hern2013Improved} proposed a multi-stage ANN to forecast load demand in two stages. Firstly, the daily peak and valley values were generated by ANN. Second, based on the peak and valley load, the whole electricity demand curve was produced. This method outperformed normal ANN for significantly reduced the MAPE. Although multiple ANNs can provide promising results for load forecasting, \cite{89aai-Hern2013Improved} revealed that the multi-stage ANN suffers from higher computational complexity than a single ANN. 

In general, ANN has \textcolor{black}{apparent} advantages for its adaptive learning and function approximation capabilities. Since ANN can deal with the high randomness and uncertainty of the time series well, it is \textcolor{black}{recommended for STPLF}. However, ANN models suffer from \textcolor{black}{long training time} and are easy to fall into local optimum. Therefore, researchers mostly focused on \textcolor{black}{optimizing} the neural network structure, such as combining with fuzzy logic \cite{63newm-5398613}\cite{25-Mandal1997Fuzzy}\cite{99aaa}, to further improve the model training efficiency and forecasting accuracy.

Many modern and advanced methods have emerged based on ANN, such as self-organizing map (SOM), recurrent neural network (RNN), and convolution neural network (CNN). In particular, long-short memory (LSTM) is one variation of RNN \cite{132-8985197}.

SOM is often used as improving method due to its unsupervised learning characteristics together with other forecast methods to produce the final results. In \cite{10d-1991Design,10-1991Design}, SOM was adopted to cluster days with similar load consumption patterns. Then, based on a feed-forward multilayer neural network, daily peak load and valley load were estimated \textcolor{black}{to} compute the desired hourly load. \cite{63newc-AMINNASERI20081302} adopted SOM to cluster load profiles, and principal components analysis (PCA) for reducing the dimensions of the data. Then, separate feed-forward neural network was trained for each cluster. The comparative analysis demonstrated the superiority of the proposed method. 

RNN and CNN are commonly used deep learning methods, which have more complex network structures, such as more hidden layers and recurrent structures. Deep learning models can better capture the dynamic characteristics of peak load to provide a more accurate and stable prediction and have more robust learning and generalization ability than the standard ANN, especially in the big data era. 
\cite{126-2019Deep} proposed a method to combine RNN with dynamic time warping (DTW) for short-term peak load demand forecast. The DTW was introduced to identify load curves with similar trends, and a bespoke gated RNN was trained to forecast daily peak load demand one month ahead based on the half-hourly load data. The proposed method achieved a satisfactory MAPE of 1.01\%. 
In addition, comparative analysis suggested that the DTW distance had the ability to \textcolor{black}{adapt to} the dynamic change of non-stationary daily peak load series. In \cite{134-8994442}, the LSTM layer was adopted to forecast weekly peak load in Korea. In this study, input variables including weekly peak load, weekly temperature, and weekly GDP of the previous year were used. The LSTM layer in this paper was proved to be able to capture more useful characteristics of the load data, and results showed good forecast accuracy with the lowest forecast error of 2.16\%.

About 60 studies have employed ANN and deep learning based methods to perform peak load demand forecasting, revealing its dominant popularity in the peak load forecast. Some representative references related to these methods are listed in TABLE \ref{t5}.

\begin{table}[!h]
	\caption{ANN/ANN-based methods for peak load demand forecast}
	\begin{center}
		\resizebox{0.85\textwidth}{0.545\textwidth}{ 
			\begin{tabular}{lllllll}
				\hline
				Reference    & Model detail                                                                                                                                                                                                                                                                                                                                                                                                                                        & Input variable                                                                                                                                                                                                                                                                                                                                                                                                                                                                                                  & Forecast horizon                                                                       & Geographic scope                                                                                                                                  & Forecast output                                                                                                                                   & Performance                                                                                                                        \\ \hline
				&                                                                                                                                                                                                                                                                                                                                                                                                                                                     &                                                                                                                                                                                                                                                                                                                                                                                                                                                                                                                 &                                                                                        &                                                                                                                                                   &                                                                                                                                                   &                                                                                                                                    \\
				\cite{11e-Saeed1991Electric}        & \begin{tabular}[c]{@{}l@{}}Peak load: 3 input neurons, \\ 5 hidden neurons and the \\ output peak load at a given day; \\ total load: 3 input neurons, \\ 5 hidden neurons and the output\\ total load at a given day; \\ hourly load: 6 input neurons, \\ 10 hidden neurons, the output \\ load at a given hour.\end{tabular}                                                                                                                      & \begin{tabular}[c]{@{}l@{}}Predicted temperatures; \\ peak load value, total \\ load of a day (a sum), \\ hourly load (1-24 hour ahead), \\ weather variables \\ (average, peak and lowest \\ temperature at predicted day)\end{tabular}                                                                                                                                                                                                                                                                        & \begin{tabular}[c]{@{}l@{}}Daily peak \\ demand forecast \\ (STPLF)\end{tabular}       & \begin{tabular}[c]{@{}l@{}}Region \\ (Seattle/Tacoma)\end{tabular}                                                                                & \begin{tabular}[c]{@{}l@{}}Load values \\ ( peak load value, \\ total load of a day \\ (a sum), hourly \\ load (1-24 hour \\ ahead))\end{tabular} & \begin{tabular}[c]{@{}l@{}}Error: \\ Peak load:\\ 2.60\%; \\ Total load: \\ 3.39\%; \\ Hourly load: \\ 1.64\%\end{tabular}         \\
				&                                                                                                                                                                                                                                                                                                                                                                                                                                                     &                                                                                                                                                                                                                                                                                                                                                                                                                                                                                                                 &                                                                                        &                                                                                                                                                   &                                                                                                                                                   &                                                                                                                                    \\
				\cite{13-Ho1992Short}           & \begin{tabular}[c]{@{}l@{}}The neural network has \\ 46 input nodes, \\ 60 hidden nodes, \\ and the one output \\ layer (peak/valley load)\end{tabular}                                                                                                                                                                                                                                                                                             & \begin{tabular}[c]{@{}l@{}}Historical peak/valley loads, \\ weather variables \\ (high/low temperature)\end{tabular}                                                                                                                                                                                                                                                                                                                                                                                            & \begin{tabular}[c]{@{}l@{}}Daily peak \\ demand forecast \\ (STPLF)\end{tabular}       & \begin{tabular}[c]{@{}l@{}}Region \\ (Taiwan)\end{tabular}                                                                                        & \begin{tabular}[c]{@{}l@{}}Load value \\ (next day’s \\ peak/valley \\ load value)\end{tabular}                                                   & \begin{tabular}[c]{@{}l@{}}Error: \\ 1.19\%\end{tabular}                                                                           \\
				&                                                                                                                                                                                                                                                                                                                                                                                                                                                     &                                                                                                                                                                                                                                                                                                                                                                                                                                                                                                                 &                                                                                        &                                                                                                                                                   &                                                                                                                                                   &                                                                                                                                    \\
				\cite{26newc-1997Cascaded}       & \begin{tabular}[c]{@{}l@{}}The lower ANNs: \\ 16 input neurons, 8 hidden \\ neuron, 3 output neurons; \\ The upper ANNs: \\ 107 neurons, 2 hidden layers \\ each contains 35 neurons, \\ 48 outputs indicating \\ 48 half-hourly loads.\end{tabular}                                                                                                                                                                                                & \begin{tabular}[c]{@{}l@{}}Historical loads, \\ weather variables \\ (maximum and minimum \\ temperatures, \\ maximum humidity), \\ calendar variables \\ (time of the day, \\ day of the week, special\\ event and holidays)\end{tabular}                                                                                                                                                                                                                                                                      & \begin{tabular}[c]{@{}l@{}}Half-hourly load \\ for the next day\\ (STPLF)\end{tabular} & \begin{tabular}[c]{@{}l@{}}Country \\ (Kuwait)\end{tabular}                                                                                       & \begin{tabular}[c]{@{}l@{}}Load values \\ (peak load value, \\ valley load value, \\ daily load, half-hourly\\  load profile.\end{tabular}        & \begin{tabular}[c]{@{}l@{}}MPE for \\ peak load:\\ 3.18\%\end{tabular}                                                             \\
				&                                                                                                                                                                                                                                                                                                                                                                                                                                                     &                                                                                                                                                                                                                                                                                                                                                                                                                                                                                                                 &                                                                                        &                                                                                                                                                   &                                                                                                                                                   &                                                                                                                                    \\
				\cite{36-2002Artificial}           & \begin{tabular}[c]{@{}l@{}}Feedforward neural \\ network based on: \\ Levenberg-Marquardt \\ back-propagation algorithm, \\ Quasi-Newton \\ back-propagation algorithm \\ + principle components\\ analysis (PCA)\end{tabular}                                                                                                                                                                                                                      & \begin{tabular}[c]{@{}l@{}}Peak load of previous day, \\ weather variables \\ (temperature of the day \\ (maximum. minimum), \\ gross minimum temperature. \\ rainfall, evaporation per day, \\ sunshine hours of \\ the previous day, \\ wind speed, the dry bulb \\ temperature, the wet bulb \\ temperature. the vapour pressure, \\ the relative humidity (\%) and \\ the soil temperature ), \\ calendar variables \\ (seasons, year number, \\ day of the year, day of the week, \\ holiday)\end{tabular} & \begin{tabular}[c]{@{}l@{}}Daily peak \\ demand forecast\\ (STPLF)\end{tabular}        & \begin{tabular}[c]{@{}l@{}}Region\\ (a 220 kV substation \\ of Haryana Vidyut \\ Prasaran Nigam Ltd. \\ (HVPNL) )\end{tabular}                    & \begin{tabular}[c]{@{}l@{}}Load values \\ (peak load value \\ of the current day/ \\ one to seven days \\ ahead)\end{tabular}                     & \begin{tabular}[c]{@{}l@{}}MAPE: \\ LMBP: \\ 2.87\%, \\ Quasi-Newton \\ BP: \\ 2.38\%-2.41\%\end{tabular}                          \\
				&                                                                                                                                                                                                                                                                                                                                                                                                                                                     &                                                                                                                                                                                                                                                                                                                                                                                                                                                                                                                 &                                                                                        &                                                                                                                                                   &                                                                                                                                                   &                                                                                                                                    \\
				\cite{37-Saini2002Artificial}           & \begin{tabular}[c]{@{}l@{}}Feedforward neural network \\ based on the conjugate \\ gradient (CG) backpropagation \\ algorithm (Fletcher–Reeves \\ conjugate gradient \\ backpropagation method \\ (FRCGBP), Polak–Ribiere \\ conjugate gradient \\ backpropagation method \\ (PRCGBP), Powell–Beale \\ conjugate gradient backpropagation\\  method (PBCGBP) and scaled \\ conjugate gradient backpropagation\\  method (SCGBP)) + PCA\end{tabular} & \begin{tabular}[c]{@{}l@{}}Peak load of previous day, \\ weather variables \\ (temperature of the day \\ (maximum. minimum), \\ gross minimum temperature. \\ rainfall, evaporation per day, \\ sunshine hours of \\ the previous day, \\ wind speed, the dry bulb \\ temperature, the wet bulb \\ temperature. the vapour pressure, \\ the relative humidity (\%) and \\ the soil temperature ), \\ calendar variables \\ (seasons, year number, \\ day of the year, day of the week, \\ holiday)\end{tabular} & \begin{tabular}[c]{@{}l@{}}Daily peak \\ demand forecast\\ (STPLF)\end{tabular}        & \begin{tabular}[c]{@{}l@{}}Region\\ (a 220 kV substation \\ of Haryana Vidyut \\ Prasaran Nigam Ltd. \\ (HVPNL) )\end{tabular}                    & \begin{tabular}[c]{@{}l@{}}Load values \\ (peak load value \\ of the current day/ \\ one to seven days \\ ahead)\end{tabular}                     & \begin{tabular}[c]{@{}l@{}}MAPE: \\ FRCGBP:\\ 2.43\%, \\ PRCGBP: \\ 2.31\%, \\ PBCGBP: \\ 2.32\%, \\ SCGBP: \\ 2.40\%\end{tabular} \\
				&                                                                                                                                                                                                                                                                                                                                                                                                                                                     &                                                                                                                                                                                                                                                                                                                                                                                                                                                                                                                 &                                                                                        &                                                                                                                                                   &                                                                                                                                                   &                                                                                                                                    \\
				\cite{63newc-AMINNASERI20081302} & \begin{tabular}[c]{@{}l@{}}Self-organizing map (SOM) + \\ Feedforward neural network \\ + PCA\end{tabular}                                                                                                                                                                                                                                                                                                                                          & \begin{tabular}[c]{@{}l@{}}Peak load, \\ weather variables \\ (temperature, wind speed, \\ cloud cover, relative humidity), \\ calendar variables \\ (day of the week, \\ week of the month, \\ month of year, year number, \\ holidays)\end{tabular}                                                                                                                                                                                                                                                           & \begin{tabular}[c]{@{}l@{}}Daily peak \\ demand forecast\\ (STPLF)\end{tabular}        & \begin{tabular}[c]{@{}l@{}}Region \\ (Tehran Regional \\ Electric Utility \\ Company)\end{tabular}                                                & \begin{tabular}[c]{@{}l@{}}Load values \\ (peak load value \\ one day ahead)\end{tabular}                                                         & \begin{tabular}[c]{@{}l@{}}MAPE: \\ 1.5\%-2.61\%\end{tabular}                                                                      \\
				&                                                                                                                                                                                                                                                                                                                                                                                                                                                     &                                                                                                                                                                                                                                                                                                                                                                                                                                                                                                                 &                                                                                        &                                                                                                                                                   &                                                                                                                                                   &                                                                                                                                    \\
				\cite{126-2019Deep}          & \begin{tabular}[c]{@{}l@{}}Dynamic time warping + \\ Bespoke gated Recurrent \\ Neural Network (RNN)\end{tabular}                                                                                                                                                                                                                                                                                                                                   & \begin{tabular}[c]{@{}l@{}}Historical load every 30 minutes, \\ weather variables \\ (daily average temperature), \\ calendar variables \\ (day of the week, holidays)\\ Forecast output:\\ single load value (daily peak load)\end{tabular}                                                                                                                                                                                                                                                                    & \begin{tabular}[c]{@{}l@{}}Daily peak \\ demand forecast\\ (STPLF)\end{tabular}        & \begin{tabular}[c]{@{}l@{}}Country \\ (competition data \\ from the European \\ Network on Intelligent \\ Technologies (EUNITE))\end{tabular} & \begin{tabular}[c]{@{}l@{}}Load values \\ (daily peak load)\end{tabular}                                                                          & \begin{tabular}[c]{@{}l@{}}MAPE: \\ 1.01\%\end{tabular}                                                                            \\
				&                                                                                                                                                                                                                                                                                                                                                                                                                                                     &                                                                                                                                                                                                                                                                                                                                                                                                                                                                                                                 &                                                                                        &                                                                                                                                                   &                                                                                                                                                   &                                                                                                                                    \\ \hline
		\end{tabular}}
		\label{t5}
	\end{center}
\end{table}

\subsubsection{Support Vector Machines}

As one popular machine learning method, support vector machines can minimize actual risk by seeking risk minimization so that to get satisfactory forecasting performance. The variation of SVMs for regression problems is represented as support vector regression (SVR) \cite{85-21}, which is efficient for large-scale regression problems \cite{85-5}.

Given a training dataset $T={(\mathbf{x_1}, y_1),... , (\mathbf{x_i}, y_i),... , (\mathbf{x_m}, y_m)}$ where $\mathbf{x_i}$$\in$$R^n$ denotes the $i$-th observation ($n$-dimensional input vector), $y_i$$\in$$R$ is the output corresponding to $\mathbf{x_i}$, and $m$ denotes the size of training set. For non-linear SVMs, the basic idea is to introduce kernel as below:
\begin{equation}\label{eq1}
y=\langle{\mathbf{w},\phi(\mathbf{x})}\rangle+ b
\end{equation} where $\phi(\mathbf{x})$ is the hypothetical higher dimensional feature space. Coefficients $\mathbf{w}$ and $b$ need to be estimated based on the structure risk minimization principle.

\cite{64-2009Forecasting} introduced local prediction based on SVM for electric daily peak load forecast. The local prediction can find the approximation function in the reconstructed embedded space. The partitioned inputs were assigned with \textcolor{black}{an} SVM model in each sub-domains, and thus local prediction could make better forecasts than the single/global model. \cite{104-2016Peak} developed a novel online-SVM model based on the standard SVM. The proposed model was used to forecast daily peak load for \textcolor{black}{the} residential building in Surrey, and results showed that the model could be a more intelligent tool for smart grid systems. \cite{115-8319611} adopted SVM to build a model for monthly peak load prediction. Firstly, feature selection was implemented based on correlation analysis. Then the training set was reconstructed by the topology network and random walk with restart (RWR) algorithm. Moreover, a feed-forward correlation was utilized to minimize the effect of unknown errors. Finally, the preprocessed training data was fed into SVM to train a model with higher accuracy.

Similar to ANN, SVMs are more suitable for STPLF and can cope with nonlinear and high dimensional data \cite{104-2016Peak}\cite{115-8319611}. The disadvantage of SVMs is also similar to that of ANN for suffering from long training time with large data sets. \textcolor{black}{Besides}, the hyperparameters of SVMs need to be manually selected, which is also a complex step that needed to be optimized.

\subsubsection{Ensemble Learning}

Ensemble learning trains multiple learners and aggregates each learner's predicted results to obtain the final output through combining strategies, which generally involve averaging, voting, and stacking \cite{Ensemble-Polikar:2009}. According to the dependencies between learners, one possible classification of the popular ensemble learning methods is as follows:

\begin{itemize}
    \item [(1)] 
    Learners have to be generated in sequence to satisfy the strong dependency between them (boosting).   
    \item [(2)]
    Learners are allowed to be \textcolor{black}{simultaneously generated} since there is no strong dependence between them (bagging and random forest).
\end{itemize}

Ensemble learning has been widely used in peak load demand forecast in recent years. Ensemble models used in the reviewed studies mainly are: boosting, bagging, and random forest.

Boosting adjusts the sample distribution according to the performance of the initial learner so that samples with the wrong prediction get more attention than others, and then it trains the next learner based on the adjusted sample. The process is iterated until a specified number of learner clusters are generated, or the aggregated learning criteria reaches the stop threshold \cite{Ensemble-Polikar:2009}. Commonly used boosting algorithms in the reviewed papers are adaptive boosting (AdaBoost), boosting tree, gradient boosting (GB) and Extreme gradient boosting (XGBoost). \cite{boosting2018-AHMAD20181008} adopted three machine learning models (ANN with nonlinear autoregressive exogenous multivariable inputs, multivariate linear regression, and AdaBoost) to predict load profile one month, one season, and one year ahead at \textcolor{black}{the} district level. During training, \textcolor{black}{datasets with different sizes were utilized for training models for different prediction intervals.} This paper also adopted feature extraction to select essential variables, and the results showed that the AdaBoost outperformed other models significantly for all prediction intervals. Moreover, for seasonal forecasting, the error range of AdaBoost was \textcolor{black}{relatively} narrow, which indicated that \textcolor{black}{the} model trained based on AdaBoost was more capable of capturing the dynamic change of load curves. \cite{boosting2019-ZHANG2019116358} conducted short-term load forecasting for southern California. In this study, different \textcolor{black}{models were adopted (multivariate linear regression, random forest, and GB) and the installed solar capacity was identified to be an important feature during the forecasting.} The comparative experiment results revealed two insights: (1) The fact that the installed solar capacity became an important feature suggested that new and clean energy resources are \textcolor{black}{important components in the system that researchers need to pay more attention to; (2) Different forecasting accuracy in different periods indicated that being able to capture the fluctuation} of load curves is important for forecast. \cite{boosting2020-LU2020117756} combined complete ensemble empirical mode decomposition with XGBoost to predict daily load consumption, daily peak load, and daily water delivery. Compared to traditional XGBoost, the hybrid model showed a lower MAPE of 5.99\% for the daily peak load demand forecast.

Bagging is based on bootstrap sampling. It carries out multiple times of put-back sampling for a given dataset and trains learners simultaneously based on the obtained sampling set. When bagging is applied to a regression task, a simple mean or median can be adopted to obtain the final output \cite{bagging2018-53}. \cite{bagging2018-DEOLIVEIRA2018776}, for the first time, utilized bagging to forecast monthly load demand for countries with different development stages. The paper combined bagging with exponential smoothing and SARIMA and then used simple mean and median to aggregate the results from single learners. A new variation of bagging, Remainder Sieve Bootstrap (RSB) was also proposed to enhance the forecasting results, and the result showed that the proposed method yielded the best MAPE for both developed and developing countries. 

Random forest (RF) can be seen as an extension of bagging, which further introduces random selection in constructing individual decision tree learners based on bagging. 

The RF firstly uses bootstrap to generate its training sets, and then a decision tree is constructed for each of the training set. Features are randomly selected and an optimization criteria is used to guide the split of nodes in constructing each decision tree learner. The prediction strategies of RF are: voting for the classification task, and averaging for the regression task \cite{Ensemble-Polikar:2009}.

As the number of learners increases, RF generally converges to a smaller generalization error than bagging. Moreover, the training efficiency of RF is often superior to bagging, benefiting from the randomness in constructing single learners. In \cite{97newd-FAN20141}, an ensemble method combining eight popular forecasting algorithms (multiple linear regression (MLR), ARIMA, SVM, RF, multi-layer perceptron, boosting tree, and multivariate adaptive regression splines) is proposed for peak load forecast. Each model in the studies was \textcolor{black}{assigned to a weight} by GA. The results showed that SVM and RF had the largest weights, which indicates that these two algorithms contributed more potential gains for enhancing peak load demand forecast accuracy. \cite{RF2018-WANG201811} adopted RF to predict hourly load usage patterns for two educational buildings in North Central Florida, and the feature importance distribution was also produced as a by-product. The proposed model was compared with \textcolor{black}{the} regression tree and SVM, and the results showed that RF had the best superiority among all the trained models. Moreover, the feature importance distribution also proved that the influential features changed depending on different education periods, which indicated that the load usage behavior of educational buildings is highly related to different semesters.

\subsubsection{Hybrid techniques}

Many novel hybrid models with satisfactory forecast performance have been proposed in the reviewed papers, and some of the models have already been discussed in the previous section. TABLE \ref{t6} summarised papers utilizing hybrid models according to combinations of methods in different forecast methods development stages.  

 \begin{table*}[!h]
    \caption{Summary of papers utilizing hybrid model}
    
    \begin{center}
        \resizebox{0.5\textwidth}{0.17\textwidth}{
            \begin{tabular}{cc}
                \hline
                Combination of the stages &
                Hybrid models with references \\ \hline
                \begin{tabular}[c]{@{}c@{}}Manual/human expert stage \\+ Classic stage\end{tabular} &
                Human knowledge + ARIMA \cite{30-Amjady2001Short} \\ \hline
                Classic stage + Classic stage &
                \begin{tabular}[c]{@{}c@{}}Decomposition + Regression \cite{9s-Barakat1990Short}\\ ARIMA + Regression \cite{63newm-5398613}\end{tabular} \\ \hline
                Classic stage + Advanced stage &
                \begin{tabular}[c]{@{}c@{}}Regression + FL \cite{28-1999The}\cite{29}\\ Regression + GA \cite{34}\cite{50-2005Comparison}\\ Regression + PCA \cite{92-6867500}\\ Exponential Smoothing + FL \cite{108newd-Laouafi2016Daily}\end{tabular} \\ \hline
                Advanced stage + Advanced stage &
                \begin{tabular}[c]{@{}c@{}}FL + ANN \cite{25-Mandal1997Fuzzy}\cite{27newdaa-Haida1998Peak}\cite{63newm-5398613}\cite{99aaa}\cite{108-7893595}\\ FL + ES \cite{21-1995Short}\cite{30newa-30Kiartzis2000A}\\ FL + SOM \cite{56-4075966}\\ FL + GA \cite{72-0The}\\ ANNs \cite{89aai-Hern2013Improved}\\ ANN + SOM \cite{48-1556396}\\ ANN + SOM + PCA \cite{63newc-AMINNASERI20081302}\\ ANN + GA \cite{81-6121762}\cite{88aaa-2013A}\cite{98-7011462}\\ GA + RBFN + SVM \cite{85aab-2012Building}\end{tabular} \\ \hline
        \end{tabular}}
        \label{t6}
    \end{center}
 \end{table*}

\textit{Manual + Classic stage.} There are a few papers that proposed hybrid models based on the combination of manual stage and classic stage, in which \cite{30-Amjady2001Short} was the earliest work among the obtained papers that utilized the combination of classic forecasting methods with human experience. In this paper, human experts’ opinions were selected as one of the initial input variables for the daily peak load demand forecast. The proposed modified ARIMA was compared with standard ARIMA, and the results revealed that the former had the best performance with the lowest MAPE of 1.01\% for predicting the daily peak load of cold Sunday to cold Wednesday.

\textit{Classic + Advanced stage.} Some papers combined methods from \textcolor{black}{the} classic stage with methods from the advanced stage. Among which, \cite{28-1999The} combined fuzzy logic with a regression model. The fuzzy set theory is good at representing the uncertainty of the data, which allows the use of additional customer information as inputs to the forecast model, and could achieve more accurate forecasts. \cite{92-6867500} used the combination of PCA and MLR to forecast weekly peak load at \textcolor{black}{the} distribution level. Firstly, the correlation analysis was utilized to select the important features, and the PCA was adopted to reduce the redundancy of the input dimensions. Finally, the output from PCA was applied to MLR to perform mid-term peak load prediction. This hybrid model was simpler than many advanced AI-based methods, yet could also achieve satisfactory forecast accuracy.

\textit{Advanced + Advanced stage.} From TABLE \ref{t6} we can see that most of the proposed hybrid models are the combinations of the \textcolor{black}{advanced stage methods}. Among which, \cite{88aaa-2013A} proposed a hybrid method to forecast daily peak load for Iran. The model was built using the combination of wavelet decomposition, NN, and GA. Historical load data and weather variables from three different cities were used to train the model. The proposed model was also compared with other advanced models, and the results showed that this model outperformed most of the models. \cite{30newa-30Kiartzis2000A}, proposed a hybrid model combining fuzzy logic with the expert system. In this study, fuzzy logic has the advantage of obtaining the uncertain and incomplete information from the \textcolor{black}{real-world} data, which will be then considered as the input of the expert system, such that the hybrid model can make more accurate predictions based on the acquired knowledge. \cite{63newm-5398613} and \cite{99aaa} both combined fuzzy logic with neural network. The advantage of the hybrid model is that \textcolor{black}{the neural network has strong self-learning ability and can make good use of the expression provided by fuzzy logic to produce forecasts with higher accuracy.} Moreover, the fuzzy neural network is effective when handling peak loads with strong fluctuations, and it is good at capturing the calendar effect than other advanced models. In \cite{85aab-2012Building}, the real-valued genetic algorithm (RGA) based neural network-SVM model was proposed. In the model, the neural network was responsible for producing the growth index for the forecast target, SVM was adopted to output the deviation value, and the RGA was adopted to select optimal parameters for the neural network and SVM. The experiment demonstrated that the proposed hybrid model had good performance on both short and mid-term load demand forecast.

\section{Discussion and summary}\label{S4}

This section will first give a comparative analysis of the peak demand forecast methods. Then, improving methods for peak load forecast models are discussed. Finally, a comprehensive summary and discussion of the papers reviewed will be presented.

\subsection{Comparative studies of different models}

Each forecasting method has its advantages and disadvantages, therefore, it is necessary to compare the performance of various forecasting methods to understand their advantages and limitations. To this end, we will summarize the existing comparative studies in the literature. Some representative comparative studies are listed in TABLE \ref{t7} including composite/combined models or comparison analysis. A composite model could refer to inter-methods composite models (i.e. hybrid models) and result-weighted composite models. Based on different development stages of forecast methods (classic stage and advanced stage), existing comparative studies could be categorized into intra-comparison (e.g. methods within the classic stage) and inter-comparison (e.g. methods from both classic and advanced stage). 

When compared with human expert opinions, the methods in the classic stage showed good performance, as presented in \cite{26aa-Masood1997EDSSF}. In the classic stage, regression, as the most popular method in the reviewed papers, are often selected as the benchmark for building hybrid models \cite{9a-1990A} \cite{37aa-2003Regional} \cite{50-2005Comparison} \cite{82aap-2011Prediction}. For instance, \cite{9a-1990A} combined the exponential smoothing with regression to forecast the daily peak load demand, and compared the results with the combination of ARIMA and regression. Results showed that the former model could alleviate the bias caused by the latter. \cite{82aap-2011Prediction} also used regression as a benchmark to compare its performance with the hybrid model (Reg-SARIMA-GARCH).

In comparing the methods in classic stage with methods in the advanced stage, \cite{37aa-2003Regional} utilized ANN and regression to forecast annual peak load demand for Taiwan and the results showed that ANN could achieve a better performance.  For instance, \cite{50-2005Comparison} combined GA with symbolic regression to build \textcolor{black}{an STPLF framework, and the results showed the hybrid model could achieve comparable performance to an ANN model.} 

\begin{table*}[htbp]
    \caption{Papers with composite model/comparative analysis}
    \begin{center}
        \resizebox{0.87\textwidth}{0.62\textwidth}{ 

            \begin{tabular}{lllll}
                \hline
                Reference    & Model/Experiment type                                                             & Detail                                                                                                                                                                                                                                        & Forecast contents                                                                                                       & Performance                                                                                                                                                                                                                                                                                                                                                                                      \\ \hline
                &                                                                                   &                                                                                                                                                                                                                                               &                                                                                                                         &                                                                                                                                                                                                                                                                                                                                                                                                  \\
                \cite{26aa-Masood1997EDSSF}         & Composite model                                                                   & \begin{tabular}[c]{@{}l@{}}Built an decision support \\ system (DSS) to compare \\ ARIMA, exponential smoothing \\ and human expert suggestions\end{tabular}                                                                                                           & \begin{tabular}[c]{@{}l@{}}Monthly peak load demand\\ for UAE\end{tabular}                                                   & \begin{tabular}[c]{@{}l@{}}Different models suited for \\ different areas (Abu-Dhabi: \\ ARIMA(1,1,1)(0,1,2), \\ Dubai: Exponential smoothing,\\ Sharjah: ARIMA(1,0,2)(0,1,3)); \\ DSS outperformed human experts\end{tabular}                                                                                                                                                                   \\
                &                                                                                   &                                                                                                                                                                                                                                               &                                                                                                                         &                                                                                                                                                                                                                                                                                                                                                                                                  \\
                \cite{37aa-2003Regional}         & Comparative analysis                                                              & ANN vs Regression                                                                                                                                                                                                                             & \begin{tabular}[c]{@{}l@{}}Annual peak load demand \\ for 4 regions of Taiwan \\ (Region of China)\end{tabular}              & \begin{tabular}[c]{@{}l@{}}MAPE of ANN vs Regression:\\ Northern: 1.06\% vs 2.45\%\\ Central: 1.73\% vs 8.52\%\\ Southern: 2.48\% vs 8.29\%\\ Eastern: 3.62\% vs 4.10\%\end{tabular}                                                                                                                                                                                                             \\
                &                                                                                   &                                                                                                                                                                                                                                               &                                                                                                                         &                                                                                                                                                                                                                                                                                                                                                                                                  \\
                \cite{9a-1990A}        & \begin{tabular}[c]{@{}l@{}}Composite model + \\ comparative analysis\end{tabular} & \begin{tabular}[c]{@{}l@{}}Applied exponential smoothing \\ to regression model. \\ Regression + ARIMA vs\\ Regression + exponential smoothing\end{tabular}                                                                                   & \begin{tabular}[c]{@{}l@{}}Hourly load, daily peak \\ demand for PG\&E\end{tabular}                                     & \begin{tabular}[c]{@{}l@{}}Regression + exponential \\ smoothing could almost eliminate \\ the negative bias caused by \\ regression + ARIMA\end{tabular}                                                                                                                                                                                                                                        \\
                &                                                                                   &                                                                                                                                                                                                                                               &                                                                                                                         &                                                                                                                                                                                                                                                                                                                                                                                                  \\
                \cite{50-2005Comparison}           & \begin{tabular}[c]{@{}l@{}}Composite model + \\ comparative analysis\end{tabular} & \begin{tabular}[c]{@{}l@{}}GP + symbolic regression\\ vs MLP (ANN)\end{tabular}                                                                                                                                                               & \begin{tabular}[c]{@{}l@{}}Daily peak load demand for \\ a distribution power system\\ in Romania\end{tabular}               & \begin{tabular}[c]{@{}l@{}}Maximum APE of GP + \\ symbolic regression vs MLP:\\ 10\% vs 11\% \\ MPE of  GP + symbolic \\ regression vs MLP:\\ 0.2\% vs 0.1\%\end{tabular}                                                                                                                                                                                                                        \\
                &                                                                                   &                                                                                                                                                                                                                                               &                                                                                                                         &                                                                                                                                                                                                                                                                                                                                                                                                  \\
                &                                                                                   &                                                                                                                                                                                                                                               &                                                                                                                         &                                                                                                                                                                                                                                                                                                                                                                                                  \\
                \cite{82aap-2011Prediction}         & \begin{tabular}[c]{@{}l@{}}Composite model +\\ comparative analysis\end{tabular}  & \begin{tabular}[c]{@{}l@{}}Piece-wise linear regression vs \\ SARIMA vs a SARIMA with \\ generalized autoregressive conditional \\ heteroskedastic errors \\ (SARIMA-GARCH)  vs \\ regression-SARIMA-GARCH \\ (Reg-SARIMA-GARCH)\end{tabular} & \begin{tabular}[c]{@{}l@{}}Daily peak load demand for \\ South Africa\end{tabular}                                           & \begin{tabular}[c]{@{}l@{}}MAPE for regression vs SARIMA \\ vs SARIMA-GARCH vs \\ Reg-SARIMA-GARCH:\\ 2.77\% vs 1.47\% vs 1.43\% \\ vs 1.42\%\end{tabular}                                                                                                                                                                                                                                       \\
                &                                                                                   &                                                                                                                                                                                                                                               &                                                                                                                         &                                                                                                                                                                                                                                                                                                                                                                                                  \\
                \cite{30newa-30Kiartzis2000A}        & \begin{tabular}[c]{@{}l@{}}Composite model +\\ comparative analysis\end{tabular}  & \begin{tabular}[c]{@{}l@{}}Fuzzy logic (FL) + expert system (ES) \\ vs NN with similar structure\end{tabular}                                                                                                                                 & \begin{tabular}[c]{@{}l@{}}Morning and afternoon \\ valley, noon and evening \\ peak for different seasons\end{tabular} & \begin{tabular}[c]{@{}l@{}}Yearly ME for FL + ES vs NN:\\ 2.45\% vs 2.56\%\end{tabular}                                                                                                                                                                                                                                                                                                          \\
                &                                                                                   &                                                                                                                                                                                                                                               &                                                                                                                         &                                                                                                                                                                                                                                                                                                                                                                                                  \\
                \cite{63newm-5398613}       & \begin{tabular}[c]{@{}l@{}}Composite model +\\ comparative analysis\end{tabular}  & ANFIS (Fuzzy logic + NN) vs ARMA                                                                                                                                                                                                              & \begin{tabular}[c]{@{}l@{}}Daily peak load demand forecast \\ for a utility company \\ in Malaysia\end{tabular}             & \begin{tabular}[c]{@{}l@{}}Average MAPE for ANFIS \\ vs ARMA:\\ Including weekends: \\ 7.26\% vs 2.45\%\\ Excluding weekends: \\ 1.95\% vs 4.67\%\end{tabular}                                                                                                                                                                                                                                   \\
                &                                                                                   &                                                                                                                                                                                                                                               &                                                                                                                         &                                                                                                                                                                                                                                                                                                                                                                                                  \\
                \cite{85aab-2012Building}         & \begin{tabular}[c]{@{}l@{}}Composite model +\\ comparative analysis\end{tabular}  & \begin{tabular}[c]{@{}l@{}}RGA-SVM (real-valued genetic \\ algorithm-SVM) vs KF vs RBF \\ vs RGA based NN-SVM \\ (NN thread to output yearly growth \\ index and SVM thread to output \\ moment-specific forecast )\end{tabular}              & \begin{tabular}[c]{@{}l@{}}Daily peak load demand \\ with occurring time\end{tabular}                                        & \begin{tabular}[c]{@{}l@{}}MAPE for RGA-SVM vs KF vs \\ RBF vs RGA based NN-SVM:\\ one week:14.26\% vs 6.63\% \\ vs 6.39\% vs 3.20\%\\ Occurring time: Hit rate of RGA \\ based NN-SVM (one-hour time \\ deviation tolerance): 91.7\%\end{tabular}                                                                                                                                                 \\
                &                                                                                   &                                                                                                                                                                                                                                               &                                                                                                                         &                                                                                                                                                                                                                                                                                                                                                                                                  \\
                \cite{128-8791587}          & Comparative analysis                                                              & \begin{tabular}[c]{@{}l@{}}Regard peak hour forecast \\ as a classification problem;\\ Naive Bayes vs SVM vs Random \\ Forest (RF) vs AdaBoost vs CNN \\ vs LSTM vs Stacked\\ AutoEncoder (SAE)\end{tabular}                                  & \begin{tabular}[c]{@{}l@{}}Daily peak hour one day ahead\\ for Ontario\end{tabular}                                     & \begin{tabular}[c]{@{}l@{}}Accuracy for Naive Bayes vs \\ SVM vs RF vs AdaBoost \\ vs CNN vs LSTM vs SAE:\\ Winter: 0.83 vs 0.97 vs 0.97 \\ vs 0.82 vs 0.97 vs 0.98 vs 0.97\\ Summer: 0.66 vs 0.95 vs 0.94 \\ vs 0.63 vs 0.94 vs 0.95 vs 0.94\end{tabular}                                                                                                                                       \\
                &                                                                                   &                                                                                                                                                                                                                                               &                                                                                                                         &                                                                                                                                                                                                                                                                                                                                                                                                  \\
                \cite{boosting2020-LU2020117756} & \begin{tabular}[c]{@{}l@{}}Composite model +\\ comparative analysis\end{tabular}  & \begin{tabular}[c]{@{}l@{}}CEEMDAN(complete ensemble \\ empirical mode decomposition)\\ -XGBoost vs CEEMDAN-RF vs \\ RBFNN vs PSO-SVM vs LSSVM \\ (least squares SVM)\end{tabular}                                                            & \begin{tabular}[c]{@{}l@{}}Daily energy consumption\\ + daily peak power +\\ daily water delivery\end{tabular}          & \begin{tabular}[c]{@{}l@{}}MAPE for CEEMDAN-XGBoost\\ vs CEEMDAN-RF vs RBFNN vs \\ PSO-SVM vs LSSVM vs XGBoost:\\ daily energy consumption: 4.85\%\\ vs 6.26\% vs 7.67\% vs 7.92 vs \\ 7.87\% vs 8.06\%\\ daily peak power: 5.99\% vs \\ 6.40\% vs 9.25\% vs 8.15\% vs \\ 8.89\% vs 9.13\%\\ daily water delivery: 5.09\% vs \\ 6.31\% vs 8.30\% vs 8.08\%\\ vs 8.37\% vs 8.32\%\end{tabular} \\
                &                                                                                   &                                                                                                                                                                                                                                               &                                                                                                                         &                                                                                                                                                                                                                                                                                                                                                                                                  \\ \hline
        \end{tabular}}
        \label{t7}
    \end{center}
\end{table*}

Considering different distribution and diversity of the data and problem, hybrid methods are not always achieving satisfactory performance and sometimes could be counterproductive. \cite{63newm-5398613} compared ARMA with a hybrid model (FL + ANN) for daily peak load demand forecast. The obtained results showed that ARMA performed better when samples were trained with weekends whereas the proposed hybrid model gained better forecast accuracy when excluding the weekends. \cite{85aab-2012Building} combined real-valued GA with SVM and ANN, and the hybrid model was then used for producing daily peak load and its occurring time. The experiment compared the proposed model with other models (real-valued GA-SVM, KF, RBFN). Under the same \textcolor{black}{experimental} conditions, surprisingly the real-valued GA-SVM model produced the worst results.

There are also a few studies that conducted the comparative analysis based on methods in the advanced stage only. For example, \cite{128-8791587} formulated peak load forecast as a classification problem and compared several methods including LSTM, SVM, RF, CNN, and Adaboost. The results showed that among all the methods, LSTM had the best performance following by SVM, RF, and CNN, whereas Adaboost produced unsatisfactory forecast results. \cite{boosting2020-LU2020117756} proposed a hybrid model (complete ensemble empirical mode decomposition (CEEMDAN) combined with XGBoost), which was then compared with other models such as CEEMDAN-RF, RBFN. The results revealed that the proposed model generated the best performance, whereas RBFN had the largest MAPE among these three models.

\subsection{Improving methods for peak load demand forecast models} \label{improving}

As aforementioned, hybrid methods by combing different methods could be an option to improve the forecast accuracy. There are some other measures that could be taken to further improve the forecast performance such as through optimizing the model inputs (data normalization, feature selection and transformation), and improving the models/algorithms (e.g. by integrating clustering methods). 

\subsubsection{\textbf{Data}}

The magnitude difference between the data set and various variables is likely to lead to the deviation prediction of the training algorithm. Many training algorithms, such as SVR, require input variables of a similar order of magnitude. Beside, in the real scenarios, load data often need to be normalized due to privacy requests \cite{73aa-GOIA2010700}. Therefore, data normalization is a necessary preprocessing step for training the model. Among the reviewed papers, the commonly used data normalization methods are: zero mean normalization (Z-score normalization) \cite{63newc-AMINNASERI20081302}\cite{104-2016Peak} and Min-Max normalization \cite{89-2014Linguistic}\cite{108newl-Julio2016Linear}.

\textcolor{black}{The} training data size is another important factor that could affect the output accuracy of the model in the training process. If the training size is too small, information learned by the model will be insufficient, and the performance will be poor as a consequence. On the other hand, too much training data will lead to \textcolor{black}{low computational efficiency}. Therefore, a good trade-off and balance between the training size and the computation time is worth investigating.  \cite{108aaa-2016Artificial} considers training data of four different lengths (one week to four weeks) in forecasting sub-hourly load usage and daily peak load. The training results showed that the larger the training size, the higher the training accuracy of the neural network model. In \cite{84aaf-2012Finding}, different training sizes were used to predict the peak load two days to one week ahead. Specifically, the training data are the hourly load from New South Wales in the past three months, six months, nine months, and one year, respectively. \textcolor{black}{The results showed that the model trained with six months of historical data is the best at predicting the peak load in the coming days among all the models.}

\subsubsection{\textbf{Feature transformation/Feature selection}}

\textcolor{black}{As aforementioned, the input variables are often numerous especially in the big data era when training a forecasting model.} However, many variables may have unrelated characteristics with the target/ response variable , and variables may also be interdependent, which may easily lead to long training time and decreased forecast performance.

Feature transformation and feature selection are usually adopted to address the problem \cite{FST-escolano2009feature}.
 
Feature transformation aims to get \textcolor{black}{transformed features} by creating a new feature space and the commonly used methods include PCA, independent component analysis (ICA), and linear discriminant analysis (LDA). Feature selection \cite{fs-DASH1997131} is to select a subset from the original feature space and commonly used methods include filtering, wrapper and embedding.  

Most of the reviewed studies utilizing feature selection on peak load forecast adopted filtering and wrapper \cite{108aaa-2016Artificial} while those using feature transformation utilized PCA. For instance, PCA was compared with correlation analysis in \cite{97newd-FAN20141}, in which the original pattern matrix of the training data is 28$\times$1095. Through correlation analysis, variables with correlation factors greater than 0.95 were selected. By applying PCA, the dimension of the input matrix was reduced to 11$\times$100. After combining the user-defined neural network to train \textcolor{black}{the} model for daily peak load demand forecast, the forecasting accuracy showed that the trained model using PCA was superior to correlation analysis both in computational time and training accuracy.

\subsubsection{\textbf{Clustering methods}}

With the installation of smart meters, high resolution distributed energy consumption data (e.g. at building levels) becomes available, which provides opportunities in studying different behaviours of forecast models under different buildings. For instance, \cite{dai2020energy} compared performance of different forecast models on different buildings and concluded that clustering buildings based on their historical load usage patterns should be considered to produce more meaning insights (e.g. to improve forecast accuracy) instead of their predefined building use types. Clustering methods divide the data into different clusters according to certain standards, such as distance criterion. After clustering, the data within the same cluster have great similarity, while the data belonging to different clusters have great difference \cite{3-2008A}. The accuracy of peak load forecast can be improved by training different models for different clusters and then obtaining the aggregated final forecasts.

Clustering methods can be divided into partition based clustering (e.g. $K$-means), hierarchical clustering, density-based clustering (e.g. density-based spatial clustering of applications with noise (DBSCAN)), and model-based clustering (e.g. Gaussian mixture models) \cite{4-2013The}. Some studies employing clustering in the peak load forecast are summarized in TABLE \ref{t8} where commonly used clustering algorithms in peak load forecast are $K$-means, hierarchical clustering, SOM, and fuzzy clustering (FC).

\begin{table}[htbp]
	\caption{Clustering methods for improving the performance of peak load forecast}
    \begin{center} 
        \begin{tabular}{ll}
            \hline
            Clustering methods         & References              \\ \hline
            K-means and its extensions & \cite{97newd-FAN20141},\cite{53-2007A},\cite{108newl-Julio2016Linear} \\
            Hierarchical clustering    & \cite{55-2006Peak},\cite{58-2007Load}            \\
            Fuzzy clustering           & \cite{108newd-Laouafi2016Daily},\cite{89-2014Linguistic},\cite{41-2004Peak},\cite{49-2005An},        \\
            SOM                        & \cite{56newd-2006Developed},\cite{10d-1991Design},\cite{63newc-AMINNASERI20081302}  \\ \hline
        \end{tabular}
        \label{t8}
    \end{center}
 \end{table}

$K$-means is a classical algorithm of partition based clustering, which has high efficiency when handling \textcolor{black}{large-scale data.} Some variants based on $K$-means, such as the entropy weighted $K$-means \cite{97newd-FAN20141} have also been used in peak load forecast. Hierarchical clustering can be classified into aggregation hierarchical clustering and splitting hierarchical clustering \cite{Cluster}. For instance, \cite{55-2006Peak} adopted hierarchical clustering to optimize the input daily data of a feed-forward neural network (FNN) to predict load usage during a peak period, and the results demonstrated that the FNN could converge more quickly and produce more accurate results. SOM is a commonly used clustering method owing to its unsupervised feature. \cite{63newc-AMINNASERI20081302} firstly utilized SOM to cluster peak loads, then each cluster was trained separately by FNN to get a specified model. Results showed that the proposed hybrid method is effective for daily peak load forecast.

The above methods belong to hard clustering since each data point can only be assigned to a single cluster. Instead, fuzzy clustering such as Fuzzy C-means is a soft clustering method where each observation can belong to multiple clusters with corresponding membership coefficients \cite{bezdek2013pattern}. For instance, \cite{41-2004Peak} adopted fuzzy clustering to cluster peak load patterns according to the working/ non-working days. \cite{49-2005An} combined fuzzy clustering with FNN to forecast load curves during peak load period.

\subsection{Summary of the reviewed studies}

TABLE \ref{t9} gives a comprehensive summary of the reviewed papers including their forecasting periods, forecast outputs, input variables, improving methods and geographical scope. 

It is worth mentioning that the classification of peak load forecast methods into three stages (i.e. manual, classic and advanced) generally aligns with the evolving of power systems. In the manual and classic stage, the traditional energy-intensive industry dominated the electricity market with a relative stable peak demand patterns. Peak load forecasting based on statistical methods were commonly used. With the development of smart grids and the changing energy landscape at both demand side (e.g. demand side management and electric vehicles) and supply side (e.g. intermittent renewable energy supply at both transmission and distribution level), peak demand patterns become more random and less predictable. As such, more advanced methods that can better take advantage of big data and capture complex patterns such as deep learning and hybrid machine learning methods are preferable choices. 

In addition, different from traditional load forecasting, the occurrence time and magnitudes of the peak demand are equally important in peak load forecasting. The peak load occurrence time is a field that may be more related to extreme value theories or quantile regression because of some rare events. Moreover, considering the uncertainty of peak load, it is also an effective forecasting method to take the peak load as anomalous data to quantify its occurrence probability and magnitude probability \cite{ProbabilisticLFFB}. 

 \begin{table*}[htbp]
    \caption{Comprehensive summary for the reviewed literature FS/FT: Feature Selection/Feature Transformation; H: Historical load data; W: Weather variables; C: Calendar variables; E/O: Economic/Other variables; V: Peak values; V+T: Peak values+Occurring time; LP: Load Profiles.}
    \begin{center}
        \resizebox{\textwidth}{0.5\textwidth}{ 
            \begin{tabular}{clllllllllllllll}
                \hline
                \multirow{2}{*}{Methods}                                                                    & \multicolumn{1}{c}{\multirow{2}{*}{\begin{tabular}[c]{@{}c@{}}Forecast \\ periods\end{tabular}}} & \multicolumn{2}{c}{Improving methods}                      & \multicolumn{4}{c}{Input variables}                                                             & \multicolumn{4}{c}{Geographic scope}                                                                                                                                    & \multicolumn{3}{c}{Output}                                               & \multicolumn{1}{c}{\multirow{2}{*}{References}}                                                                                                                                                                                                      \\ \cline{3-15}
                & \multicolumn{1}{c}{}                                                                             & \multicolumn{1}{c}{Clustering} & \multicolumn{1}{c}{FS/FT} & \multicolumn{1}{c}{H} & \multicolumn{1}{c}{W} & \multicolumn{1}{c}{C} & \multicolumn{1}{c}{E/O} & \multicolumn{1}{c}{Region} & \multicolumn{1}{c}{Country} & \multicolumn{1}{c}{City} & \multicolumn{1}{c}{\begin{tabular}[c]{@{}c@{}}Building/\\ Household\end{tabular}} & \multicolumn{1}{c}{V} & \multicolumn{1}{c}{V+T} & \multicolumn{1}{c}{LP} & \multicolumn{1}{c}{}                                                                                                                                                                                                                                 \\ \hline
                \multirow{3}{*}{Regression}                                                                 & STPLF                                                                                            & $\checkmark$                   &                           & $\checkmark$          & $\checkmark$          & $\checkmark$          &                         & $\checkmark$               & $\checkmark$                &                          & $\checkmark$                                                                      & $\checkmark$          & $\checkmark$            &                        & \begin{tabular}[c]{@{}l@{}}\cite{6},\cite{9s-Barakat1990Short},\cite{9a-1990A},\cite{19-2002Regression}, \cite{26news-1997Short}, \cite{27newdaa-Haida1998Peak}, \cite{28-1999The},\\ \cite{28newr-1999reg},\cite{34},\cite{41-2004Peak},\cite{58-2007Load},\cite{73aa-GOIA2010700},\cite{83-6345263},\cite{94-6934706},\\ \cite{95-2014Non},\cite{96-7022373},\cite{97newd-FAN20141}, \cite{108newl-Julio2016Linear},\cite{130-8810766}, \cite{118-2018Distribution} \end{tabular}                                                                                                                  \\
                & MTPLF                                                                                            & $\checkmark$                   & $\checkmark$              & $\checkmark$          & $\checkmark$          & $\checkmark$          & $\checkmark$            & $\checkmark$               &                             & $\checkmark$             &                                                                                   & $\checkmark$          &                         &                        & \cite{8-Turner2012Regression},\cite{92-6867500},\cite{93-2014Peak},\cite{124-8671682},\cite{90-2014Long}                                                                                                                                                                                                                                        \\
                & LTPLF                                                                                            &                                &                           & $\checkmark$          & $\checkmark$          & $\checkmark$          & $\checkmark$            & $\checkmark$               & $\checkmark$                & $\checkmark$             &                                                                                   & $\checkmark$          &                         &                        & \cite{27},\cite{106-2016Robust},\cite{117-0Long},\cite{70-2008Density}                                                                                                                                                                                                                                        \\ \hline
                \multirow{3}{*}{\begin{tabular}[c]{@{}c@{}}Stochastic \\ time series\\ models\end{tabular}} & STPLF                                                                                            &                                &                           & $\checkmark$          & $\checkmark$          & $\checkmark$          &                         & $\checkmark$               & $\checkmark$                &                          & $\checkmark$                                                                      & $\checkmark$          &                         &                        & \begin{tabular}[c]{@{}l@{}}\cite{84aaf-2012Finding},\cite{99aaa},\cite{102-7399017},\cite{30-Amjady2001Short},\cite{56aad-4202249},\cite{57aat-article}, \cite{8-78},\cite{63newm-5398613},\\ \cite{133-2020The},\cite{9s-Barakat1990Short}\cite{82aap-2011Prediction},\cite{73-5697708},\cite{99-7117006} \end{tabular}                                                                                                                                      \\
                & MTPLF {}                                                                                           &                                &                           & $\checkmark$          & $\checkmark$          & $\checkmark$          &                         & $\checkmark$               & $\checkmark$                & $\checkmark$             &                                                                                   & $\checkmark$          &                         &                        & \cite{62-2008Electricity},\cite{57aam-2007Monthly},\cite{12-1992New},\cite{7-IEEE},\cite{5-Fong2011The},\cite{80-2011Discrete},\cite{83aaf-2012Forecasting}                                                                                                                                                                                                                            \\
                & LTPLF                                                                                            &                                &                           &                       &                       &                       &                         &                            &                             &                          &                                                                                   &                       &                         &                        &                                                                                                                                                                                                                                                      \\ \hline
                \multirow{3}{*}{\begin{tabular}[c]{@{}c@{}}Time series\\ decomposition\end{tabular}}        & STPLF                                                                                            &                                &                           & $\checkmark$          & $\checkmark$          & $\checkmark$          &                         & $\checkmark$               & $\checkmark$                &                          &                                                                                   & $\checkmark$          &                         & $\checkmark$           & \cite{66-2009Long},\cite{88aaa-2013A},\cite{22-Choi1996A},\cite{108-7893595}                                                                                                                                                                                                                                    \\
                & MTPLF                                                                                            &                                &                           & $\checkmark$          & $\checkmark$          &                       &                         & $\checkmark$               &                             & $\checkmark$             &                                                                                   & $\checkmark$          &                         &                        & \cite{2-Gupta1971A},\cite{5-Fong2011The},\cite{8-Barakat1989Forecasting},\cite{68-2009The},\cite{80-2011Discrete}                                                                                                                                                                                                                                          \\
                & LTPLF                                                                                            &                                &                           &                       &                       &                       &                         &                            &                             &                          &                                                                                   &                       &                         &                        &                                                                                                                                                                                                                                                      \\ \hline
                \multirow{3}{*}{\begin{tabular}[c]{@{}c@{}}Exponential \\ smoothing\end{tabular}}           & STPLF                                                                                            &                                &                           & $\checkmark$          & $\checkmark$          &                       &                         & $\checkmark$               & $\checkmark$                &                          &                                                                                   & $\checkmark$          &                         &                        & \cite{9a-1990A},\cite{38aa-J2017Short},\cite{108newd-Laouafi2016Daily}                                                                                                                                                                                                                                         \\
                & MTPLF                                                                                            &                                &                           & $\checkmark$          & $\checkmark$          & $\checkmark$          &                         & $\checkmark$               & $\checkmark$                &                          &                                                                                   & $\checkmark$          &                         &                        & \cite{9s-Barakat1990Short},\cite{12-1992New},\cite{26aa-Masood1997EDSSF}                                                                                                                                                                                                                                         \\
                & LTPLF                                                                                            &                                &                           &                       &                       &                       &                         &                            &                             &                          &                                                                                   &                       &                         &                        &                                                                                                                                                                                                                                                      \\ \hline
                \multirow{3}{*}{\begin{tabular}[c]{@{}c@{}}Kalman\\ filtering\end{tabular}}                 & STPLF                                                                                            &                                &                           & $\checkmark$          & $\checkmark$          & $\checkmark$          & $\checkmark$            & $\checkmark$               &                             &                          &                                                                                   & $\checkmark$          & $\checkmark$            &                        & \cite{21-1995Short},\cite{85aab-2012Building}                                                                                                                                                                                                                                              \\
                & MTPLF                                                                                            &                                &                           &                       &                       &                       &                         &                            &                             &                          &                                                                                   &                       &                         &                        &                                                                                                                                                                                                                                                      \\
                & LTPLF                                                                                            &                                &                           &                       &                       &                       &                         &                            &                             &                          &                                                                                   &                       &                         &                        &                                                                                                                                                                                                                                                      \\ \hline
                \multirow{3}{*}{\begin{tabular}[c]{@{}c@{}}Grey \\ prediction\end{tabular}}                 & STPLF                                                                                            &                                &                           & $\checkmark$          & $\checkmark$          & $\checkmark$          &                         & $\checkmark$               &                             &                          &                                                                                   & $\checkmark$          &                         &                        & \cite{61-2008The}                                                                                                                                                                                                                                                  \\
                & MTPLF                                                                                            &                                &                           & $\checkmark$          &                       &                       &                         &                            &                             & $\checkmark$             &                                                                                   & $\checkmark$          &                         &                        & \cite{62-2008Electricity}                                                                                                                                                                                                                                                   \\
                & LTPLF                                                                                            &                                &                           & $\checkmark$          & $\checkmark$          &                       &                         & $\checkmark$               &                             &                          &                                                                                   &                       & $\checkmark$            &                        & \cite{20}                                                                                                                                                                                                                                                   \\ \hline
                \multirow{3}{*}{ANN}                                                                        & STPLF                                                                                            &                                & $\checkmark$              & $\checkmark$          & $\checkmark$          & $\checkmark$          &                         & $\checkmark$               & $\checkmark$                & $\checkmark$             & $\checkmark$                                                                      & $\checkmark$          & $\checkmark$            & $\checkmark$           & \begin{tabular}[c]{@{}l@{}}\cite{27newdaa-Haida1998Peak},\cite{37aa-2003Regional},\cite{54-2006Mutual},\cite{89aai-Hern2013Improved},\cite{99aaa},\cite{108aaa-2016Artificial},\cite{112-8104558},\\ \cite{120-8955339},\cite{130-8810766},\cite{131-8927078},\cite{10d-1991Design},\cite{10-1991Design},\cite{11-1991Artificial},\cite{11e-Saeed1991Electric},\cite{12s-1992Short},\\ \cite{13-Ho1992Short},\cite{28s-Drezga1999Short},\cite{31-Aboulmagd2001An},\cite{32-Matsui2002Peak},\cite{49-2005An},\cite{50-2005Comparison},\cite{55-2006Peak},\cite{48-1556396}, \\\cite{82-2011Daily},\cite{84-2012ANN},\cite{98-7011462},\cite{105-2016Load},\cite{108-7893595},\cite{110-2017Forecasting},\cite{123-2018Forecasting},\\ \cite{133-2020The},\cite{15-Onoda1993Next},\cite{16-Morioka1993Next},\cite{17-Mizukami1993Maximum},\cite{22-Choi1996A},\cite{25-Mandal1997Fuzzy}, \\ \cite{36-2002Artificial},\cite{88-2013Forecasting},\cite{101-7359274},\cite{26newc-1997Cascaded},\cite{56-4075966},\cite{63newc-AMINNASERI20081302},\cite{36-2002Artificial}\\ \cite{92-6867500},\cite{69-2010Peak},\cite{74aa-article},\cite{85aab-2012Building} \end{tabular} \\
                & MTPLF                                                                                            & $\checkmark$                   &                           & $\checkmark$          & $\checkmark$          & $\checkmark$          & $\checkmark$            & $\checkmark$               & $\checkmark$                & $\checkmark$             &                                                                                   & $\checkmark$          &                         &                        & \cite{57aal-Otavio2007Long},\cite{3-2008A},\cite{40-1372805},\cite{44-1412874},\cite{45-1414771},\cite{48-1556396}                                                                                                                                                                                                                           \\
                & LTPLF                                                                                            &                                & $\checkmark$              & $\checkmark$          & $\checkmark$          &                       & $\checkmark$            & $\checkmark$               & $\checkmark$                & $\checkmark$             &                                                                                   & $\checkmark$          &                         &                        & \cite{37-Saini2002Artificial}                                                                                                                                                                                                                                                   \\ \hline
                \multirow{3}{*}{\begin{tabular}[c]{@{}c@{}}Expert\\ system\end{tabular}}                    & STPLF                                                                                            &                                &                           & $\checkmark$          & $\checkmark$          &                       &                         & $\checkmark$               & $\checkmark$                &                          &                                                                                   & $\checkmark$          &                         &                        & \cite{13aa-Hsu1992Fuzzy},\cite{30newa-30Kiartzis2000A}                                                                                                                                                                                                                                         \\
                & MTPLF                                                                                            &                                &                           &                       &                       &                       &                         &                            &                             &                          &                                                                                   &                       &                         &                        &                                                                                                                                                                                                                                                      \\
                & LTPLF                                                                                            &                                &                           & $\checkmark$          & $\checkmark$          & $\checkmark$          & $\checkmark$            & $\checkmark$               &                             &                          &                                                                                   & $\checkmark$          &                         &                        & \cite{36newL-2002Long}                                                                                                                                                                                                                                               \\ \hline
                \multirow{3}{*}{\begin{tabular}[c]{@{}c@{}}Fuzzy \\ logic\end{tabular}}                     & STPLF                                                                                            & $\checkmark$                   &                           & $\checkmark$          & $\checkmark$          & $\checkmark$          &                         & $\checkmark$               & $\checkmark$                &                          &                                                                                   & $\checkmark$          &                         &                        & \begin{tabular}[c]{@{}l@{}}\cite{13aa-Hsu1992Fuzzy},\cite{25-Mandal1997Fuzzy},\cite{27newdaa-Haida1998Peak},\cite{28-1999The},\cite{29},\cite{30-Amjady2001Short},\cite{38-1244703}, \\\cite{63newm-5398613},\cite{89-2014Linguistic},\cite{99aaa},\cite{108newd-Laouafi2016Daily},\cite{125aav-Jamaaluddin_2018} \end{tabular}                                                                                                                                                           \\
                & MTPLF                                                                                            &                                &                           & $\checkmark$          & $\checkmark$          & $\checkmark$          &                         &                            &                             &                          & $\checkmark$                                                                      & $\checkmark$          &                         &                        & \cite{72-0The}                                                                                                                                                                                                                                                   \\
                & LTPLF                                                                                            &                                &                           &                       &                       &                       &                         &                            &                             &                          &                                                                                   &                       &                         &                        &                                                                                                                                                                                                                                                      \\ \hline
                \multirow{3}{*}{\begin{tabular}[c]{@{}c@{}}Genetic \\ algorithm\end{tabular}}               & STPLF                                                                                            &                                &                           & $\checkmark$          & $\checkmark$          & $\checkmark$          &                         & $\checkmark$               & $\checkmark$                &                          & $\checkmark$                                                                      & $\checkmark$          &                         &                        & \cite{34},\cite{50-2005Comparison},\cite{72-0The},\cite{81-6121762},\cite{85aab-2012Building},\cite{88aaa-2013A},\cite{97newd-FAN20141}                                                                                                                                                                                                                           \\
                & MTPLF                                                                                            &                                &                           &                       &                       &                       &                         &                            &                             &                          &                                                                                   &                       &                         &                        &                                                                                                                                                                                                                                                      \\
                & LTPLF                                                                                            &                                &                           & $\checkmark$          &                       &                       &                         &                            & $\checkmark$                &                          &                                                                                   & $\checkmark$          &                         &                        & \cite{87aaa-El2013Electric}                                                                                                                                                                                                                                                 \\ \hline
                \multirow{3}{*}{SVMs}                                                                       & STPLF                                                                                            &                                & $\checkmark$              & $\checkmark$          & $\checkmark$          & $\checkmark$          &                         & $\checkmark$               & $\checkmark$                & $\checkmark$             & $\checkmark$                                                                      & $\checkmark$          & $\checkmark$            &                        & \cite{60-2008Special},\cite{61-2008The},\cite{64-2009Forecasting},\cite{104-2016Peak},\cite{111-2017Combining},\cite{115-8319611},\cite{116-8330143},\cite{128-8791587}                                                                                                                                                                                                                        \\
                & MTPLF                                                                                            &                                &                           &                       &                       &                       &                         &                            &                             &                          &                                                                                   &                       &                         &                        &                                                                                                                                                                                                                                                      \\
                & LTPLF                                                                                            &                                &                           &                       &                       &                       &                         &                            &                             &                          &                                                                                   &                       &                         &                        &                                                                                                                                                                                                                                                      \\ \hline
                \multirow{3}{*}{Boosting}                                                                   & STPLF                                                                                            &                                & $\checkmark$              & $\checkmark$          & $\checkmark$          & $\checkmark$          &                         & $\checkmark$               &                             &                          & $\checkmark$                                                                      & $\checkmark$          &                         & $\checkmark$           & \cite{boosting2019-ZHANG2019116358},\cite{boosting2020-LU2020117756}                                                                                                                                                                                                                                    \\
                & MTPLF                                                                                            &                                & $\checkmark$              & $\checkmark$          & $\checkmark$          & $\checkmark$          &                         & $\checkmark$               &                             &                          &                                                                                   &                       &                         & $\checkmark$           & \cite{boosting2018-AHMAD20181008}                                                                                                                                                                                                                                        \\
                & LTPLF                                                                                            &                                & $\checkmark$              & $\checkmark$          & $\checkmark$          & $\checkmark$          &                         & $\checkmark$               &                             &                          &                                                                                   &                       &                         & $\checkmark$           & \cite{boosting2018-AHMAD20181008}                                                                                                                                                                                                                                         \\ \hline
                \multirow{3}{*}{Bagging}                                                                    & STPLF                                                                                            &                                &                           &                       &                       &                       &                         &                            &                             &                          &                                                                                   &                       &                         &                        &                                                                                                                                                                                                                                                      \\
                & MTPLF                                                                                            &                                &                           & $\checkmark$          & $\checkmark$          & $\checkmark$          & $\checkmark$            &                            & $\checkmark$                &                          &                                                                                   & $\checkmark$          &                         &                        & \cite{bagging2018-DEOLIVEIRA2018776}                                                                                                                                                                                                                                          \\
                & LTPLF                                                                                            &                                &                           &                       &                       &                       &                         &                            &                             &                          &                                                                                   &                       &                         &                        &                                                                                                                                                                                                                                                      \\ \hline
                \multirow{3}{*}{RF}                                                                         & STPLF                                                                                            & $\checkmark$                   & $\checkmark$              & $\checkmark$          & $\checkmark$          & $\checkmark$          & $\checkmark$            &                            &                             &                          & $\checkmark$                                                                      & $\checkmark$          &                         & $\checkmark$           & \cite{97newd-FAN20141},\cite{RF2018-WANG201811},\cite{RF2020-SATREMELOY2020114246}                                                                                                                                                                                                                                   \\
                & MTPLF                                                                                            &                                &                           &                       &                       &                       &                         &                            &                             &                          &                                                                                   &                       &                         &                        &                                                                                                                                                                                                                                                      \\
                & LTPLF                                                                                            &                                &                           &                       &                       &                       &                         &                            &                             &                          &                                                                                   &                       &                         &                        &                                                                                                                                                                                                                                                      \\ \hline
                \multirow{3}{*}{CNN}                                                                        & STPLF                                                                                            &                                &                           & $\checkmark$          & $\checkmark$          & $\checkmark$          &                         & $\checkmark$               &                             &                          &                                                                                   &                       & $\checkmark$            &                        & \cite{128-8791587},\cite{129-8881305}                                                                                                                                                                                                                                              \\
                & MTPLF                                                                                            &                                &                           &                       &                       &                       &                         &                            &                             &                          &                                                                                   &                       &                         &                        &                                                                                                                                                                                                                                                      \\
                & LTPLF                                                                                            &                                &                           &                       &                       &                       &                         &                            &                             &                          &                                                                                   &                       &                         &                        &                                                                                                                                                                                                                                                      \\ \hline
                \multirow{3}{*}{RNN}                                                                        & STPLF                                                                                            &                                &                           & $\checkmark$          & $\checkmark$          & $\checkmark$          &                         & $\checkmark$               &                             &                          & $\checkmark$                                                                      & $\checkmark$          & $\checkmark$            &                        & \cite{126-2019Deep},\cite{127-2019Evolutionary},\cite{128-8791587}                                                                                                                                                                                                                                          \\
                & MTPLF                                                                                            &                                &                           & $\checkmark$          & $\checkmark$          & $\checkmark$          & $\checkmark$            & $\checkmark$               & $\checkmark$                &                          &                                                                                   & $\checkmark$          &                         &                        & \cite{132-8985197},\cite{134-8994442}                                                                                                                                                                                                                                              \\
                & LTPLF                                                                                            &                                &                           &                       &                       &                       &                         &                            &                             &                          &                                                                                   &                       &                         &                        &                                                                                                                                                                                                                                                      \\ \hline
        \end{tabular}}
        \label{t9}
    \end{center}
 \end{table*}

For the input variables, historical load data, weather variables, and calendar variables were usually used in STPLF. On the other hand, economic and other variables such as population growth rate were frequently used in MTPLF and LTPLF. Moreover, due to the high randomness of peak load, the forecast model is greatly affected by small probability events such as extreme weather and accidental events. Accidental events vary among individuals/ entities, whose impact is often limited to a small range and therefore difficult to forecast. Although extreme weather also belongs to the small probability event, its impact is usually well studied.  For instance, it is necessary to focus on climatic factors such as the maximum (lowest) temperature, the duration of the high (low) temperature, and the humidity. The maximum (lowest) temperature determines the peak load value. The duration of high (low) temperature affects the peak occurring time range, and the humidity further aggravates the difference between the physical and the actual temperature, thus affecting consumers' electricity usage decisions.

Most of the forecast methods utilizing improving techniques (e.g. clustering) belong to the advanced stage of peak load demand forecast, which indicates that research attention to this field is increasing. However, it is worth pointing out that although there are many clustering methods available, the current load curvy-based clustering heavily relies on additional physical and social-economic information from entities/users (in other words, the domain knowledge) in order to properly interpret clustering results. More efforts on how to effectively incorporate domain knowledge into the forecast methods and improving techniques are needed. 

As for the forecast geographic scope, researchers usually consider peak load forecast over a wide range of regions or countries during the classic stage. With the development of smart grids and the installation of smart meters at the local level, there are more high resolution temporal and spatial data becoming available  \cite{4-2013The}. In addition, the increasing penetration of distributed energy resources (e.g. electric vehicles and microgrids) coupled with distributed intelligence and local energy applications \cite{5-Fong2011The} brings the operation and maintenance of power systems into a new era of disaggregated environment. From the perspective of peak load forecast , the highly random human activities  will have higher impact on the forecast performance in small geographic areas (e.g. community level) than aggregated level (e.g. region/country)  \cite{107-2016Distribution}. Therefore, more research efforts on the interaction between electricity usage decisions of end users and disaggregated load forecasting are needed in the future. 
                                                                                 
\section{Conclusion and future work}\label{S5}

 Motived by the importance of peak load demand forecast from the perspectives of electricity market stakeholders, this paper carries out a systematic review of the peak load demand forecast, which aims to summarize existing studies on the topic and provide guidance for future research. First, we aim to provide an unified problem definition for peak demand load forecast. Then, the peak load demand forecast methods were categorized into three stages based on their development timeline, and a thorough review of relevant methods in each stage was conducted. Moreover, a comparative analysis of different forecast methods was presented, and useful improving techniques for enhancing the forecast performance were discussed. Finally,  a comprehensive summary \textcolor{black}{of} reviewed papers on the peak load forecast framework was presented with possible future research directions.

With high-resolution load data (e.g. residential smart meter data) becoming increasingly available, data privacy is an important issue that needs to be addressed. In the new digital era, \textcolor{black}{using} private encryption algorithms to protect the consumers' data has become an essential task that researchers must deal with \cite{6-Lim2016Security}. There are challenges in terms of electricity data transmission and storage compliance, security and privacy protection \cite{privacy1}. In addition, it is well known that the data size and quality usually determines the training quality of machine learning models. However, in practice relevant data that are deemed necessary for a forecast task might be owned by different organizations.  To make accurate predictions, it is necessary to combine diverse data sources from different organizations in building the model. This could be achieved by aggregating all the data sources into a third party central database, however, it may face inevitable security risks because of the central distribution of the data \cite{privacy2}. Therefore, \textcolor{black}{designing} the forecasting framework under the premise of meeting data privacy, security, and regulatory requirements (e.g. through federated learning \cite{yang2019federated}) is an important future research trend on peak load demand forecast.

\bibliographystyle{cas-model1-num-names}
\bibliography{EPDF}

\begin{thebibliography}{178}
\expandafter\ifx\csname natexlab\endcsname\relax\def\natexlab#1{#1}\fi
\providecommand{\url}[1]{\texttt{#1}}
\providecommand{\href}[2]{#2}
\providecommand{\path}[1]{#1}
\providecommand{\DOIprefix}{doi:}
\providecommand{\ArXivprefix}{arXiv:}
\providecommand{\URLprefix}{URL: }
\providecommand{\Pubmedprefix}{pmid:}
\providecommand{\doi}[1]{\href{http://dx.doi.org/#1}{\path{#1}}}
\providecommand{\Pubmed}[1]{\href{pmid:#1}{\path{#1}}}
\providecommand{\bibinfo}[2]{#2}
\ifx\xfnm\relax \def\xfnm[#1]{\unskip,\space#1}\fi
\bibitem[{Zhou and Brown(2017)}]{AR-5}
\bibinfo{author}{S.~Zhou}, \bibinfo{author}{M.~A. Brown},
\newblock \bibinfo{title}{Smart meter deployment in europe: A comparative case
  study on the impacts of national policy schemes},
\newblock \bibinfo{journal}{Journal of Cleaner Production}
  \bibinfo{volume}{144} (\bibinfo{year}{2017}) \bibinfo{pages}{22--32}.
\bibitem[{AR-(2020)}]{AR-6}
\bibinfo{title}{Office of gas and electricity markets (ofgem), transition to
  smart meters.},
  \bibinfo{howpublished}{https://www.ofgem.gov.uk/gas/retail-market/metering/transition-smart-meters/},
  \bibinfo{year}{2020}. \bibinfo{note}{[accessed 18 December 2020]}.
\bibitem[{{Rahimi} and {Ipakchi}(2010)}]{Areview-1}
\bibinfo{author}{F.~{Rahimi}}, \bibinfo{author}{A.~{Ipakchi}},
\newblock \bibinfo{title}{Demand response as a market resource under the smart
  grid paradigm},
\newblock \bibinfo{journal}{IEEE Transactions on Smart Grid}
  \bibinfo{volume}{1} (\bibinfo{year}{2010}) \bibinfo{pages}{82--88}.
\bibitem[{Sheffrin et~al.(2008)Sheffrin, Yoshimura, LaPlante, and
  Neenan}]{reason-6}
\bibinfo{author}{A.~Sheffrin}, \bibinfo{author}{H.~Yoshimura},
  \bibinfo{author}{D.~LaPlante}, \bibinfo{author}{B.~Neenan},
\newblock \bibinfo{title}{Harnessing the power of demand},
\newblock \bibinfo{journal}{The Electricity Journal} \bibinfo{volume}{21}
  (\bibinfo{year}{2008}) \bibinfo{pages}{39--50}.
\bibitem[{Espinoza et~al.(2007)Espinoza, Suykens, Belmans, and
  De~Moor}]{reason-10}
\bibinfo{author}{M.~J.~C. Espinoza}, \bibinfo{author}{J.~U. Suykens},
  \bibinfo{author}{R.~U. Belmans}, \bibinfo{author}{B.~U.~J. De~Moor},
\newblock \bibinfo{title}{Electric load forecasting - using kernel based
  modeling for nonlinear system identification},
\newblock \bibinfo{journal}{IEEE Control Systems Magazine} \bibinfo{volume}{27}
  (\bibinfo{year}{2007}) \bibinfo{pages}{43--57}.
\bibitem[{Koutsopoulos and Tassiulas(2011)}]{reason-4}
\bibinfo{author}{I.~Koutsopoulos}, \bibinfo{author}{L.~Tassiulas},
\newblock \bibinfo{title}{Control and optimization meet the smart power grid:
  Scheduling of power demands for optimal energy management},
\newblock in: \bibinfo{booktitle}{Proceedings of the 2nd International
  Conference on Energy-efficient Computing and Networking},
  \bibinfo{year}{2011}, pp. \bibinfo{pages}{41--50}.
\bibitem[{Saebi et~al.(2010)Saebi, Taheri, Mohammadi, and
  Nayer}]{saebi2010demand}
\bibinfo{author}{J.~Saebi}, \bibinfo{author}{H.~Taheri},
  \bibinfo{author}{J.~Mohammadi}, \bibinfo{author}{S.~S. Nayer},
\newblock \bibinfo{title}{Demand bidding/buyback modeling and its impact on
  market clearing price},
\newblock in: \bibinfo{booktitle}{2010 IEEE International Energy Conference},
  \bibinfo{organization}{IEEE}, \bibinfo{year}{2010}, pp.
  \bibinfo{pages}{791--796}.
\bibitem[{Cappers et~al.(2010)Cappers, Goldman, and Kathan}]{cappers2010demand}
\bibinfo{author}{P.~Cappers}, \bibinfo{author}{C.~Goldman},
  \bibinfo{author}{D.~Kathan},
\newblock \bibinfo{title}{Demand response in us electricity markets: Empirical
  evidence},
\newblock \bibinfo{journal}{Energy} \bibinfo{volume}{35} (\bibinfo{year}{2010})
  \bibinfo{pages}{1526--1535}.
\bibitem[{Koh and Lim(2015)}]{koh2015evaluating}
\bibinfo{author}{S.~L. Koh}, \bibinfo{author}{Y.~S. Lim},
\newblock \bibinfo{title}{Evaluating the economic benefits of peak load
  shifting for building owners and grid operator},
\newblock in: \bibinfo{booktitle}{2015 International Conference on Smart Grid
  and Clean Energy Technologies (ICSGCE)}, \bibinfo{organization}{IEEE},
  \bibinfo{year}{2015}, pp. \bibinfo{pages}{30--34}.
\bibitem[{Sardi et~al.(2017)Sardi, Mithulananthan, Gallagher, and
  Hung}]{sardi2017multiple}
\bibinfo{author}{J.~Sardi}, \bibinfo{author}{N.~Mithulananthan},
  \bibinfo{author}{M.~Gallagher}, \bibinfo{author}{D.~Q. Hung},
\newblock \bibinfo{title}{Multiple community energy storage planning in
  distribution networks using a cost-benefit analysis},
\newblock \bibinfo{journal}{Applied energy} \bibinfo{volume}{190}
  (\bibinfo{year}{2017}) \bibinfo{pages}{453--463}.
\bibitem[{Mao et~al.(2009)Mao, Zeng, Leng, Zhai, and Keane}]{mao2009short}
\bibinfo{author}{H.~Mao}, \bibinfo{author}{X.-J. Zeng},
  \bibinfo{author}{G.~Leng}, \bibinfo{author}{Y.-J. Zhai},
  \bibinfo{author}{J.~A. Keane},
\newblock \bibinfo{title}{Short-term and midterm load forecasting using a
  bilevel optimization model},
\newblock \bibinfo{journal}{IEEE Transactions on Power Systems}
  \bibinfo{volume}{24} (\bibinfo{year}{2009}) \bibinfo{pages}{1080--1090}.
\bibitem[{SD(2020)}]{SD}
\bibinfo{title}{Sciencedirect (sd)},
  \bibinfo{howpublished}{https://www.sciencedirect.com/}, \bibinfo{year}{2020}.
  \bibinfo{note}{[accessed 16 August 2020]}.
\bibitem[{7-I(2020)}]{7-IEEE}
\bibinfo{title}{the institution of electrical and electronics engineers
  (ieee)}, \bibinfo{howpublished}{https://www.ieee.org/}, \bibinfo{year}{2020}.
  \bibinfo{note}{[accessed 16 August 2020]}.
\bibitem[{Mcsharry et~al.(2005)Mcsharry, Bouwman, and
  Bloemhof}]{46-2005Probabilistic}
\bibinfo{author}{P.~E. Mcsharry}, \bibinfo{author}{S.~Bouwman},
  \bibinfo{author}{G.~Bloemhof},
\newblock \bibinfo{title}{Probabilistic forecasts of the magnitude and timing
  of peak electricity demand},
\newblock \bibinfo{journal}{IEEE Transactions on Power Systems}
  (\bibinfo{year}{2005}) \bibinfo{pages}{1166--1172}.
\bibitem[{Wang(2006)}]{56newd-2006Developed}
\bibinfo{author}{Z.~Wang},
\newblock \bibinfo{title}{Developed case-based reasoning system for short-term
  load forecasting},
\newblock in: \bibinfo{booktitle}{2006 IEEE Power Engineering Society General
  Meeting}, \bibinfo{organization}{IEEE}, \bibinfo{year}{2006}, pp.
  \bibinfo{pages}{6--pp}.
\bibitem[{Hyndman and Fan(2009)}]{70-2008Density}
\bibinfo{author}{R.~J. Hyndman}, \bibinfo{author}{S.~Fan},
\newblock \bibinfo{title}{Density forecasting for long-term peak electricity
  demand},
\newblock \bibinfo{journal}{IEEE Transactions on Power Systems}
  \bibinfo{volume}{25} (\bibinfo{year}{2009}) \bibinfo{pages}{1142--1153}.
\bibitem[{Khan et~al.(2011)Khan, Khan, and Ullah}]{81-6121762}
\bibinfo{author}{G.~M. Khan}, \bibinfo{author}{S.~Khan},
  \bibinfo{author}{F.~Ullah},
\newblock \bibinfo{title}{Short-term daily peak load forecasting using fast
  learning neural network},
\newblock in: \bibinfo{booktitle}{2011 11th International Conference on
  Intelligent Systems Design and Applications}, \bibinfo{organization}{IEEE},
  \bibinfo{year}{2011}, pp. \bibinfo{pages}{843--848}.
\bibitem[{Yalcinoz and Eminoglu(2005)}]{46news-2005Short}
\bibinfo{author}{T.~Yalcinoz}, \bibinfo{author}{U.~Eminoglu},
\newblock \bibinfo{title}{Short term and medium term power distribution load
  forecasting by neural networks},
\newblock \bibinfo{journal}{Energy Conversion and Management}
  \bibinfo{volume}{46} (\bibinfo{year}{2005}) \bibinfo{pages}{1393--1405}.
\bibitem[{Stoll and Garver(1989)}]{46new1}
\bibinfo{author}{H.~G. Stoll}, \bibinfo{author}{L.~J. Garver},
  \bibinfo{title}{Least-cost electric utility planning},
  \bibinfo{publisher}{Wiley New York}, \bibinfo{year}{1989}.
\bibitem[{Yu et~al.(2019)Yu, Niu, Tang, and Wu}]{126-2019Deep}
\bibinfo{author}{Z.~Yu}, \bibinfo{author}{Z.~Niu}, \bibinfo{author}{W.~Tang},
  \bibinfo{author}{Q.~Wu},
\newblock \bibinfo{title}{Deep learning for daily peak load forecasting--a
  novel gated recurrent neural network combining dynamic time warping},
\newblock \bibinfo{journal}{Ieee Access} \bibinfo{volume}{7}
  (\bibinfo{year}{2019}) \bibinfo{pages}{17184--17194}.
\bibitem[{Amjady(2001)}]{30-Amjady2001Short}
\bibinfo{author}{N.~Amjady},
\newblock \bibinfo{title}{Short-term hourly load forecasting using time-series
  modeling with peak load estimation capability},
\newblock \bibinfo{journal}{IEEE Transactions on Power Systems}
  \bibinfo{volume}{16} (\bibinfo{year}{2001}) \bibinfo{pages}{798--805}.
\bibitem[{Phimphachan et~al.(2004)Phimphachan, Chamnongthai, Kumhom,
  Jittiwarangkul, and Sangswang}]{44-1412874}
\bibinfo{author}{S.~Phimphachan}, \bibinfo{author}{K.~Chamnongthai},
  \bibinfo{author}{P.~Kumhom}, \bibinfo{author}{N.~Jittiwarangkul},
  \bibinfo{author}{A.~Sangswang},
\newblock \bibinfo{title}{Energy and peak load forecast models using neural
  network for fast developing area},
\newblock in: \bibinfo{booktitle}{IEEE International Symposium on
  Communications and Information Technology, 2004. ISCIT 2004.},
  volume~\bibinfo{volume}{1}, \bibinfo{organization}{IEEE},
  \bibinfo{year}{2004}, pp. \bibinfo{pages}{389--393}.
\bibitem[{Saini and Soni(2002)}]{36-2002Artificial}
\bibinfo{author}{L.~Saini}, \bibinfo{author}{M.~Soni},
\newblock \bibinfo{title}{Artificial neural network based peak load forecasting
  using levenberg--marquardt and quasi-newton methods},
\newblock \bibinfo{journal}{IEE Proceedings-Generation, Transmission and
  Distribution} \bibinfo{volume}{149} (\bibinfo{year}{2002})
  \bibinfo{pages}{578--584}.
\bibitem[{Nagasaka and Al~Mamun(2004)}]{40-1372805}
\bibinfo{author}{K.~Nagasaka}, \bibinfo{author}{M.~Al~Mamun},
\newblock \bibinfo{title}{Long-term peak demand prediction of 9 japanese power
  utilities using radial basis function networks},
\newblock in: \bibinfo{booktitle}{IEEE Power Engineering Society General
  Meeting, 2004.}, \bibinfo{organization}{IEEE}, \bibinfo{year}{2004}, pp.
  \bibinfo{pages}{315--322}.
\bibitem[{Broadwater and Sargent(1997)}]{26-1997Estimating}
\bibinfo{author}{R.~P. Broadwater}, \bibinfo{author}{A.~Sargent},
\newblock \bibinfo{title}{Estimating substation peaks from load research data},
\newblock \bibinfo{journal}{IEEE Transactions on Power Delivery}
  \bibinfo{volume}{12} (\bibinfo{year}{1997}) \bibinfo{pages}{451--456}.
\bibitem[{Wang et~al.(2007)Wang, Wu, and Lu}]{60-2008Special}
\bibinfo{author}{J.-Z. Wang}, \bibinfo{author}{L.~Wu}, \bibinfo{author}{H.-Y.
  Lu},
\newblock \bibinfo{title}{Special periods peak load analysis and superior
  forecasting method based on ls-svm},
\newblock in: \bibinfo{booktitle}{2007 International Conference on Wavelet
  Analysis and Pattern Recognition}, volume~\bibinfo{volume}{1},
  \bibinfo{organization}{IEEE}, \bibinfo{year}{2007}, pp.
  \bibinfo{pages}{249--253}.
\bibitem[{{Contaxi} et~al.(2006){Contaxi}, {Delkis}, {Kavatza}, and
  {Vournas}}]{56-4075966}
\bibinfo{author}{E.~{Contaxi}}, \bibinfo{author}{C.~{Delkis}},
  \bibinfo{author}{S.~{Kavatza}}, \bibinfo{author}{C.~{Vournas}},
\newblock \bibinfo{title}{The effect of humidity in a weather-sensitive peak
  load forecasting model},
\newblock in: \bibinfo{booktitle}{2006 IEEE PES Power Systems Conference and
  Exposition}, \bibinfo{year}{2006}, pp. \bibinfo{pages}{1528--1534}.
\bibitem[{Ramanathan et~al.(1997)Ramanathan, Engle, Granger, Vahid-Araghi, and
  Brace}]{26news-1997Short}
\bibinfo{author}{R.~Ramanathan}, \bibinfo{author}{R.~Engle},
  \bibinfo{author}{C.~W.~J. Granger}, \bibinfo{author}{F.~Vahid-Araghi},
  \bibinfo{author}{C.~Brace},
\newblock \bibinfo{title}{Short-run forecasts of electricity loads and peaks},
\newblock \bibinfo{journal}{International Journal of Forecasting}
  \bibinfo{volume}{13} (\bibinfo{year}{1997}) \bibinfo{pages}{161--174}.
\bibitem[{Nuchprayoon(2016)}]{108-7893595}
\bibinfo{author}{S.~Nuchprayoon},
\newblock \bibinfo{title}{Forecasting of daily load curve on monthly peak day
  using load research data and harmonics model},
\newblock in: \bibinfo{booktitle}{2016 6th IEEE International Conference on
  Control System, Computing and Engineering (ICCSCE)}, \bibinfo{year}{2016},
  pp. \bibinfo{pages}{338--342}.
\bibitem[{Zhengyuan(2008)}]{62-2008Electricity}
\bibinfo{author}{J.~Zhengyuan},
\newblock \bibinfo{title}{Electricity consumption forecasting in peak load
  month based on variable weight combination forecasting model},
\newblock in: \bibinfo{booktitle}{2008 IEEE International Conference on
  Automation and Logistics}, \bibinfo{organization}{IEEE},
  \bibinfo{year}{2008}, pp. \bibinfo{pages}{1265--1269}.
\bibitem[{{Liu} and {Brown}(2019)}]{128-8791587}
\bibinfo{author}{J.~{Liu}}, \bibinfo{author}{L.~E. {Brown}},
\newblock \bibinfo{title}{Prediction of hour of coincident daily peak load},
\newblock in: \bibinfo{booktitle}{2019 IEEE Power Energy Society Innovative
  Smart Grid Technologies Conference (ISGT)}, \bibinfo{year}{2019}, pp.
  \bibinfo{pages}{1--5}.
\bibitem[{Goia et~al.(2010)Goia, May, and Fusai}]{73aa-GOIA2010700}
\bibinfo{author}{A.~Goia}, \bibinfo{author}{C.~May},
  \bibinfo{author}{G.~Fusai},
\newblock \bibinfo{title}{Functional clustering and linear regression for peak
  load forecasting},
\newblock \bibinfo{journal}{International Journal of Forecasting}
  \bibinfo{volume}{26} (\bibinfo{year}{2010}) \bibinfo{pages}{700 -- 711}.
\bibitem[{Xue et~al.(2012)Xue, Xu, and Watada}]{85aab-2012Building}
\bibinfo{author}{J.~Xue}, \bibinfo{author}{Z.~Xu}, \bibinfo{author}{J.~Watada},
\newblock \bibinfo{title}{Building an integrated hybrid model for short-term
  and mid-term load forecasting with genetic optimization},
\newblock \bibinfo{journal}{International journal of innovative computing,
  information \& control: IJICIC} \bibinfo{volume}{8} (\bibinfo{year}{2012})
  \bibinfo{pages}{7381--7391}.
\bibitem[{Yang et~al.(1995)Yang, Liang, Shih, and Huang}]{20}
\bibinfo{author}{H.-T. Yang}, \bibinfo{author}{T.-C. Liang},
  \bibinfo{author}{K.-R. Shih}, \bibinfo{author}{C.-L. Huang},
\newblock \bibinfo{title}{Power system yearly peak load forecasting: a grey
  system modeling approach},
\newblock in: \bibinfo{booktitle}{Proceedings 1995 International Conference on
  Energy Management and Power Delivery EMPD'95}, volume~\bibinfo{volume}{1},
  \bibinfo{organization}{IEEE}, \bibinfo{year}{1995}, pp.
  \bibinfo{pages}{261--266}.
\bibitem[{Alfuhaid et~al.(1997)Alfuhaid, El-Sayed, and
  Mahmoud}]{26newc-1997Cascaded}
\bibinfo{author}{A.~S. Alfuhaid}, \bibinfo{author}{M.~A. El-Sayed},
  \bibinfo{author}{M.~S. Mahmoud},
\newblock \bibinfo{title}{Cascaded artificial neural networks for short-term
  load forecasting},
\newblock \bibinfo{journal}{Power Systems IEEE Transactions on}
  \bibinfo{volume}{12} (\bibinfo{year}{1997}) \bibinfo{pages}{1524--1529}.
\bibitem[{Hyndman and Athanasopoulos(2014)}]{7-30}
\bibinfo{author}{R.~J. Hyndman}, \bibinfo{author}{G.~Athanasopoulos},
\newblock \bibinfo{title}{Forecasting: Principles and practice},
\newblock \bibinfo{journal}{London: Bowker-Saur. Pharo}
  (\bibinfo{year}{2014}).
\bibitem[{Gillies et~al.(1956)Gillies, Bernholtz, and Sandiford}]{1-1956A}
\bibinfo{author}{D.~K.~A. Gillies}, \bibinfo{author}{B.~Bernholtz},
  \bibinfo{author}{P.~J. Sandiford},
\newblock \bibinfo{title}{A new approach to forecasting daily peak loads
  [includes discussion]},
\newblock \bibinfo{journal}{Power Apparatus \& Systems Part III Transactions of
  the American Institute of Electrical Engineers} \bibinfo{volume}{75}
  (\bibinfo{year}{1956}).
\bibitem[{Turner et~al.(2012)Turner, Downing, and
  Bogard}]{8-Turner2012Regression}
\bibinfo{author}{J.~E. Turner}, \bibinfo{author}{D.~J. Downing},
  \bibinfo{author}{J.~S. Bogard}, \bibinfo{title}{Statistical methods in
  radiation physics}, \bibinfo{publisher}{Wiley Online Library},
  \bibinfo{year}{2012}.
\bibitem[{Papalexopoulos(1990)}]{9a-1990A}
\bibinfo{author}{A.~D. Papalexopoulos},
\newblock \bibinfo{title}{A regression-based approach to short-term system load
  forecasting},
\newblock \bibinfo{journal}{IEEE Transactions on Power Systems}
  \bibinfo{volume}{5} (\bibinfo{year}{1990}) \bibinfo{pages}{1535--1547}.
\bibitem[{{Mtembo} et~al.(2014){Mtembo}, {Taylor}, and {Ekwue}}]{94-6934706}
\bibinfo{author}{V.~{Mtembo}}, \bibinfo{author}{G.~A. {Taylor}},
  \bibinfo{author}{A.~{Ekwue}},
\newblock \bibinfo{title}{A novel econometric model for peak demand
  forecasting},
\newblock in: \bibinfo{booktitle}{2014 49th International Universities Power
  Engineering Conference (UPEC)}, \bibinfo{year}{2014}, pp.
  \bibinfo{pages}{1--6}.
\bibitem[{Alfares and Nazeeruddin(1999)}]{28newr-1999reg}
\bibinfo{author}{H.~Alfares}, \bibinfo{author}{M.~Nazeeruddin},
\newblock \bibinfo{title}{Regression-based methodology for daily peak load
  forecasting},
\newblock in: \bibinfo{booktitle}{Proceedings of the 2nd International
  Conference on Operations and Quantitative Management},
  volume~\bibinfo{volume}{3}, \bibinfo{year}{1999}, pp.
  \bibinfo{pages}{468--471}.
\bibitem[{Haida and Muto(1994)}]{19-2002Regression}
\bibinfo{author}{T.~Haida}, \bibinfo{author}{S.~Muto},
\newblock \bibinfo{title}{Regression based peak load forecasting using a
  transformation technique},
\newblock \bibinfo{journal}{IEEE Transactions on Power Systems}
  \bibinfo{volume}{9} (\bibinfo{year}{1994}) \bibinfo{pages}{1788--1794}.
\bibitem[{Atsawathawichok et~al.(2014)Atsawathawichok, Teekaput, and
  Ploysuwan}]{90-2014Long}
\bibinfo{author}{P.~Atsawathawichok}, \bibinfo{author}{P.~Teekaput},
  \bibinfo{author}{T.~Ploysuwan},
\newblock \bibinfo{title}{Long term peak load forecasting in thailand using
  multiple kernel gaussian process},
\newblock in: \bibinfo{booktitle}{2014 11th International Conference on
  Electrical Engineering/Electronics, Computer, Telecommunications and
  Information Technology (ECTI-CON)}, \bibinfo{organization}{IEEE},
  \bibinfo{year}{2014}, pp. \bibinfo{pages}{1--4}.
\bibitem[{Bichpuriya et~al.(2014)Bichpuriya, Soman, and
  Subramanyam}]{95-2014Non}
\bibinfo{author}{Y.~K. Bichpuriya}, \bibinfo{author}{S.~Soman},
  \bibinfo{author}{A.~Subramanyam},
\newblock \bibinfo{title}{Non-parametric probability density forecast of an
  hourly peak load during a month},
\newblock in: \bibinfo{booktitle}{2014 Power Systems Computation Conference},
  \bibinfo{organization}{IEEE}, \bibinfo{year}{2014}, pp.
  \bibinfo{pages}{1--6}.
\bibitem[{Haida et~al.(1998)Haida, Muto, Takahashi, and
  Ishi}]{27newdaa-Haida1998Peak}
\bibinfo{author}{T.~Haida}, \bibinfo{author}{S.~Muto},
  \bibinfo{author}{Y.~Takahashi}, \bibinfo{author}{Y.~Ishi},
\newblock \bibinfo{title}{Peak load forecasting using multiple-year data with
  trend data processing techniques},
\newblock \bibinfo{journal}{Electrical Engineering in Japan}
  \bibinfo{volume}{124} (\bibinfo{year}{1998}) \bibinfo{pages}{7--16}.
\bibitem[{Barakat et~al.(1990)Barakat, Qayyum, Hamed, and
  Al~Rashed}]{9s-Barakat1990Short}
\bibinfo{author}{E.~Barakat}, \bibinfo{author}{M.~Qayyum},
  \bibinfo{author}{M.~Hamed}, \bibinfo{author}{S.~Al~Rashed},
\newblock \bibinfo{title}{Short-term peak demand forecasting in fast developing
  utility with inherit dynamic load characteristics. i. application of
  classical time-series methods. ii. improved modelling of system dynamic load
  characteristics},
\newblock \bibinfo{journal}{IEEE Transactions on Power Systems}
  \bibinfo{volume}{5} (\bibinfo{year}{1990}) \bibinfo{pages}{813--824}.
\bibitem[{Barakat and Eissa(1989)}]{8-Barakat1989Forecasting}
\bibinfo{author}{E.~Barakat}, \bibinfo{author}{M.~Eissa},
\newblock \bibinfo{title}{Forecasting monthly peak demand in fast growing
  electric utility using a composite multiregression-decomposition model},
\newblock in: \bibinfo{booktitle}{IEE Proceedings C (Generation, Transmission
  and Distribution)}, volume \bibinfo{volume}{136},
  \bibinfo{organization}{IET}, \bibinfo{year}{1989}, pp.
  \bibinfo{pages}{35--41}.
\bibitem[{Nazarko and Zalewski(1999)}]{28-1999The}
\bibinfo{author}{J.~Nazarko}, \bibinfo{author}{W.~Zalewski},
\newblock \bibinfo{title}{The fuzzy regression approach to peak load estimation
  in power distribution systems},
\newblock \bibinfo{journal}{IEEE Transactions on Power Systems}
  \bibinfo{volume}{14} (\bibinfo{year}{1999}) \bibinfo{pages}{809--814}.
\bibitem[{Gupta(1971)}]{2-Gupta1971A}
\bibinfo{author}{P.~Gupta},
\newblock \bibinfo{title}{A stochastic approach to peak power-demand
  forecasting in electric utility systems},
\newblock \bibinfo{journal}{IEEE Transactions on Power Apparatus and Systems}
  (\bibinfo{year}{1971}) \bibinfo{pages}{824--832}.
\bibitem[{Fong and Yang(2011)}]{5-Fong2011The}
\bibinfo{author}{S.~Fong}, \bibinfo{author}{H.~Yang},
\newblock \bibinfo{title}{The six technical gaps between intelligent
  applications and real-time data mining: a critical review},
\newblock \bibinfo{journal}{journal of emerging technologies in web
  intelligence} \bibinfo{volume}{3} (\bibinfo{year}{2011})
  \bibinfo{pages}{63--73}.
\bibitem[{Dongxiao et~al.(2009)Dongxiao, Yunyun, and Jinpeng}]{68-2009The}
\bibinfo{author}{N.~Dongxiao}, \bibinfo{author}{Z.~Yunyun},
  \bibinfo{author}{L.~Jinpeng},
\newblock \bibinfo{title}{The application of time series seasonal
  multiplicative model and garch error amending model on forecasting the
  monthly peak load},
\newblock in: \bibinfo{booktitle}{2009 International Forum on Computer
  Science-Technology and Applications}, volume~\bibinfo{volume}{3},
  \bibinfo{organization}{IEEE}, \bibinfo{year}{2009}, pp.
  \bibinfo{pages}{135--138}.
\bibitem[{Choi et~al.(1996)Choi, Park, Kim, and Kim}]{22-Choi1996A}
\bibinfo{author}{J.-G. Choi}, \bibinfo{author}{J.-K. Park},
  \bibinfo{author}{K.-H. Kim}, \bibinfo{author}{J.-C. Kim},
\newblock \bibinfo{title}{A daily peak load forecasting system using a chaotic
  time series},
\newblock in: \bibinfo{booktitle}{Proceedings of International Conference on
  Intelligent System Application to Power Systems},
  \bibinfo{organization}{IEEE}, \bibinfo{year}{1996}, pp.
  \bibinfo{pages}{283--287}.
\bibitem[{Moazzami et~al.(2013)Moazzami, Khodabakhshian, and
  Hooshmand}]{88aaa-2013A}
\bibinfo{author}{M.~Moazzami}, \bibinfo{author}{A.~Khodabakhshian},
  \bibinfo{author}{R.~Hooshmand},
\newblock \bibinfo{title}{A new hybrid day-ahead peak load forecasting method
  for iran's national grid},
\newblock \bibinfo{journal}{Applied Energy} \bibinfo{volume}{101}
  (\bibinfo{year}{2013}) \bibinfo{pages}{489--501}.
\bibitem[{Mahmoud et~al.(2009)Mahmoud, Mhamdi, and
  Jaidane-Saidane}]{66-2009Long}
\bibinfo{author}{M.~O.~M. Mahmoud}, \bibinfo{author}{F.~Mhamdi},
  \bibinfo{author}{M.~Jaidane-Saidane},
\newblock \bibinfo{title}{Long term multi-scale analysis of the daily peak load
  based on the empirical mode decomposition},
\newblock in: \bibinfo{booktitle}{2009 IEEE Bucharest PowerTech},
  \bibinfo{organization}{IEEE}, \bibinfo{year}{2009}, pp.
  \bibinfo{pages}{1--6}.
\bibitem[{Beiraghi and Ranjbar(2011)}]{80-2011Discrete}
\bibinfo{author}{M.~Beiraghi}, \bibinfo{author}{A.~Ranjbar},
\newblock \bibinfo{title}{Discrete fourier transform based approach to forecast
  monthly peak load},
\newblock in: \bibinfo{booktitle}{2011 Asia-Pacific Power and Energy
  Engineering Conference}, \bibinfo{organization}{IEEE}, \bibinfo{year}{2011},
  pp. \bibinfo{pages}{1--5}.
\bibitem[{El-Razaz and Al-Mohawes(1986)}]{5-El1986Weekly}
\bibinfo{author}{Z.~El-Razaz}, \bibinfo{author}{N.~Al-Mohawes},
\newblock \bibinfo{title}{Weekly peak load forecasting for fast-developing
  cities},
\newblock \bibinfo{journal}{Canadian Electrical Engineering Journal}
  \bibinfo{volume}{11} (\bibinfo{year}{1986}) \bibinfo{pages}{184--187}.
\bibitem[{Barakat et~al.(1992)Barakat, Al-Qassim, and Al~Rashed}]{12-1992New}
\bibinfo{author}{E.~Barakat}, \bibinfo{author}{J.~Al-Qassim},
  \bibinfo{author}{S.~Al~Rashed},
\newblock \bibinfo{title}{New model for peak demand forecasting applied to
  highly complex load characteristics of a fast developing area},
\newblock in: \bibinfo{booktitle}{IEE Proceedings C (Generation, Transmission
  and Distribution)}, volume \bibinfo{volume}{139},
  \bibinfo{organization}{IET}, \bibinfo{year}{1992}, pp.
  \bibinfo{pages}{136--140}.
\bibitem[{{Fadhilah} et~al.(2009){Fadhilah}, {Suriawati}, {Amir}, {Izham}, and
  {Mahendran}}]{63newm-5398613}
\bibinfo{author}{A.~R. {Fadhilah}}, \bibinfo{author}{S.~{Suriawati}},
  \bibinfo{author}{H.~H. {Amir}}, \bibinfo{author}{Z.~A. {Izham}},
  \bibinfo{author}{S.~{Mahendran}},
\newblock \bibinfo{title}{Malaysian day-type load forecasting},
\newblock in: \bibinfo{booktitle}{2009 3rd International Conference on Energy
  and Environment (ICEE)}, \bibinfo{year}{2009}, pp. \bibinfo{pages}{408--411}.
\bibitem[{Mahmud et~al.(2020)Mahmud, Ravishankar, Hossain, and
  Dong}]{133-2020The}
\bibinfo{author}{K.~Mahmud}, \bibinfo{author}{J.~Ravishankar},
  \bibinfo{author}{M.~J. Hossain}, \bibinfo{author}{Z.~Y. Dong},
\newblock \bibinfo{title}{The impact of prediction errors in the domestic peak
  power demand management},
\newblock \bibinfo{journal}{IEEE Transactions on Industrial Informatics}
  \bibinfo{volume}{16} (\bibinfo{year}{2020}) \bibinfo{pages}{4567--4579}.
\bibitem[{Sigauke and Chikobvu(2011)}]{82aap-2011Prediction}
\bibinfo{author}{C.~Sigauke}, \bibinfo{author}{D.~Chikobvu},
\newblock \bibinfo{title}{Prediction of daily peak electricity demand in south
  africa using volatility forecasting models},
\newblock \bibinfo{journal}{Energy Economics} \bibinfo{volume}{33}
  (\bibinfo{year}{2011}) \bibinfo{pages}{882--888}.
\bibitem[{As'~ad(2012)}]{84aaf-2012Finding}
\bibinfo{author}{M.~As'~ad},
\newblock \bibinfo{title}{Finding the best arima model to forecast daily peak
  electricity demand},
\newblock in: \bibinfo{booktitle}{Proceedings of the Fifth Annual ASEARC
  Conference - Looking to the future - Programme and Proceedings},
  \bibinfo{organization}{University of Wollongong}, \bibinfo{year}{2012}, pp.
  \bibinfo{pages}{1--4}.
\bibitem[{Roken and Badri(2006)}]{57aat-article}
\bibinfo{author}{R.~M. Roken}, \bibinfo{author}{M.~A. Badri},
\newblock \bibinfo{title}{Time series models for forecasting monthly
  electricity peak-load for dubai},
\newblock \bibinfo{journal}{Chancellor's Undergraduate Research Award}
  (\bibinfo{year}{2006}) \bibinfo{pages}{1--14}.
\bibitem[{Kareem and Majeed(2006)}]{57aam-2007Monthly}
\bibinfo{author}{Y.~Kareem}, \bibinfo{author}{A.~R. Majeed},
\newblock \bibinfo{title}{Monthly peak-load demand forecasting for sulaimany
  governorate using sarima.},
\newblock in: \bibinfo{booktitle}{2006 IEEE/PES Transmission \& Distribution
  Conference and Exposition: Latin America}, \bibinfo{organization}{IEEE},
  \bibinfo{year}{2006}, pp. \bibinfo{pages}{1--5}.
\bibitem[{Rallapalli and Ghosh(2012)}]{83aaf-2012Forecasting}
\bibinfo{author}{S.~R. Rallapalli}, \bibinfo{author}{S.~Ghosh},
\newblock \bibinfo{title}{Forecasting monthly peak demand of electricity in
  india—a critique},
\newblock \bibinfo{journal}{Energy Policy} \bibinfo{volume}{45}
  (\bibinfo{year}{2012}) \bibinfo{pages}{516--520}.
\bibitem[{Ghosh(2008)}]{8-78}
\bibinfo{author}{S.~Ghosh},
\newblock \bibinfo{title}{Univariate time-series forecasting of monthly peak
  demand of electricity in northern india},
\newblock \bibinfo{journal}{International Journal of Indian Culture and
  Business Management} \bibinfo{volume}{1} (\bibinfo{year}{2008})
  \bibinfo{pages}{466--474}.
\bibitem[{{Razak} et~al.(2010){Razak}, {Hashim}, {Abidin}, and
  {Shitan}}]{73-5697708}
\bibinfo{author}{F.~A. {Razak}}, \bibinfo{author}{A.~H. {Hashim}},
  \bibinfo{author}{I.~Z. {Abidin}}, \bibinfo{author}{M.~{Shitan}},
\newblock \bibinfo{title}{Moving holidays' effects on the malaysian peak daily
  load},
\newblock in: \bibinfo{booktitle}{2010 IEEE International Conference on Power
  and Energy}, \bibinfo{year}{2010}, pp. \bibinfo{pages}{906--910}.
\bibitem[{{Heylman} et~al.(2015){Heylman}, {Kim}, and {Wang}}]{99-7117006}
\bibinfo{author}{C.~{Heylman}}, \bibinfo{author}{Y.~G. {Kim}},
  \bibinfo{author}{J.~{Wang}},
\newblock \bibinfo{title}{Forecasting energy trends and peak usage at the
  university of virginia},
\newblock in: \bibinfo{booktitle}{2015 Systems and Information Engineering
  Design Symposium}, \bibinfo{year}{2015}, pp. \bibinfo{pages}{362--368}.
\bibitem[{{Ananthasingam} and {Atputharajah}(2015)}]{102-7399017}
\bibinfo{author}{A.~{Ananthasingam}}, \bibinfo{author}{A.~{Atputharajah}},
\newblock \bibinfo{title}{Forecast daily night peak electric power demand in
  sri lankan power system},
\newblock in: \bibinfo{booktitle}{2015 IEEE 10th International Conference on
  Industrial and Information Systems (ICIIS)}, \bibinfo{year}{2015}, pp.
  \bibinfo{pages}{238--243}.
\bibitem[{{Hor} et~al.(2006){Hor}, {Watson}, and {Majithia}}]{56aad-4202249}
\bibinfo{author}{C.~{Hor}}, \bibinfo{author}{S.~J. {Watson}},
  \bibinfo{author}{S.~{Majithia}},
\newblock \bibinfo{title}{Daily load forecasting and maximum demand estimation
  using arima and garch},
\newblock in: \bibinfo{booktitle}{2006 International Conference on
  Probabilistic Methods Applied to Power Systems}, \bibinfo{year}{2006}, pp.
  \bibinfo{pages}{1--6}.
\bibitem[{Ostertagov{\'a} and Ostertag(2011)}]{7-64-incollection}
\bibinfo{author}{E.~Ostertagov{\'a}}, \bibinfo{author}{O.~Ostertag},
\newblock \bibinfo{title}{The simple exponential smoothing model},
\newblock in: \bibinfo{booktitle}{The 4th International Conference on Modelling
  of Mechanical and Mechatronic Systems, Technical University of Ko{\v{s}}ice,
  Slovak Republic, Proceedings of conference}, \bibinfo{year}{2011}, pp.
  \bibinfo{pages}{380--384}.
\bibitem[{Gardner~Jr(1985)}]{26aa-27}
\bibinfo{author}{E.~S. Gardner~Jr},
\newblock \bibinfo{title}{Exponential smoothing: The state of the art},
\newblock \bibinfo{journal}{Journal of forecasting} \bibinfo{volume}{4}
  (\bibinfo{year}{1985}) \bibinfo{pages}{1--28}.
\bibitem[{Yaffee and McGee(2000)}]{17du12-Yaffee2000Introduction}
\bibinfo{author}{R.~A. Yaffee}, \bibinfo{author}{M.~McGee}, \bibinfo{title}{An
  introduction to time series analysis and forecasting: with applications of
  SAS{\textregistered} and SPSS{\textregistered}},
  \bibinfo{publisher}{Elsevier}, \bibinfo{year}{2000}.
\bibitem[{Badri et~al.(1997)Badri, Al-Mutawa, Davis, and
  Davis}]{26aa-Masood1997EDSSF}
\bibinfo{author}{M.~A. Badri}, \bibinfo{author}{A.~Al-Mutawa},
  \bibinfo{author}{D.~Davis}, \bibinfo{author}{D.~Davis},
\newblock \bibinfo{title}{Edssf: A decision support system (dss) for
  electricity peak-load forecasting},
\newblock \bibinfo{journal}{Energy} \bibinfo{volume}{22} (\bibinfo{year}{1997})
  \bibinfo{pages}{579--589}.
\bibitem[{Taylor(2003)}]{38aa-J2017Short}
\bibinfo{author}{J.~W. Taylor},
\newblock \bibinfo{title}{Short-term electricity demand forecasting using
  double seasonal exponential smoothing},
\newblock \bibinfo{journal}{Journal of the Operational Research Society}
  \bibinfo{volume}{54} (\bibinfo{year}{2003}) \bibinfo{pages}{799--805}.
\bibitem[{Billah et~al.(2006)Billah, King, Snyder, and
  Koehler}]{billah2006exponential}
\bibinfo{author}{B.~Billah}, \bibinfo{author}{M.~L. King},
  \bibinfo{author}{R.~D. Snyder}, \bibinfo{author}{A.~B. Koehler},
\newblock \bibinfo{title}{Exponential smoothing model selection for
  forecasting},
\newblock \bibinfo{journal}{International journal of forecasting}
  \bibinfo{volume}{22} (\bibinfo{year}{2006}) \bibinfo{pages}{239--247}.
\bibitem[{Julier and Uhlmann(1997)}]{Kalman1}
\bibinfo{author}{S.~J. Julier}, \bibinfo{author}{J.~K. Uhlmann},
\newblock \bibinfo{title}{New extension of the kalman filter to nonlinear
  systems},
\newblock in: \bibinfo{booktitle}{Signal processing, sensor fusion, and target
  recognition VI}, volume \bibinfo{volume}{3068},
  \bibinfo{organization}{International Society for Optics and Photonics},
  \bibinfo{year}{1997}, pp. \bibinfo{pages}{182--193}.
\bibitem[{Dash et~al.(1995)Dash, Satpathy, and Rahman}]{21-1995Short}
\bibinfo{author}{P.~Dash}, \bibinfo{author}{H.~Satpathy},
  \bibinfo{author}{S.~Rahman},
\newblock \bibinfo{title}{Short term daily average and peak load predications
  using a hybrid intelligent approach},
\newblock in: \bibinfo{booktitle}{Proceedings 1995 International Conference on
  Energy Management and Power Delivery EMPD'95}, volume~\bibinfo{volume}{2},
  \bibinfo{organization}{IEEE}, \bibinfo{year}{1995}, pp.
  \bibinfo{pages}{565--570}.
\bibitem[{Ran and Chaoyun(2008)}]{61-2008The}
\bibinfo{author}{L.~Ran}, \bibinfo{author}{G.~Chaoyun},
\newblock \bibinfo{title}{The relevance analysis between electrical day peak
  load and meteorological index based on wavelet denoising and svm},
\newblock in: \bibinfo{booktitle}{2008 Third International Conference on
  Electric Utility Deregulation and Restructuring and Power Technologies},
  \bibinfo{organization}{IEEE}, \bibinfo{year}{2008}, pp.
  \bibinfo{pages}{788--793}.
\bibitem[{Hsu and Ho(1992)}]{13aa-Hsu1992Fuzzy}
\bibinfo{author}{Y.-Y. Hsu}, \bibinfo{author}{K.-L. Ho},
\newblock \bibinfo{title}{Fuzzy expert systems: an application to short-term
  load forecasting},
\newblock in: \bibinfo{booktitle}{IEE Proceedings C (Generation, Transmission
  and Distribution)}, volume \bibinfo{volume}{139},
  \bibinfo{organization}{IET}, \bibinfo{year}{1992}, pp.
  \bibinfo{pages}{471--477}.
\bibitem[{Kiartzis et~al.(2000)Kiartzis, Bakirtzis, Theocharis, and
  Tsagas}]{30newa-30Kiartzis2000A}
\bibinfo{author}{S.~Kiartzis}, \bibinfo{author}{A.~Bakirtzis},
  \bibinfo{author}{J.~Theocharis}, \bibinfo{author}{G.~Tsagas},
\newblock \bibinfo{title}{A fuzzy expert system for peak load forecasting.
  application to the greek power system},
\newblock in: \bibinfo{booktitle}{2000 10th Mediterranean Electrotechnical
  Conference. Information Technology and Electrotechnology for the
  Mediterranean Countries. Proceedings. MeleCon 2000 (Cat. No. 00CH37099)},
  volume~\bibinfo{volume}{3}, \bibinfo{organization}{IEEE},
  \bibinfo{year}{2000}, pp. \bibinfo{pages}{1097--1100}.
\bibitem[{El-naggar and Al-rumaih(2005)}]{87aaa-El2013Electric}
\bibinfo{author}{K.~M. El-naggar}, \bibinfo{author}{K.~A. Al-rumaih},
\newblock \bibinfo{title}{Electric load forecasting using genetic based
  algorithm, optimal filter estimator and least error squares technique:
  Comparative study},
\newblock \bibinfo{journal}{World Academy of Science, Engineering and
  Technology} \bibinfo{volume}{6} (\bibinfo{year}{2005})
  \bibinfo{pages}{138--142}.
\bibitem[{Agatonovic-Kustrin and Beresford(2000)}]{17du20-Agatonovic2000Basic}
\bibinfo{author}{S.~Agatonovic-Kustrin}, \bibinfo{author}{R.~Beresford},
\newblock \bibinfo{title}{Basic concepts of artificial neural network (ann)
  modeling and its application in pharmaceutical research},
\newblock \bibinfo{journal}{Journal of Pharmaceutical \& Biomedical Analysis}
  \bibinfo{volume}{22} (\bibinfo{year}{2000}) \bibinfo{pages}{717--727}.
\bibitem[{Ghomi et~al.(2010)Ghomi, Goodarzi, and Goodarzi}]{69-2010Peak}
\bibinfo{author}{M.~Ghomi}, \bibinfo{author}{M.~Goodarzi},
  \bibinfo{author}{M.~Goodarzi},
\newblock \bibinfo{title}{Peak load forecasting of electric utilities for west
  province of iran by using neural network without weather information},
\newblock in: \bibinfo{booktitle}{2010 12th International Conference on
  Computer Modelling and Simulation}, \bibinfo{organization}{IEEE},
  \bibinfo{year}{2010}, pp. \bibinfo{pages}{28--32}.
\bibitem[{Saeed~Madani(1991)}]{11e-Saeed1991Electric}
\bibinfo{author}{S.~Saeed~Madani},
\newblock \bibinfo{title}{Electric load forecasting using an artificial neural
  network},
\newblock \bibinfo{journal}{IEEE Transactions on Power Systems}
  \bibinfo{volume}{6} (\bibinfo{year}{1991}) \bibinfo{pages}{442--449}.
\bibitem[{Hsu and Chen(2003)}]{37aa-2003Regional}
\bibinfo{author}{C.~C. Hsu}, \bibinfo{author}{C.~Y. Chen},
\newblock \bibinfo{title}{Regional load forecasting in taiwan––applications
  of artificial neural networks},
\newblock \bibinfo{journal}{Energy Conversion \& Management}
  \bibinfo{volume}{44} (\bibinfo{year}{2003}) \bibinfo{pages}{1941--1949}.
\bibitem[{Saini(2008)}]{63newp-Saini2008Peak}
\bibinfo{author}{L.~M. Saini},
\newblock \bibinfo{title}{Peak load forecasting using bayesian regularization,
  resilient and adaptive backpropagation learning based artificial neural
  networks},
\newblock \bibinfo{journal}{Electric Power Systems Research}
  \bibinfo{volume}{78} (\bibinfo{year}{2008}) \bibinfo{pages}{1302--1310}.
\bibitem[{Chae et~al.(2016)Chae, Horesh, Hwang, and
  Lee}]{108aaa-2016Artificial}
\bibinfo{author}{Y.~T. Chae}, \bibinfo{author}{R.~Horesh},
  \bibinfo{author}{Y.~Hwang}, \bibinfo{author}{Y.~M. Lee},
\newblock \bibinfo{title}{Artificial neural network model for forecasting
  sub-hourly electricity usage in commercial buildings},
\newblock \bibinfo{journal}{Energy \& Buildings} \bibinfo{volume}{111}
  (\bibinfo{year}{2016}) \bibinfo{pages}{184--194}.
\bibitem[{Ghods and Kalantar(2010)}]{74aa-article}
\bibinfo{author}{L.~Ghods}, \bibinfo{author}{M.~Kalantar},
\newblock \bibinfo{title}{Long-term peak demand forecasting by using radial
  basis function neural networks},
\newblock \bibinfo{journal}{Iranian Journal of Electrical and Electronic
  Engineering} \bibinfo{volume}{6} (\bibinfo{year}{2010})
  \bibinfo{pages}{175--182}.
\bibitem[{Temraz et~al.(1998)Temraz, El-Nagar, and Salama}]{27}
\bibinfo{author}{H.~K. Temraz}, \bibinfo{author}{K.~M. El-Nagar},
  \bibinfo{author}{M.~M.~A. Salama},
\newblock \bibinfo{title}{Applications of noniterative least absolute value
  estimation for forecasting annual peak electric power demand},
\newblock \bibinfo{journal}{Canadian Journal of Electrical and Computer
  Engineering} \bibinfo{volume}{23} (\bibinfo{year}{1998})
  \bibinfo{pages}{141--146}.
\bibitem[{Hern{\'a}ndez et~al.(2013)Hern{\'a}ndez, Balad{\'o}n, Aguiar,
  Calavia, Carro, S{\'a}nchez-Esguevillas, Sanju{\'a}n, Gonz{\'a}lez, and
  Lloret}]{89aai-Hern2013Improved}
\bibinfo{author}{L.~Hern{\'a}ndez}, \bibinfo{author}{C.~Balad{\'o}n},
  \bibinfo{author}{J.~Aguiar}, \bibinfo{author}{L.~Calavia},
  \bibinfo{author}{B.~Carro}, \bibinfo{author}{A.~S{\'a}nchez-Esguevillas},
  \bibinfo{author}{J.~Sanju{\'a}n}, \bibinfo{author}{{\'a}.~Gonz{\'a}lez},
  \bibinfo{author}{J.~Lloret},
\newblock \bibinfo{title}{Improved short-term load forecasting based on
  two-stage predictions with artificial neural networks in a microgrid
  environment},
\newblock \bibinfo{journal}{Energies} \bibinfo{volume}{6}
  (\bibinfo{year}{2013}) \bibinfo{pages}{4489--4507}.
\bibitem[{Mandal and Agrawal(1997)}]{25-Mandal1997Fuzzy}
\bibinfo{author}{S.~Mandal}, \bibinfo{author}{A.~Agrawal},
\newblock \bibinfo{title}{Fuzzy-neural network based short term peak and
  average load forecasting (stpalf) system with network security},
\newblock in: \bibinfo{booktitle}{IECEC-97 Proceedings of the Thirty-Second
  Intersociety Energy Conversion Engineering Conference (Cat. No. 97CH6203)},
  volume~\bibinfo{volume}{3}, \bibinfo{organization}{IEEE},
  \bibinfo{year}{1997}, pp. \bibinfo{pages}{2193--2196}.
\bibitem[{Tavassoli-Hojati et~al.(2020)Tavassoli-Hojati, Ghaderi, Iranmanesh,
  Hilber, and Shayesteh}]{99aaa}
\bibinfo{author}{Z.~Tavassoli-Hojati}, \bibinfo{author}{S.~Ghaderi},
  \bibinfo{author}{H.~Iranmanesh}, \bibinfo{author}{P.~Hilber},
  \bibinfo{author}{E.~Shayesteh},
\newblock \bibinfo{title}{A self-partitioning local neuro fuzzy model for
  short-term load forecasting in smart grids},
\newblock \bibinfo{journal}{Energy} \bibinfo{volume}{199}
  (\bibinfo{year}{2020}) \bibinfo{pages}{117514}.
\bibitem[{Ermatita et~al.(2019)Ermatita, Pahendra, Darnila, Sadli, Sinambela,
  and Fuadi}]{132-8985197}
\bibinfo{author}{Ermatita}, \bibinfo{author}{I.~Pahendra},
  \bibinfo{author}{E.~Darnila}, \bibinfo{author}{M.~Sadli},
  \bibinfo{author}{M.~Sinambela}, \bibinfo{author}{W.~Fuadi},
\newblock \bibinfo{title}{Peak load forecasting based on long short term
  memory},
\newblock in: \bibinfo{booktitle}{2019 International Conference on Informatics,
  Multimedia, Cyber and Information System (ICIMCIS)}, \bibinfo{year}{2019},
  pp. \bibinfo{pages}{137--140}.
\bibitem[{Hsu and Yang(1991{\natexlab{a}})}]{10d-1991Design}
\bibinfo{author}{Y.-Y. Hsu}, \bibinfo{author}{C.-C. Yang},
\newblock \bibinfo{title}{Design of artificial neural networks for short-term
  load forecasting. part 1: Self-organising feature maps for day type
  identification},
\newblock in: \bibinfo{booktitle}{IEE Proceedings C (Generation, Transmission
  and Distribution)}, volume \bibinfo{volume}{138},
  \bibinfo{organization}{IET}, \bibinfo{year}{1991}{\natexlab{a}}, pp.
  \bibinfo{pages}{407--413}.
\bibitem[{Hsu and Yang(1991{\natexlab{b}})}]{10-1991Design}
\bibinfo{author}{Y.-Y. Hsu}, \bibinfo{author}{C.-C. Yang},
\newblock \bibinfo{title}{Design of artificial neural networks for short-term
  load forecasting. part 2: Multilayer feedforward networks for peak load and
  valley load forecasting},
\newblock in: \bibinfo{booktitle}{IEE Proceedings C (Generation, Transmission
  and Distribution)}, volume \bibinfo{volume}{138},
  \bibinfo{organization}{IET}, \bibinfo{year}{1991}{\natexlab{b}}, pp.
  \bibinfo{pages}{414--418}.
\bibitem[{Amin-Naseri and Soroush(2008)}]{63newc-AMINNASERI20081302}
\bibinfo{author}{M.~Amin-Naseri}, \bibinfo{author}{A.~Soroush},
\newblock \bibinfo{title}{Combined use of unsupervised and supervised learning
  for daily peak load forecasting},
\newblock \bibinfo{journal}{Energy Conversion and Management}
  \bibinfo{volume}{49} (\bibinfo{year}{2008}) \bibinfo{pages}{1302--1308}.
\bibitem[{{Kwon} et~al.(2019){Kwon}, {Park}, and {Song}}]{134-8994442}
\bibinfo{author}{B.~{Kwon}}, \bibinfo{author}{R.~{Park}},
  \bibinfo{author}{K.~{Song}},
\newblock \bibinfo{title}{Weekly peak load forecasting for 104 weeks using deep
  learning algorithm},
\newblock in: \bibinfo{booktitle}{2019 IEEE PES Asia-Pacific Power and Energy
  Engineering Conference (APPEEC)}, \bibinfo{year}{2019}, pp.
  \bibinfo{pages}{1--4}.
\bibitem[{Ho et~al.(1992)Ho, Hsu, and Yang}]{13-Ho1992Short}
\bibinfo{author}{K.~Ho}, \bibinfo{author}{Y.-Y. Hsu}, \bibinfo{author}{C.-C.
  Yang},
\newblock \bibinfo{title}{Short term load forecasting using a multilayer neural
  network with an adaptive learning algorithm},
\newblock \bibinfo{journal}{IEEE Transactions on Power Systems}
  \bibinfo{volume}{7} (\bibinfo{year}{1992}) \bibinfo{pages}{141--149}.
\bibitem[{Saini and Soni(2002)}]{37-Saini2002Artificial}
\bibinfo{author}{L.~M. Saini}, \bibinfo{author}{M.~K. Soni},
\newblock \bibinfo{title}{Artificial neural network-based peak load forecasting
  using conjugate gradient methods},
\newblock \bibinfo{journal}{IEEE Transactions on Power Systems}
  \bibinfo{volume}{17} (\bibinfo{year}{2002}) \bibinfo{pages}{907--912}.
\bibitem[{Drucker et~al.(1997)Drucker, Burges, Kaufman, Smola, Vapnik
  et~al.}]{85-21}
\bibinfo{author}{H.~Drucker}, \bibinfo{author}{C.~J. Burges},
  \bibinfo{author}{L.~Kaufman}, \bibinfo{author}{A.~Smola},
  \bibinfo{author}{V.~Vapnik}, et~al.,
\newblock \bibinfo{title}{Support vector regression machines},
\newblock \bibinfo{journal}{Advances in neural information processing systems}
  \bibinfo{volume}{9} (\bibinfo{year}{1997}) \bibinfo{pages}{155--161}.
\bibitem[{Coello and Mezura-Montes(2002)}]{85-5}
\bibinfo{author}{C.~Coello}, \bibinfo{author}{E.~Mezura-Montes},
\newblock \bibinfo{title}{Constraint-handling in genetic algorithms through the
  use of dominance-based tournament selection},
\newblock \bibinfo{journal}{Advanced Engineering Informatics}
  \bibinfo{volume}{16} (\bibinfo{year}{2002}) \bibinfo{pages}{193--203}.
\bibitem[{El-Attar et~al.(2009)El-Attar, Goulermas, and
  Wu}]{64-2009Forecasting}
\bibinfo{author}{E.~El-Attar}, \bibinfo{author}{J.~Goulermas},
  \bibinfo{author}{Q.~Wu},
\newblock \bibinfo{title}{Forecasting electric daily peak load based on local
  prediction},
\newblock in: \bibinfo{booktitle}{2009 IEEE Power \& Energy Society General
  Meeting}, \bibinfo{organization}{IEEE}, \bibinfo{year}{2009}, pp.
  \bibinfo{pages}{1--6}.
\bibitem[{Dhillon et~al.(2016)Dhillon, Rahman, Ahmad, and
  Hossain}]{104-2016Peak}
\bibinfo{author}{J.~Dhillon}, \bibinfo{author}{S.~A. Rahman},
  \bibinfo{author}{S.~U. Ahmad}, \bibinfo{author}{M.~J. Hossain},
\newblock \bibinfo{title}{Peak electricity load forecasting using online
  support vector regression},
\newblock in: \bibinfo{booktitle}{2016 IEEE Canadian Conference on Electrical
  and Computer Engineering (CCECE)}, \bibinfo{organization}{IEEE},
  \bibinfo{year}{2016}, pp. \bibinfo{pages}{1--4}.
\bibitem[{{Zeng} and {Jin}(2018)}]{115-8319611}
\bibinfo{author}{P.~{Zeng}}, \bibinfo{author}{M.~{Jin}},
\newblock \bibinfo{title}{Peak load forecasting based on multi-source data and
  day-to-day topological network},
\newblock \bibinfo{journal}{IET Generation, Transmission Distribution}
  \bibinfo{volume}{12} (\bibinfo{year}{2018}) \bibinfo{pages}{1374--1381}.
\bibitem[{Polikar(2012)}]{Ensemble-Polikar:2009}
\bibinfo{author}{R.~Polikar},
\newblock \bibinfo{title}{Ensemble learning},
\newblock in: \bibinfo{booktitle}{Ensemble machine learning},
  \bibinfo{publisher}{Springer}, \bibinfo{year}{2012}, pp.
  \bibinfo{pages}{1--34}.
\bibitem[{Ahmad and Chen(2018)}]{boosting2018-AHMAD20181008}
\bibinfo{author}{T.~Ahmad}, \bibinfo{author}{H.~Chen},
\newblock \bibinfo{title}{Potential of three variant machine-learning models
  for forecasting district level medium-term and long-term energy demand in
  smart grid environment},
\newblock \bibinfo{journal}{Energy} \bibinfo{volume}{160}
  (\bibinfo{year}{2018}) \bibinfo{pages}{1008--1020}.
\bibitem[{Zhang et~al.(2019)Zhang, Li, Zou, and
  Quiring}]{boosting2019-ZHANG2019116358}
\bibinfo{author}{N.~Zhang}, \bibinfo{author}{Z.~Li}, \bibinfo{author}{X.~Zou},
  \bibinfo{author}{S.~M. Quiring},
\newblock \bibinfo{title}{Comparison of three short-term load forecast models
  in southern california},
\newblock \bibinfo{journal}{Energy} \bibinfo{volume}{189}
  (\bibinfo{year}{2019}) \bibinfo{pages}{116358}.
\bibitem[{Lu et~al.(2020)Lu, Cheng, Ma, and Hu}]{boosting2020-LU2020117756}
\bibinfo{author}{H.~Lu}, \bibinfo{author}{F.~Cheng}, \bibinfo{author}{X.~Ma},
  \bibinfo{author}{G.~Hu},
\newblock \bibinfo{title}{Short-term prediction of building energy consumption
  employing an improved extreme gradient boosting model: a case study of an
  intake tower},
\newblock \bibinfo{journal}{Energy} \bibinfo{volume}{203}
  (\bibinfo{year}{2020}) \bibinfo{pages}{117756}.
\bibitem[{Rumelhart et~al.(1988)Rumelhart, Hinton, and
  Williams}]{bagging2018-53}
\bibinfo{author}{D.~E. Rumelhart}, \bibinfo{author}{G.~E. Hinton},
  \bibinfo{author}{R.~J. Williams}, \bibinfo{title}{Learning Internal
  Representations by Error Propagation}, \bibinfo{publisher}{MIT Press},
  \bibinfo{year}{1988}.
\bibitem[{{de Oliveira} and {Cyrino
  Oliveira}(2018)}]{bagging2018-DEOLIVEIRA2018776}
\bibinfo{author}{E.~M. {de Oliveira}}, \bibinfo{author}{F.~L. {Cyrino
  Oliveira}},
\newblock \bibinfo{title}{Forecasting mid-long term electric energy consumption
  through bagging arima and exponential smoothing methods},
\newblock \bibinfo{journal}{Energy} \bibinfo{volume}{144}
  (\bibinfo{year}{2018}) \bibinfo{pages}{776--788}.
\bibitem[{Fan et~al.(2014)Fan, Xiao, and Wang}]{97newd-FAN20141}
\bibinfo{author}{C.~Fan}, \bibinfo{author}{F.~Xiao}, \bibinfo{author}{S.~Wang},
\newblock \bibinfo{title}{Development of prediction models for next-day
  building energy consumption and peak power demand using data mining
  techniques},
\newblock \bibinfo{journal}{Applied Energy} \bibinfo{volume}{127}
  (\bibinfo{year}{2014}) \bibinfo{pages}{1--10}.
\bibitem[{Wang et~al.(2018)Wang, Wang, Zeng, Srinivasan, and
  Ahrentzen}]{RF2018-WANG201811}
\bibinfo{author}{Z.~Wang}, \bibinfo{author}{Y.~Wang},
  \bibinfo{author}{R.~Zeng}, \bibinfo{author}{R.~S. Srinivasan},
  \bibinfo{author}{S.~Ahrentzen},
\newblock \bibinfo{title}{Random forest based hourly building energy
  prediction},
\newblock \bibinfo{journal}{Energy and Buildings} \bibinfo{volume}{171}
  (\bibinfo{year}{2018}) \bibinfo{pages}{11--25}.
\bibitem[{Cartina et~al.(2000)Cartina, Alexandrescu, Grigoras, and Moshe}]{29}
\bibinfo{author}{G.~Cartina}, \bibinfo{author}{V.~Alexandrescu},
  \bibinfo{author}{G.~Grigoras}, \bibinfo{author}{M.~Moshe},
\newblock \bibinfo{title}{Peak load estimation in distribution networks by
  fuzzy regression approach},
\newblock in: \bibinfo{booktitle}{2000 10th Mediterranean Electrotechnical
  Conference. Information Technology and Electrotechnology for the
  Mediterranean Countries. Proceedings. MeleCon 2000 (Cat. No. 00CH37099)},
  volume~\bibinfo{volume}{3}, \bibinfo{organization}{IEEE},
  \bibinfo{year}{2000}, pp. \bibinfo{pages}{907--910}.
\bibitem[{Kato et~al.(2002)Kato, Yukita, Goto, and Ichiyanagi}]{34}
\bibinfo{author}{S.~Kato}, \bibinfo{author}{K.~Yukita},
  \bibinfo{author}{Y.~Goto}, \bibinfo{author}{K.~Ichiyanagi},
\newblock \bibinfo{title}{Study of daily peak load forecasting by structured
  representation on genetic algorithms for function fitting},
\newblock in: \bibinfo{booktitle}{IEEE/PES Transmission and Distribution
  Conference and Exhibition}, volume~\bibinfo{volume}{3},
  \bibinfo{organization}{IEEE}, \bibinfo{year}{2002}, pp.
  \bibinfo{pages}{1686--1690}.
\bibitem[{Gavrilas et~al.(2005)Gavrilas, Sfintes, and
  Ivanov}]{50-2005Comparison}
\bibinfo{author}{M.~Gavrilas}, \bibinfo{author}{C.~V. Sfintes},
  \bibinfo{author}{O.~Ivanov},
\newblock \bibinfo{title}{Comparison of neural and evolutionary approaches to
  peak load estimation in distribution systems},
\newblock in: \bibinfo{booktitle}{EUROCON 2005-The International Conference on"
  Computer as a Tool"}, volume~\bibinfo{volume}{2},
  \bibinfo{organization}{IEEE}, \bibinfo{year}{2005}, pp.
  \bibinfo{pages}{1461--1464}.
\bibitem[{{Torkzadeh} et~al.(2014){Torkzadeh}, {Mirzaei}, {Mirjalili},
  {Anaraki}, {Sehhati}, and {Behdad}}]{92-6867500}
\bibinfo{author}{R.~{Torkzadeh}}, \bibinfo{author}{A.~{Mirzaei}},
  \bibinfo{author}{M.~M. {Mirjalili}}, \bibinfo{author}{A.~S. {Anaraki}},
  \bibinfo{author}{M.~R. {Sehhati}}, \bibinfo{author}{F.~{Behdad}},
\newblock \bibinfo{title}{Medium term load forecasting in distribution systems
  based on multi linear regression principal component analysis: A novel
  approach},
\newblock in: \bibinfo{booktitle}{2014 19th Conference on Electrical Power
  Distribution Networks (EPDC)}, \bibinfo{year}{2014}, pp.
  \bibinfo{pages}{66--70}.
\bibitem[{Laouafi et~al.(2016)Laouafi, Mordjaoui, Laouafi, and
  Boukelia}]{108newd-Laouafi2016Daily}
\bibinfo{author}{A.~Laouafi}, \bibinfo{author}{M.~Mordjaoui},
  \bibinfo{author}{F.~Laouafi}, \bibinfo{author}{T.~E. Boukelia},
\newblock \bibinfo{title}{Daily peak electricity demand forecasting based on an
  adaptive hybrid two-stage methodology},
\newblock \bibinfo{journal}{International Journal of Electrical Power \& Energy
  Systems} \bibinfo{volume}{77} (\bibinfo{year}{2016})
  \bibinfo{pages}{136--144}.
\bibitem[{Lee et~al.(2010)Lee, Wang, Yi-Yu, Tai, and Shi}]{72-0The}
\bibinfo{author}{M.-R. Lee}, \bibinfo{author}{S.-J. Wang},
  \bibinfo{author}{L.~Yi-Yu}, \bibinfo{author}{L.-I. Tai},
  \bibinfo{author}{H.-J. Shi},
\newblock \bibinfo{title}{The maximum power demand forecasting with fuzzy
  theory},
\newblock in: \bibinfo{booktitle}{2010 International Symposium on Computer,
  Communication, Control and Automation (3CA)}, volume~\bibinfo{volume}{2},
  \bibinfo{organization}{IEEE}, \bibinfo{year}{2010}, pp.
  \bibinfo{pages}{419--422}.
\bibitem[{Carpinteiro et~al.(2005)Carpinteiro, Leme, de~Souza
  et~al.}]{48-1556396}
\bibinfo{author}{O.~A. Carpinteiro}, \bibinfo{author}{R.~C. Leme},
  \bibinfo{author}{A.~de~Souza}, et~al.,
\newblock \bibinfo{title}{A hierarchical hybrid neural model with time
  integrators in long-term peak-load forecasting},
\newblock in: \bibinfo{booktitle}{Proceedings. 2005 IEEE International Joint
  Conference on Neural Networks, 2005.}, volume~\bibinfo{volume}{5},
  \bibinfo{organization}{IEEE}, \bibinfo{year}{2005}, pp.
  \bibinfo{pages}{2960--2965}.
\bibitem[{{Ivanov} and {Gavrilaş}(2014)}]{98-7011462}
\bibinfo{author}{O.~{Ivanov}}, \bibinfo{author}{M.~{Gavrilaş}},
\newblock \bibinfo{title}{Multilayer perceptron architecture optimization for
  peak load estimation},
\newblock in: \bibinfo{booktitle}{12th Symposium on Neural Network Applications
  in Electrical Engineering (NEUREL)}, \bibinfo{year}{2014}, pp.
  \bibinfo{pages}{67--72}.
\bibitem[{Yu et~al.(2013)Yu, Lee, Jeong, and Kim}]{89-2014Linguistic}
\bibinfo{author}{J.~Yu}, \bibinfo{author}{H.~Lee}, \bibinfo{author}{Y.~Jeong},
  \bibinfo{author}{S.~Kim},
\newblock \bibinfo{title}{Linguistic fuzzy modeling approach for daily peak
  load forecasting},
\newblock in: \bibinfo{booktitle}{2013 International Conference on Fuzzy Theory
  and Its Applications (iFUZZY)}, \bibinfo{organization}{IEEE},
  \bibinfo{year}{2013}, pp. \bibinfo{pages}{116--121}.
\bibitem[{Sarduy et~al.(2016)Sarduy, Santo, and
  Saidel}]{108newl-Julio2016Linear}
\bibinfo{author}{J.~R.~G. Sarduy}, \bibinfo{author}{K.~G.~D. Santo},
  \bibinfo{author}{M.~A. Saidel},
\newblock \bibinfo{title}{Linear and non-linear methods for prediction of peak
  load at university of são paulo},
\newblock \bibinfo{journal}{Measurement} \bibinfo{volume}{78}
  (\bibinfo{year}{2016}) \bibinfo{pages}{187--201}.
\bibitem[{Escolano et~al.(2009)Escolano, Suau, and
  Bonev}]{FST-escolano2009feature}
\bibinfo{author}{F.~Escolano}, \bibinfo{author}{P.~Suau},
  \bibinfo{author}{B.~Bonev},
\newblock \bibinfo{title}{Feature selection and transformation},
\newblock \bibinfo{journal}{Information Theory in Computer Vision and Pattern
  Recognition}  (\bibinfo{year}{2009}) \bibinfo{pages}{211--269}.
\bibitem[{Dash and Liu(1997)}]{fs-DASH1997131}
\bibinfo{author}{M.~Dash}, \bibinfo{author}{H.~Liu},
\newblock \bibinfo{title}{Feature selection for classification},
\newblock \bibinfo{journal}{Intelligent data analysis} \bibinfo{volume}{1}
  (\bibinfo{year}{1997}) \bibinfo{pages}{131--156}.
\bibitem[{Dai and Meng(2020)}]{dai2020energy}
\bibinfo{author}{S.~Dai}, \bibinfo{author}{F.~Meng},
\newblock \bibinfo{title}{Energy forecasting with building characteristics
  analysis},
\newblock in: \bibinfo{booktitle}{2020 International Joint Conference on Neural
  Networks (IJCNN)}, \bibinfo{organization}{IEEE}, \bibinfo{year}{2020}, pp.
  \bibinfo{pages}{1--7}.
\bibitem[{Fu et~al.(2018)Fu, Zeng, Feng, and Cai}]{3-2008A}
\bibinfo{author}{X.~Fu}, \bibinfo{author}{X.-J. Zeng},
  \bibinfo{author}{P.~Feng}, \bibinfo{author}{X.~Cai},
\newblock \bibinfo{title}{Clustering-based short-term load forecasting for
  residential electricity under the increasing-block pricing tariffs in china},
\newblock \bibinfo{journal}{Energy} \bibinfo{volume}{165}
  (\bibinfo{year}{2018}) \bibinfo{pages}{76--89}.
\bibitem[{Jain et~al.(1999)Jain, Murty, and Flynn}]{4-2013The}
\bibinfo{author}{A.~K. Jain}, \bibinfo{author}{M.~N. Murty},
  \bibinfo{author}{P.~J. Flynn},
\newblock \bibinfo{title}{Data clustering: a review},
\newblock \bibinfo{journal}{ACM computing surveys (CSUR)} \bibinfo{volume}{31}
  (\bibinfo{year}{1999}) \bibinfo{pages}{264--323}.
\bibitem[{Salimi-Beni et~al.(2006)Salimi-Beni, Farrokhzad, Fotuhi-Firuzabad,
  and Alemohammad}]{53-2007A}
\bibinfo{author}{A.~Salimi-Beni}, \bibinfo{author}{D.~Farrokhzad},
  \bibinfo{author}{M.~Fotuhi-Firuzabad}, \bibinfo{author}{S.~Alemohammad},
\newblock \bibinfo{title}{A new approach to determine base, intermediate and
  peak-demand in an electric power system},
\newblock in: \bibinfo{booktitle}{2006 International Conference on Power System
  Technology}, \bibinfo{organization}{IEEE}, \bibinfo{year}{2006}, pp.
  \bibinfo{pages}{1--5}.
\bibitem[{Jin et~al.(2006)Jin, Feng, and Jilai}]{55-2006Peak}
\bibinfo{author}{L.~Jin}, \bibinfo{author}{Y.~Feng},
  \bibinfo{author}{Y.~Jilai},
\newblock \bibinfo{title}{Peak load forecasting using hierarchical clustering
  and rprop neural network},
\newblock in: \bibinfo{booktitle}{2006 IEEE PES Power Systems Conference and
  Exposition}, \bibinfo{organization}{IEEE}, \bibinfo{year}{2006}, pp.
  \bibinfo{pages}{1535--1540}.
\bibitem[{Jin et~al.(2007)Jin, Jilai, and Zhuo}]{58-2007Load}
\bibinfo{author}{L.~Jin}, \bibinfo{author}{Y.~Jilai},
  \bibinfo{author}{L.~Zhuo},
\newblock \bibinfo{title}{Load forecast for system operation in peak load
  period},
\newblock in: \bibinfo{booktitle}{2007 IEEE Power Engineering Society General
  Meeting}, \bibinfo{organization}{IEEE}, \bibinfo{year}{2007}, pp.
  \bibinfo{pages}{1--6}.
\bibitem[{Jin et~al.(2004)Jin, Lai, and Long}]{41-2004Peak}
\bibinfo{author}{L.~Jin}, \bibinfo{author}{Y.~J. Lai}, \bibinfo{author}{T.~X.
  Long},
\newblock \bibinfo{title}{Peak load forecasting based on robust regression
  model},
\newblock in: \bibinfo{booktitle}{2004 International Conference on
  Probabilistic Methods Applied to Power Systems},
  \bibinfo{organization}{IEEE}, \bibinfo{year}{2004}, pp.
  \bibinfo{pages}{123--128}.
\bibitem[{Jin et~al.(2005)Jin, Ziyang, Jingbo, and Xinying}]{49-2005An}
\bibinfo{author}{L.~Jin}, \bibinfo{author}{L.~Ziyang},
  \bibinfo{author}{S.~Jingbo}, \bibinfo{author}{S.~Xinying},
\newblock \bibinfo{title}{An efficient method for peak load forecasting},
\newblock in: \bibinfo{booktitle}{2005 International Power Engineering
  Conference}, \bibinfo{organization}{IEEE}, \bibinfo{year}{2005}, pp.
  \bibinfo{pages}{1--52}.
\bibitem[{Ding and He(2002)}]{Cluster}
\bibinfo{author}{C.~Ding}, \bibinfo{author}{X.~He},
\newblock \bibinfo{title}{Cluster merging and splitting in hierarchical
  clustering algorithms},
\newblock in: \bibinfo{booktitle}{2002 IEEE International Conference on Data
  Mining, 2002. Proceedings.}, \bibinfo{organization}{IEEE},
  \bibinfo{year}{2002}, pp. \bibinfo{pages}{139--146}.
\bibitem[{Bezdek(2013)}]{bezdek2013pattern}
\bibinfo{author}{J.~C. Bezdek}, \bibinfo{title}{Pattern recognition with fuzzy
  objective function algorithms}, \bibinfo{publisher}{Springer Science \&
  Business Media}, \bibinfo{year}{2013}.
\bibitem[{Xu et~al.(2019)Xu, Wang, and Tang}]{ProbabilisticLFFB}
\bibinfo{author}{L.~Xu}, \bibinfo{author}{S.~Wang}, \bibinfo{author}{R.~Tang},
\newblock \bibinfo{title}{Probabilistic load forecasting for buildings
  considering weather forecasting uncertainty and uncertain peak load},
\newblock \bibinfo{journal}{Applied Energy} \bibinfo{volume}{237}
  (\bibinfo{year}{2019}) \bibinfo{pages}{180--195}.
\bibitem[{Stetson and Stark(1988)}]{6}
\bibinfo{author}{L.~E. Stetson}, \bibinfo{author}{G.~L. Stark},
\newblock \bibinfo{title}{Peak electrical demands of individuals and groups of
  rural residential customers},
\newblock \bibinfo{journal}{IEEE Transactions on Industry Applications}
  \bibinfo{volume}{24} (\bibinfo{year}{1988}) \bibinfo{pages}{772--776}.
\bibitem[{Huang et~al.(2012)Huang, Li, and Liu}]{83-6345263}
\bibinfo{author}{J.~Huang}, \bibinfo{author}{Y.~Li}, \bibinfo{author}{Y.~Liu},
\newblock \bibinfo{title}{Summer daily peak load forecasting considering
  accumulation effect and abrupt change of temperature},
\newblock in: \bibinfo{booktitle}{2012 IEEE Power and Energy Society General
  Meeting}, \bibinfo{organization}{IEEE}, \bibinfo{year}{2012}, pp.
  \bibinfo{pages}{1--4}.
\bibitem[{Ploysuwan(2014)}]{96-7022373}
\bibinfo{author}{T.~Ploysuwan},
\newblock \bibinfo{title}{Spectral mixture kernel for pattern discovery and
  time series forecasting of electricity peak load},
\newblock in: \bibinfo{booktitle}{TENCON 2014-2014 IEEE Region 10 Conference},
  \bibinfo{organization}{IEEE}, \bibinfo{year}{2014}, pp.
  \bibinfo{pages}{1--5}.
\bibitem[{Ogihara and Urano(2019)}]{130-8810766}
\bibinfo{author}{K.~Ogihara}, \bibinfo{author}{S.~Urano},
\newblock \bibinfo{title}{A study of risk reduction for daily peak load demand
  forecasting},
\newblock in: \bibinfo{booktitle}{2019 IEEE Milan PowerTech},
  \bibinfo{organization}{IEEE}, \bibinfo{year}{2019}, pp.
  \bibinfo{pages}{1--6}.
\bibitem[{Tairen et~al.(2018)Tairen, Lehr, Olga, and
  Manel}]{118-2018Distribution}
\bibinfo{author}{C.~Tairen}, \bibinfo{author}{J.~M. Lehr},
  \bibinfo{author}{L.~Olga}, \bibinfo{author}{M.-R. Manel},
\newblock \bibinfo{title}{Distribution feeder-level day-ahead peak load
  forecasting methods and comparative study},
\newblock \bibinfo{journal}{IET Generation Transmission \& Distribution}
  \bibinfo{volume}{12} (\bibinfo{year}{2018}) \bibinfo{pages}{3270--3278}.
\bibitem[{Ploysuwan et~al.(2014)Ploysuwan, Atsawathawichok, and
  Teekaput}]{93-2014Peak}
\bibinfo{author}{T.~Ploysuwan}, \bibinfo{author}{P.~Atsawathawichok},
  \bibinfo{author}{P.~Teekaput},
\newblock \bibinfo{title}{Peak load forecasting of electricity generating
  authority of thailand by gaussian process},
\newblock in: \bibinfo{booktitle}{2014 International Electrical Engineering
  Congress (iEECON)}, \bibinfo{organization}{IEEE}, \bibinfo{year}{2014}, pp.
  \bibinfo{pages}{1--4}.
\bibitem[{{Shabbir} et~al.(2018){Shabbir}, {Ali}, {Liang}, and
  {Chowdhury}}]{124-8671682}
\bibinfo{author}{M.~N. S.~K. {Shabbir}}, \bibinfo{author}{M.~Z. {Ali}},
  \bibinfo{author}{X.~{Liang}}, \bibinfo{author}{M.~S.~A. {Chowdhury}},
\newblock \bibinfo{title}{A probabilistic approach considering contingency
  parameters for peak load demand forecasting},
\newblock \bibinfo{journal}{Canadian Journal of Electrical and Computer
  Engineering} \bibinfo{volume}{41} (\bibinfo{year}{2018})
  \bibinfo{pages}{224--233}.
\bibitem[{Bichpuriya et~al.(2016)Bichpuriya, Soman, and
  Subramanyam}]{106-2016Robust}
\bibinfo{author}{Y.~K. Bichpuriya}, \bibinfo{author}{S.~Soman},
  \bibinfo{author}{A.~Subramanyam},
\newblock \bibinfo{title}{Robust probability density forecasts of yearly peak
  load using non-parametric model},
\newblock in: \bibinfo{booktitle}{2016 IEEE Power and Energy Society General
  Meeting (PESGM)}, \bibinfo{organization}{IEEE}, \bibinfo{year}{2016}, pp.
  \bibinfo{pages}{1--5}.
\bibitem[{Khorsheed(2018)}]{117-0Long}
\bibinfo{author}{E.~Khorsheed},
\newblock \bibinfo{title}{Long-term energy peak load forecasting models: A
  hybrid statistical approach},
\newblock in: \bibinfo{booktitle}{2018 Advances in Science and Engineering
  Technology International Conferences (ASET)}, \bibinfo{organization}{IEEE},
  \bibinfo{year}{2018}, pp. \bibinfo{pages}{1--6}.
\bibitem[{Wang and Cao(2006)}]{54-2006Mutual}
\bibinfo{author}{Z.~Wang}, \bibinfo{author}{Y.~Cao},
\newblock \bibinfo{title}{Mutual information and non-fixed anns for daily peak
  load forecasting},
\newblock in: \bibinfo{booktitle}{2006 IEEE PES Power Systems Conference and
  Exposition}, \bibinfo{organization}{IEEE}, \bibinfo{year}{2006}, pp.
  \bibinfo{pages}{1523--1527}.
\bibitem[{{Gajowniczek} et~al.(2017){Gajowniczek}, {Nafkha}, and
  {Zabkowski}}]{112-8104558}
\bibinfo{author}{K.~{Gajowniczek}}, \bibinfo{author}{R.~{Nafkha}},
  \bibinfo{author}{T.~{Zabkowski}},
\newblock \bibinfo{title}{Electricity peak demand classification with
  artificial neural networks},
\newblock in: \bibinfo{booktitle}{2017 Federated Conference on Computer Science
  and Information Systems (FedCSIS)}, \bibinfo{year}{2017}, pp.
  \bibinfo{pages}{307--315}.
\bibitem[{Gajowniczek et~al.(2018)Gajowniczek, Nafkha, and
  Zabkowski}]{120-8955339}
\bibinfo{author}{K.~Gajowniczek}, \bibinfo{author}{R.~Nafkha},
  \bibinfo{author}{T.~Zabkowski},
\newblock \bibinfo{title}{Seasonal peak demand classification with machine
  learning techniques},
\newblock in: \bibinfo{booktitle}{2018 International Conference on Applied
  Mathematics Computer Science (ICAMCS)}, \bibinfo{year}{2018}, pp.
  \bibinfo{pages}{101--1014}.
\bibitem[{{Chemetova} et~al.(2019){Chemetova}, {Santos}, and
  {Pires}}]{131-8927078}
\bibinfo{author}{S.~{Chemetova}}, \bibinfo{author}{P.~{Santos}},
  \bibinfo{author}{A.~J. {Pires}},
\newblock \bibinfo{title}{Peak load forecasting in electrical deregulated
  market environment - the dynamic tariffs},
\newblock in: \bibinfo{booktitle}{IECON 2019 - 45th Annual Conference of the
  IEEE Industrial Electronics Society}, \bibinfo{year}{2019}, pp.
  \bibinfo{pages}{2227--2232}.
\bibitem[{Park and Mohammed(1991)}]{11-1991Artificial}
\bibinfo{author}{D.~C. Park}, \bibinfo{author}{O.~Mohammed},
\newblock \bibinfo{title}{Artificial neural network based electric peak load
  forecasting},
\newblock in: \bibinfo{booktitle}{IEEE Proceedings of the SOUTHEASTCON'91},
  \bibinfo{organization}{IEEE}, \bibinfo{year}{1991}, pp.
  \bibinfo{pages}{225--228}.
\bibitem[{Lee and Cha(1992)}]{12s-1992Short}
\bibinfo{author}{K.~Y. Lee}, \bibinfo{author}{Y.~T. Cha},
\newblock \bibinfo{title}{Short-term load forecasting using an artificial
  neural network},
\newblock \bibinfo{journal}{IEEE Transactions on Power Systems}
  \bibinfo{volume}{7} (\bibinfo{year}{1992}) \bibinfo{pages}{124--132}.
\bibitem[{Drezga and Rahman(1999)}]{28s-Drezga1999Short}
\bibinfo{author}{I.~Drezga}, \bibinfo{author}{S.~Rahman},
\newblock \bibinfo{title}{Short-term load forecasting with local ann
  predictors},
\newblock \bibinfo{journal}{IEEE Transactions on Power Systems}
  \bibinfo{volume}{14} (\bibinfo{year}{1999}) \bibinfo{pages}{844--850}.
\bibitem[{Aboul-Magd et~al.(2001)}]{31-Aboulmagd2001An}
\bibinfo{author}{M.~A. Aboul-Magd}, et~al.,
\newblock \bibinfo{title}{An artificial neural network model for electrical
  daily peak load forecasting with an adjustment for holidays},
\newblock in: \bibinfo{booktitle}{LESCOPE 01. 2001 Large Engineering Systems
  Conference on Power Engineering. Conference Proceedings. Theme: Powering
  Beyond 2001 (Cat. No. 01ex490)}, \bibinfo{organization}{IEEE},
  \bibinfo{year}{2001}, pp. \bibinfo{pages}{105--113}.
\bibitem[{Matsui et~al.(2001)Matsui, Iizaka, and Fukuyama}]{32-Matsui2002Peak}
\bibinfo{author}{T.~Matsui}, \bibinfo{author}{T.~Iizaka},
  \bibinfo{author}{Y.~Fukuyama},
\newblock \bibinfo{title}{Peak load forecasting using analyzable structured
  neural network},
\newblock in: \bibinfo{booktitle}{2001 IEEE Power Engineering Society Winter
  Meeting. Conference Proceedings (Cat. No. 01CH37194)},
  volume~\bibinfo{volume}{2}, \bibinfo{organization}{IEEE},
  \bibinfo{year}{2001}, pp. \bibinfo{pages}{405--410}.
\bibitem[{Tasre et~al.(2011)Tasre, Bedekar, and Ghate}]{82-2011Daily}
\bibinfo{author}{M.~B. Tasre}, \bibinfo{author}{P.~P. Bedekar},
  \bibinfo{author}{V.~N. Ghate},
\newblock \bibinfo{title}{Daily peak load forecasting using ann},
\newblock in: \bibinfo{booktitle}{2011 Nirma University International
  Conference on Engineering}, \bibinfo{organization}{IEEE},
  \bibinfo{year}{2011}, pp. \bibinfo{pages}{1--6}.
\bibitem[{Milojkovi{\'c} et~al.(2012)Milojkovi{\'c}, Litovski, and
  Litovski}]{84-2012ANN}
\bibinfo{author}{J.~Milojkovi{\'c}}, \bibinfo{author}{I.~Litovski},
  \bibinfo{author}{V.~Litovski},
\newblock \bibinfo{title}{Ann application for the next day peak electricity
  load prediction},
\newblock in: \bibinfo{booktitle}{11th Symposium on Neural Network Applications
  in Electrical Engineering}, \bibinfo{organization}{IEEE},
  \bibinfo{year}{2012}, pp. \bibinfo{pages}{237--241}.
\bibitem[{Chemetova et~al.(2016)Chemetova, Santos, and
  Ventim-Neves}]{105-2016Load}
\bibinfo{author}{S.~Chemetova}, \bibinfo{author}{P.~Santos},
  \bibinfo{author}{M.~Ventim-Neves},
\newblock \bibinfo{title}{Load peak forecasting in different load patterns
  situations},
\newblock in: \bibinfo{booktitle}{2016 10th International Conference on
  Compatibility, Power Electronics and Power Engineering (CPE-POWERENG)},
  \bibinfo{organization}{IEEE}, \bibinfo{year}{2016}, pp.
  \bibinfo{pages}{148--151}.
\bibitem[{Jarndal and Hamdan(2017)}]{110-2017Forecasting}
\bibinfo{author}{A.~Jarndal}, \bibinfo{author}{S.~Hamdan},
\newblock \bibinfo{title}{Forecasting of peak electricity demand using annga
  and ann-pso approaches},
\newblock in: \bibinfo{booktitle}{2017 7th International Conference on
  Modeling, Simulation, and Applied Optimization (ICMSAO)},
  \bibinfo{organization}{IEEE}, \bibinfo{year}{2017}, pp.
  \bibinfo{pages}{1--5}.
\bibitem[{Leak and Venayagamoorthy(2018)}]{123-2018Forecasting}
\bibinfo{author}{M.~H. Leak}, \bibinfo{author}{G.~K. Venayagamoorthy},
\newblock \bibinfo{title}{Forecasting peak daily load in distribution feeders},
\newblock in: \bibinfo{booktitle}{2018 Clemson University Power Systems
  Conference (PSC)}, \bibinfo{organization}{IEEE}, \bibinfo{year}{2018}, pp.
  \bibinfo{pages}{1--8}.
\bibitem[{Onoda(1992)}]{15-Onoda1993Next}
\bibinfo{author}{T.~Onoda},
\newblock \bibinfo{title}{Next day's peak load forecasting using an artificial
  neural network},
\newblock in: \bibinfo{booktitle}{[1993] Proceedings of the Second
  International Forum on Applications of Neural Networks to Power Systems},
  \bibinfo{organization}{IEEE}, \bibinfo{year}{1992}, pp.
  \bibinfo{pages}{284--289}.
\bibitem[{Morioka et~al.(1992)Morioka, Sakurai, Yokoyama, and
  Sekine}]{16-Morioka1993Next}
\bibinfo{author}{Y.~Morioka}, \bibinfo{author}{K.~Sakurai},
  \bibinfo{author}{A.~Yokoyama}, \bibinfo{author}{Y.~Sekine},
\newblock \bibinfo{title}{Next day peak load forecasting using a multilayer
  neural network with an additional learning},
\newblock in: \bibinfo{booktitle}{[1993] Proceedings of the Second
  International Forum on Applications of Neural Networks to Power Systems},
  \bibinfo{organization}{IEEE}, \bibinfo{year}{1992}, pp.
  \bibinfo{pages}{60--65}.
\bibitem[{Mizukami and Nishimori(1992)}]{17-Mizukami1993Maximum}
\bibinfo{author}{Y.~Mizukami}, \bibinfo{author}{T.~Nishimori},
\newblock \bibinfo{title}{Maximum electric power demand prediction by neural
  network},
\newblock in: \bibinfo{booktitle}{[1993] Proceedings of the Second
  International Forum on Applications of Neural Networks to Power Systems},
  \bibinfo{organization}{IEEE}, \bibinfo{year}{1992}, pp.
  \bibinfo{pages}{296--301}.
\bibitem[{Abdellah and Djamel(2013)}]{88-2013Forecasting}
\bibinfo{author}{D.~Abdellah}, \bibinfo{author}{L.~Djamel},
\newblock \bibinfo{title}{Forecasting the algerian load peak profile using time
  series model based on backpropagation neural networks},
\newblock in: \bibinfo{booktitle}{4th International Conference on Power
  Engineering, Energy and Electrical Drives}, \bibinfo{organization}{IEEE},
  \bibinfo{year}{2013}, pp. \bibinfo{pages}{1734--1737}.
\bibitem[{{Takiyar}(2015)}]{101-7359274}
\bibinfo{author}{S.~{Takiyar}},
\newblock \bibinfo{title}{Grid reliability enhancement by peak load forecasting
  with a pso hybridized ann model},
\newblock in: \bibinfo{booktitle}{2015 4th International Conference on
  Reliability, Infocom Technologies and Optimization (ICRITO) (Trends and
  Future Directions)}, \bibinfo{year}{2015}, pp. \bibinfo{pages}{1--6}.
\bibitem[{Carpinteiro et~al.(2006)Carpinteiro, Leme, Souza, Pinheiro, and
  Moreira}]{57aal-Otavio2007Long}
\bibinfo{author}{O.~A.~S. Carpinteiro}, \bibinfo{author}{R.~C. Leme},
  \bibinfo{author}{A.~C. Z.~D. Souza}, \bibinfo{author}{C.~A.~M. Pinheiro},
  \bibinfo{author}{E.~M. Moreira},
\newblock \bibinfo{title}{Long-term load forecasting via a hierarchical neural
  model with time integrators},
\newblock \bibinfo{journal}{Electric Power Systems Research}
  \bibinfo{volume}{77} (\bibinfo{year}{2006}) \bibinfo{pages}{371--378}.
\bibitem[{Phimphachanh et~al.(2004)Phimphachanh, Chammongthai, Kumhom, and
  Sangswang}]{45-1414771}
\bibinfo{author}{S.~Phimphachanh}, \bibinfo{author}{K.~Chammongthai},
  \bibinfo{author}{P.~Kumhom}, \bibinfo{author}{A.~Sangswang},
\newblock \bibinfo{title}{Using neural network for long term peak load
  forecasting in vientiane municipality},
\newblock in: \bibinfo{booktitle}{2004 IEEE Region 10 Conference TENCON 2004.},
  volume \bibinfo{volume}{100}, \bibinfo{organization}{IEEE},
  \bibinfo{year}{2004}, pp. \bibinfo{pages}{319--322}.
\bibitem[{Kandil et~al.(2002)Kandil, El-Debeiky, and
  Hasanien}]{36newL-2002Long}
\bibinfo{author}{M.~S. Kandil}, \bibinfo{author}{S.~M. El-Debeiky},
  \bibinfo{author}{N.~E. Hasanien},
\newblock \bibinfo{title}{Long-term load forecasting for fast developing
  utility using a knowledge-based expert system},
\newblock \bibinfo{journal}{IEEE Power Engineering Review} \bibinfo{volume}{22}
  (\bibinfo{year}{2002}) \bibinfo{pages}{78--78}.
\bibitem[{{Rashidi} et~al.(2003){Rashidi}, {Rashidi}, and
  {Monavar}}]{38-1244703}
\bibinfo{author}{M.~{Rashidi}}, \bibinfo{author}{F.~{Rashidi}},
  \bibinfo{author}{H.~{Monavar}},
\newblock \bibinfo{title}{Peak load forecasting in power systems using
  emotional learning based fuzzy logic},
\newblock in: \bibinfo{booktitle}{SMC'03 Conference Proceedings. 2003 IEEE
  International Conference on Systems, Man and Cybernetics. Conference Theme -
  System Security and Assurance (Cat. No.03CH37483)}, \bibinfo{year}{2003}, pp.
  \bibinfo{pages}{1985--1988}.
\bibitem[{Jamaaluddin et~al.(2018)Jamaaluddin, Hadidjaja, Sulistiyowati,
  Suprayitno, Anshory, and Syahrorini}]{125aav-Jamaaluddin_2018}
\bibinfo{author}{J.~Jamaaluddin}, \bibinfo{author}{D.~Hadidjaja},
  \bibinfo{author}{I.~Sulistiyowati}, \bibinfo{author}{E.~Suprayitno},
  \bibinfo{author}{I.~Anshory}, \bibinfo{author}{S.~Syahrorini},
\newblock \bibinfo{title}{Very short term load forecasting peak load time using
  fuzzy logic},
\newblock in: \bibinfo{booktitle}{IOP Conference Series: Materials Science and
  Engineering}, volume \bibinfo{volume}{403}, \bibinfo{organization}{IOP
  Publishing}, \bibinfo{year}{2018}, p. \bibinfo{pages}{012070}.
\bibitem[{Panapakidis et~al.(2017)Panapakidis, Christoforidis, Asimopoulos, and
  Dagoumas}]{111-2017Combining}
\bibinfo{author}{I.~P. Panapakidis}, \bibinfo{author}{G.~C. Christoforidis},
  \bibinfo{author}{N.~Asimopoulos}, \bibinfo{author}{A.~S. Dagoumas},
\newblock \bibinfo{title}{Combining wavelet transform and support vector
  regression model for day-ahead peak load forecasting in the greek power
  system},
\newblock in: \bibinfo{booktitle}{2017 IEEE International Conference on
  Environment and Electrical Engineering and 2017 IEEE Industrial and
  Commercial Power Systems Europe (EEEIC/I\&CPS Europe)},
  \bibinfo{organization}{IEEE}, \bibinfo{year}{2017}, pp.
  \bibinfo{pages}{1--6}.
\bibitem[{{Azad} et~al.(2018){Azad}, {Uddin}, and {Takruri}}]{116-8330143}
\bibinfo{author}{M.~K. {Azad}}, \bibinfo{author}{S.~{Uddin}},
  \bibinfo{author}{M.~{Takruri}},
\newblock \bibinfo{title}{Support vector regression based electricity peak load
  forecasting},
\newblock in: \bibinfo{booktitle}{2018 11th International Symposium on
  Mechatronics and its Applications (ISMA)}, \bibinfo{year}{2018}, pp.
  \bibinfo{pages}{1--5}.
\bibitem[{Satre-Meloy et~al.(2020)Satre-Meloy, Diakonova, and
  Gr{\"u}newald}]{RF2020-SATREMELOY2020114246}
\bibinfo{author}{A.~Satre-Meloy}, \bibinfo{author}{M.~Diakonova},
  \bibinfo{author}{P.~Gr{\"u}newald},
\newblock \bibinfo{title}{Cluster analysis and prediction of residential peak
  demand profiles using occupant activity data},
\newblock \bibinfo{journal}{Applied Energy} \bibinfo{volume}{260}
  (\bibinfo{year}{2020}) \bibinfo{pages}{114246}.
\bibitem[{{Liu} and {Brown}(2019)}]{129-8881305}
\bibinfo{author}{J.~{Liu}}, \bibinfo{author}{L.~E. {Brown}},
\newblock \bibinfo{title}{Effect of forecast accuracy on day ahead prediction
  of coincident peak days},
\newblock in: \bibinfo{booktitle}{2019 IEEE Innovative Smart Grid Technologies
  - Asia (ISGT Asia)}, \bibinfo{year}{2019}, pp. \bibinfo{pages}{661--666}.
\bibitem[{Ai et~al.(2019)Ai, Chakravorty, and Rong}]{127-2019Evolutionary}
\bibinfo{author}{S.~Ai}, \bibinfo{author}{A.~Chakravorty},
  \bibinfo{author}{C.~Rong},
\newblock \bibinfo{title}{Evolutionary ensemble lstm based household peak
  demand prediction},
\newblock in: \bibinfo{booktitle}{2019 International Conference on Artificial
  Intelligence in Information and Communication (ICAIIC)},
  \bibinfo{organization}{IEEE}, \bibinfo{year}{2019}, pp.
  \bibinfo{pages}{1--6}.
\bibitem[{Chen et~al.(2016)Chen, Lehr, Lavrova, and
  Martinez-Ramonz}]{107-2016Distribution}
\bibinfo{author}{T.~Chen}, \bibinfo{author}{J.~Lehr},
  \bibinfo{author}{O.~Lavrova}, \bibinfo{author}{M.~Martinez-Ramonz},
\newblock \bibinfo{title}{Distribution-level peak load prediction based on
  bayesian additive regression trees},
\newblock in: \bibinfo{booktitle}{2016 IEEE Power and Energy Society General
  Meeting (PESGM)}, \bibinfo{organization}{IEEE}, \bibinfo{year}{2016}, pp.
  \bibinfo{pages}{1--5}.
\bibitem[{Lim et~al.(2016)Lim, Doh, and Chae}]{6-Lim2016Security}
\bibinfo{author}{J.~Lim}, \bibinfo{author}{I.~Doh}, \bibinfo{author}{K.~Chae},
\newblock \bibinfo{title}{Security system architecture for data integrity based
  on a virtual smart meter overlay in a smart grid system},
\newblock \bibinfo{journal}{Soft Computing} \bibinfo{volume}{20}
  (\bibinfo{year}{2016}) \bibinfo{pages}{1829--1840}.
\bibitem[{Banisar and Davies(1999)}]{privacy1}
\bibinfo{author}{D.~Banisar}, \bibinfo{author}{S.~Davies},
\newblock \bibinfo{title}{Global trends in privacy protection: An international
  survey of privacy, data protection, and surveillance laws and developments},
\newblock \bibinfo{journal}{J. Marshall J. Computer \& Info. L.}
  \bibinfo{volume}{18} (\bibinfo{year}{1999}) \bibinfo{pages}{1}.
\bibitem[{Hu et~al.(2011)Hu, Qiu, Li, Grant, Taylor, Mccaleb, Butler, and
  Hamner}]{privacy2}
\bibinfo{author}{F.~Hu}, \bibinfo{author}{M.~Qiu}, \bibinfo{author}{J.~Li},
  \bibinfo{author}{T.~Grant}, \bibinfo{author}{D.~Taylor},
  \bibinfo{author}{S.~Mccaleb}, \bibinfo{author}{L.~Butler},
  \bibinfo{author}{R.~Hamner},
\newblock \bibinfo{title}{A review on cloud computing: Design challenges in
  architecture and security},
\newblock \bibinfo{journal}{Journal of Computing \& Information Technology}
  \bibinfo{volume}{19} (\bibinfo{year}{2011}) \bibinfo{pages}{25--55}.
\bibitem[{Yang et~al.(2019)Yang, Liu, Chen, and Tong}]{yang2019federated}
\bibinfo{author}{Q.~Yang}, \bibinfo{author}{Y.~Liu}, \bibinfo{author}{T.~Chen},
  \bibinfo{author}{Y.~Tong},
\newblock \bibinfo{title}{Federated machine learning: Concept and
  applications},
\newblock \bibinfo{journal}{ACM Transactions on Intelligent Systems and
  Technology (TIST)} \bibinfo{volume}{10} (\bibinfo{year}{2019})
  \bibinfo{pages}{1--19}.

\end{thebibliography}





\end{document}